\documentclass[a4paper,11pt]{article}
\usepackage{jheppub} 

\usepackage{scalefnt}
\usepackage{hyperref}
\usepackage{todonotes}
\usepackage{graphics}
\usepackage{graphicx}
\usepackage[mathscr]{eucal}
\usepackage{amssymb}
\usepackage{amsmath}
\usepackage{tabls}
\usepackage{wasysym}
\usepackage{float}
\usepackage{listings}

\newcommand{\rwzh}[1]{R_{W\!Z\!#1}}
\newcommand{\mass}[1]{m_{#1}}
\newcommand{\vphi}{V\!\phi}
\newcommand{\wphi}{W\!\phi}
\newcommand{\zphi}{Z\!\phi}
\newcommand{\vh}{V\!h}
\newcommand{\wh}{W\!h}
\newcommand{\zh}{Z\!h}
\newcommand{\vH}{V\!H}
\newcommand{\wH}{W\!H}
\newcommand{\zH}{Z\!H}
\newcommand{\vHo}{V\!H^0}
\newcommand{\wHo}{W\!H^0}
\newcommand{\zHo}{Z\!H^0}

\newcommand{\wA}{W\!A}
\newcommand{\zA}{Z\!A}
\newcommand{\ggzh}{gg\zh}
\newcommand{\ggzH}{gg\zH}
\newcommand{\ggzHo}{gg\zHo}
\newcommand{\ggzA}{gg\zA}
\newcommand{\ggzphi}{gg\zphi}
\newcommand{\bbzh}{b\bar{b}\zh}
\newcommand{\bbzH}{b\bar{b}\zH}
\newcommand{\bbzHo}{b\bar{b}\zHo}
\newcommand{\bbzA}{b\bar{b}\zA}
\newcommand{\bbzphi}{b\bar{b}\zphi}
\newcommand{\lhc}{{\abbrev LHC}}
\newcommand{\sm}{{\abbrev SM}}
\newcommand{\thdm}{{\abbrev 2HDM}}
\newcommand{\mssm}{{\abbrev MSSM}}

\newcommand{\lo}{{\abbrev LO}}
\newcommand{\nlo}{{\abbrev NLO}}

\newcommand{\nnlo}{{\abbrev NNLO}}

\newcommand{\qcd}{{\abbrev QCD}}
\newcommand{\ew}{{\abbrev EW}}
\newcommand{\dy}{{\abbrev DY}}
\newcommand{\cp}{{\abbrev CP}}
\newcommand{\msbar}{\overline{\text{MS}}}

\newcommand{\abbrev}{\rm\scalefont{.9}}
\newcommand{\ptH}{p_T^H}
\newcommand{\ptphi}{p_T^\phi}

\newcommand{\mhiggs}{\mass{H}}

\newcommand{\pdf}{{\abbrev PDF}}

\newcommand{\eqn}[1]{Eq.\,(\ref{#1})}

\newcommand{\fig}[1]{Fig.\,\ref{#1}}
\newcommand{\figs}[1]{Figs.\,\ref{#1}}
\newcommand{\tab}[1]{Tab.\,\ref{#1}}
\newcommand{\sct}[1]{Section~\ref{#1}}

\newcommand{\order}[1]{{\cal O}(#1)}

\newcommand{\citere}[1]{Ref.\,\cite{#1}}
\newcommand{\citeres}[1]{Refs.\,\cite{#1}}

\newcommand{\vhnnlo}{{\tt vh@nnlo}}

\renewcommand{\thispagestyle}[1]{} 

\title{\vspace*{-7.6em}\begin{flushright}
\mbox{{\sf\small June 2014 --- WUB/13-12}}
\end{flushright}\vspace*{2em}
Higgs Strahlung at the Large Hadron Collider in the 2-Higgs-Doublet Model}

\author{Robert V. Harlander, Stefan Liebler, and Tom Zirke}

\affiliation{Fachbereich C, Bergische Universit\"at Wuppertal,
42097 Wuppertal, Germany}

\emailAdd{harlander@physik.uni-wuppertal.de}
\emailAdd{liebler@uni-wuppertal.de}
\emailAdd{t.zirke@uni-wuppertal.de}

\abstract{ We present a calculation of all relevant contributions to
  associated production of a Higgs boson with a weak gauge boson in the
  2-Higgs-Doublet Model (\thdm) at the \lhc{}, $pp \rightarrow \vphi$,
  with $\phi\in\lbrace h,H^0,A\rbrace$ and $V\in\{W,Z\}$. While for the
  $\wphi{}$ mode, this mostly amounts to a simple rescaling of the
  Standard Model (\sm{})
  cross section, the $\zphi$ cross section depends on several \thdm{}
  parameters. The ratio $\sigma^{\wphi}/\sigma^{\zphi}$, for which we
  present the currently most complete \sm{} prediction, therefore
  appears to be a sensitive probe of possible New Physics effects. We
  study its numerical dependence on the top and bottom Yukawa couplings,
  including their sign. Furthermore, we consider the $\wphi/\zphi$ ratio
  in exemplary \thdm{} scenarios and briefly address the effects in the
  boosted regime.  Analogous studies for other \thdm{} scenarios will
  become possible with an upcoming version of the program \vhnnlo{}
  which incorporates the \thdm{} effects.  }

\begin{document} 
\maketitle
\flushbottom

\section{Introduction}
\label{sec:introduction}

After the discovery of a Higgs boson at the Large Hadron Collider
(\lhc{})~\cite{Aad:2012tfa,Chatrchyan:2012ufa}, one of the top priorities
of the experiments is to measure its properties as precisely as
possible. Any deviation from the Standard Model (\sm{}) predictions
could be valuable information about a possible extended
theory.\footnote{For a recent review of theory predictions for
  Higgs physics at the \lhc, see \citere{Heinemeyer:2013tqa}.}

However, precision physics at hadron colliders like the \lhc{} typically
suffers from large experimental and theoretical uncertainties.  The
latter are dominated by the truncation of the perturbative series and
insufficient knowledge of the input parameters such as the strong
coupling and the parton density functions.  Observables where these
uncertainties are absent or suppressed are therefore highly desirable.

In this paper, we consider the Higgs Strahlung process, where a Higgs
boson is produced in association with a $W$ or a $Z$ boson. In the
region below about $\mhiggs=125$\,GeV, this was the dominant search mode
at the Tevatron, with the Higgs decaying into $b\bar b$ and the gauge
boson decaying leptonically~\cite{TEVNPH:2012ab}. The data show an
excess above the expected background which is, however, compatible with
a background fluctuation. At the \lhc{}, {\abbrev ATLAS} has searched
for a signal from Higgs Strahlung in the $H\to
WW^\star$~\cite{ATLAS-CONF-2013-075}, $H\to b\bar
b$~\cite{ATLAS-CONF-2013-079}, and $H\to$\,invisible
modes~\cite{ATLAS-CONF-2013-011}, providing an upper limit on the cross
section of 1.4 times the \sm{} expectation for $\mhiggs=125$\,GeV
(driven by the $H\to b\bar b$ search). {\abbrev CMS} observes a
2.1$\sigma$ excess at $\mhiggs=125$\,GeV in the $\vH\to Vb\bar b$ mode
which is compatible with the \sm{}
expectation~\cite{CMS-PAS-HIG-13-012}. Searches for
$H\to$\,invisible~\cite{CMS-PAS-HIG-13-018} and $H\to WW^{(*)}$
\cite{CMS-PAS-HIG-13-017} from Higgs Strahlung are compatible with the
background; the latter provides an upper limit on the cross section of
around ten times the \sm{} expectation. In summary, there are first
hints of the Higgs Strahlung process, but presumably it will only be
after the current shut-down of the \lhc{} when this process can be
firmly established.

The theory prediction for its total inclusive cross section within the
framework of the \sm{} is known through next-to-next-to-leading order
(\nnlo{}) \qcd{}~\cite{Han:1991ia,Brein:2003wg} and \nlo{} electroweak
(\ew{}) corrections~\cite{Ciccolini:2003jy}. Soft gluon resummation
effects are negligibly small compared to the fixed order \nnlo{}
result~\cite{Dawson:2012gs}, indicating an excellent convergence of the
perturbative series. Fully differential predictions for Higgs Strahlung
at \nnlo{} \qcd{}~\cite{Ferrera:2011bk} and including the \nlo{}
electro-weak effects~\cite{Denner:2011id} are also available (see also
\citere{Dittmaier:2012vm,Heinemeyer:2013tqa}). Through \nlo{} \qcd{},
the ratio of the cross sections for the $\wH$ and the $\zH$ process is
practically free from \qcd{} corrections. Clearly, this ratio is quite
insensitive to uncertainties due to the parton density functions
(\pdf{}s) as well, since both processes have very similar initial
states.

At \nnlo{}, however, the ratio of the $\wH$ to $\zH$ rate depends on the
top and bottom quark Yukawa couplings, due to additional contributions
specific to $\zH$ production. They are mostly due to gluon initiated
processes~\cite{Dicus:1988yh,Kniehl:1990iv,Brein:2003wg} which involve
closed third-generation quark loops.  The $\wH/\zH$ ratio should thus be
a useful quantity to study at the \lhc{}, providing direct access to
some of the most crucial parameters of the Higgs Lagrangian.

The 2-Higgs-Doublet Model (\thdm{}) is one of the simplest extensions of
the \sm{} Higgs sector (for reviews, see
\citeres{Gunion:1989we,Akeroyd:1996he,Akeroyd:1998ui,Aoki:2009ha,Branco:2011iw,Craig:2012vn}).
Recently, a number of studies have been performed in order to scan the
parameter space of a \thdm{} for regions which are still allowed in the
light of the existence of a Higgs boson with
$m_H=125$\,GeV\,\cite{Chen:2013rba,Alves:2012ez,Craig:2012vn,Craig:2012pu,
  Bai:2012ex,Azatov:2012qz,Ferreira:2012nv,Celis:2013rcs,Grinstein:2013npa,
  Krawczyk:2013gia,Altmannshofer:2012ar,Chiang:2013ixa,
  Barroso:2013zxa,Craig:2013hca,Chen:2013kt,Eberhardt:2013uba,
  Ferreira:2013qua,ATLAS:2013zla}.
In this paper, we assume \cp{} conservation in the Higgs sector and no
tree-level flavor-changing neutral currents.  The spectrum of neutral
physical Higgs bosons of the \thdm{} then consists of two \cp{}-even
Higgs bosons $h$ and $H^0$ (where, by definition, $m_h<m_{H^0}$), and
one \cp{}-odd Higgs boson $A$, often referred to as the pseudoscalar
Higgs. According to the structure of the Yukawa couplings, one
distinguishes four types of \thdm{}s, only two of which lead to
different results in the following (see Appendix~\ref{app:2HDM}).

A short review of Higgs Strahlung within the \thdm{} can be found in
\citere{Djouadi:2005gj}.  As it is well-known, a simple reweighting of
the \sm{} Higgs Strahlung cross sections according to the different
couplings of the Higgs bosons $\phi\in\{h,H^0,A\}$ to the gauge bosons
$V\in\{W,Z\}$ is not sufficient. This is due to higher order
contributions which involve top and bottom Yukawa couplings.  In this
paper, we thoroughly combine all contributions to the Higgs Strahlung
cross section, taking into account \thdm{} effects. We will show that
modified top and bottom Yukawa couplings mostly affect the $\zphi$
production cross section and thus can lead to significant deviations of
the ratio of the $\wphi$/$\zphi$ cross sections with respect to its
\sm{} value.  In addition, the final state $\zphi$ can also be produced
via $s$-channel exchange of a (virtual) scalar $\phi'\neq\phi$ in the
\thdm{} which is produced through either $b\bar b$ or $gg$ annihilation.
This can enhance the $\zphi$ production cross section dramatically. Such
effects have been studied for $\zA$ production in
\citeres{Kao:1991xg,Yin:2002sq,Kao:2003jw,Kao:2004vp,Yang:2003kr,
  Li:2005qna,Kniehl:2011aa} and for $\zh$ and $\zHo$ production in
\citere{Kniehl:2011aa}. Most of these studies were carried out in the
framework of the Minimal Supersymmetric \sm{} (\mssm{}) though, where
the parameters of the Higgs sector are tightly constraint, thus
restricting the possible effects with respect to a fully general
\thdm{}.  For our numerical analysis, we extended the program
\vhnnlo{}~\cite{Brein:2003wg,Brein:2012ne} for the calculation of
associated $\vH$ production to $\vphi$ production in the \thdm{}.

While the main results of the paper are derived from predictions for the
total inclusive cross section, we also study the influence of lower cuts
on the Higgs' transverse momentum $\ptphi$, motivated by the increased
signal-to-background ratio in the so-called boosted
regime~\cite{Butterworth:2008iy}.  We will argue that, while the
sensitivity to the Yukawa couplings is generally increased in that case,
effects induced by additional (virtual) scalar particles can be strongly
reduced. This suggests that events at low-$\ptphi$ should not be
discarded in experimental analyses.

The remainder of the paper is structured as follows: \sct{sec:theory}
reviews Higgs Strahlung in the \sm{}, before discussing the various
cross section contributions in the \thdm{}. Their implementation within
\vhnnlo{} is described, and the $\wphi$ to $\zphi$ ratio of production
cross sections as precision observable is motivated.  We provide the
most up-to-date numerical value for this ratio in the \sm{}.  In
\sct{sec:SMYukawas}, the effect of Yukawa couplings on Higgs Strahlung
in the \sm{} is discussed. \sct{sec:2HDMdiscussion} describes different
aspects of Higgs Strahlung in the \thdm{} and highlights their effects
on the $\wphi$ to $\zphi$ ratio. Differences of the observed effects in the
boosted regime are discussed in \sct{sec:boost}.
Conclusions and an outlook for possible
future studies are given in \sct{sec:conclusions}. Appendix~\ref{app:2HDM}
summarizes some information on the \thdm{} which is relevant for our
presentation.


\section{Theoretical description of the Higgs Strahlung process}
\label{sec:theory}


\subsection{Standard Model}
\label{sec:theorySM}

\begin{figure}[ht]
\begin{center}
\begin{tabular}{ccc}
\includegraphics[height=0.15\textwidth]{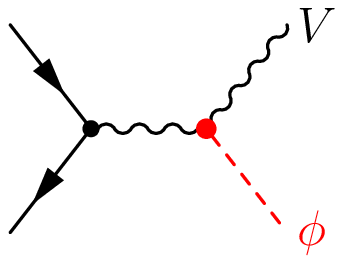}\hspace{1cm}
&\includegraphics[height=0.15\textwidth]{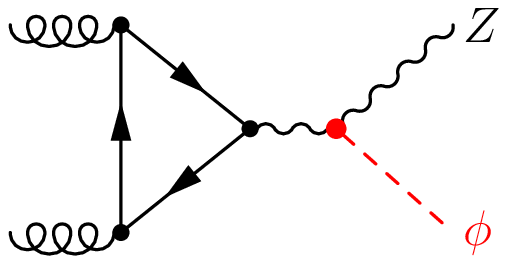}\hspace{1cm}
&\includegraphics[height=0.15\textwidth]{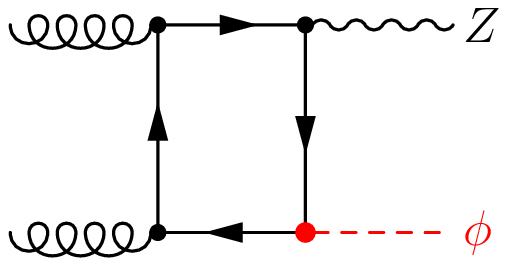}
\\
(a) & (b) & (c)
\end{tabular}
\end{center}
\vspace{-0.6cm}
\caption{Leading order Feynman diagrams: (a) Drell-Yan-like contribution
  $\sigma_{\dy}^{\vphi}$; (b,c) gluon-initiated contributions
  $\sigma_{{\ggzphi}}$.  Couplings which differ between \sm{} and
  \thdm{} are highlighted.}
\label{fig:vhinSM}
\end{figure}

The cross section for Higgs Strahlung in the \sm{} through \nnlo{}
\qcd{} can be written as
\begin{equation}
\begin{split}
\sigma^{\vH} = \sigma^{\vH}_{\dy} + \sigma_{\text{I}}^{\vH}
+\delta_{V\!Z}\cdot\left(\sigma_{\text{II}}^{\zH} + \sigma_{\ggzH}\right)\,.
\label{eq:sigmaparts}
\end{split}
\end{equation}
Through \nlo{} \qcd{}, i.e.\,$\order{\alpha_s}$, only the so-called
``Drell-Yan-like'' (\dy{}) terms $\sigma_{\dy}^{\vH}$ contribute. They
are given by the production of a virtual gauge boson $V^\ast$ and its
subsequent decay into a real gauge boson plus a Higgs; the leading order
diagram is shown in \fig{fig:vhinSM}\,(a). We exclude contributions from
the definition of $\sigma_{\dy}^{\vH}$ which involve the coupling of the
gauge boson to a closed quark loop, see for example \fig{fig:vhinSM}\,(b),
which will be attributed to $\sigma_{\ggzH}$.  The
\qcd{} corrections to $\sigma_{\dy}^{\vH}$ are therefore simply given by
the \qcd{} corrections to the Drell-Yan process $q\bar q\to V^\ast$, and
they are the same for $\wH$ and $\zH$ production at any order of
perturbation theory.  At \nnlo{}, they have been obtained in
\citere{Brein:2003wg} on the basis of \citere{Hamberg:1991np}.

\begin{figure}[ht]
\begin{center}
\begin{tabular}{cc}
\includegraphics[height=0.15\textwidth]{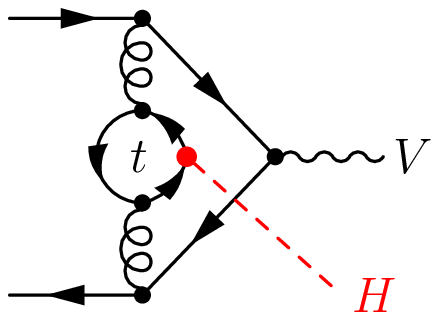}\hspace{1cm}
&\includegraphics[height=0.15\textwidth]{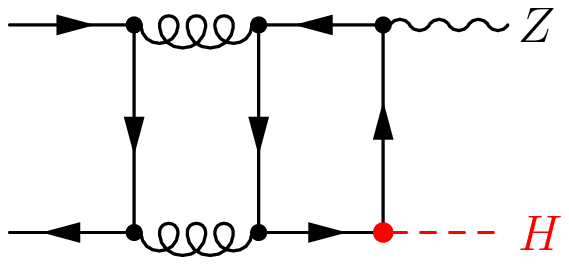}
\\
(a) & (b)
\end{tabular}
\end{center}
\vspace{-0.6cm}
\caption{Contribution to (a) $\sigma_{\text{I}}^{\vH}$ and
(b) $\sigma_{\text{II}}^{\zH}$.
}
\label{fig:vhinSMhigher}
\end{figure}

The other terms of \eqn{eq:sigmaparts} only contribute at
$\order{\alpha_s^2}$ or higher. They all involve top- or bottom-quark
loops.  The term $\sigma_{\text{I}}^{\vH}$ is very similar for $\wH$ and
$\zH$ production. Its Feynman diagrams can be obtained by inserting a
top- or bottom-quark loop into the gluon lines of (real and virtual)
\nlo{} \qcd{} diagrams for $pp\to V$ production, and radiating the Higgs
boson off this loop. A sample diagram is shown in \fig{fig:vhinSMhigher} (a);
it is their interference with \dy{}-diagrams which leads to
$\sigma_{\text{I}}^{\vH}$. The numerical effect of this contribution was
evaluated in the heavy-top limit and by neglecting the bottom Yukawa
coupling; it was found to be of the order of 1-2\% of the total cross
section~\cite{Brein:2011vx}.

The terms $\sigma_{\text{II}}^{\zH}$ and $\sigma_{\ggzH}$ collect
contributions where the $Z$ boson couples to a closed top- or
bottom-quark loop. The analogous terms are absent for $\wH$ production
as indicated by the Kronecker symbol $\delta_{V\!Z}$ in
\eqn{eq:sigmaparts}. The Higgs boson may then be emitted either from the
$Z$ boson, or from the closed quark loop.  In
$\sigma_{\text{II}}^{\zH}$, the quark loop is connected via gluons to an
external quark line. This results in either two-loop diagrams for $q\bar
q\to \zH$, see \fig{fig:vhinSMhigher}\,(b), or in one-loop diagrams with
two quarks, one gluon, and $\zH$ as external states. The resulting
amplitudes have to be interfered with the corresponding lowest order
\dy{}-contribution. In \citere{Brein:2011vx}, $\sigma_{\text{II}}^{\zH}$
was found to be even smaller than $\sigma_{\text{I}}^{\vH}$, typically
at the sub-percent level.

Of particular importance for our analysis is the contribution
$\sigma_{\ggzH}$, where the closed top- or bottom-quark loop is
connected to two initial state gluons, the $Z$ is radiated off that
loop, and the Higgs is emitted from the $Z$ or from the quark loop. The
lowest order, i.e.\,$\order{\alpha_s^2}$, was calculated in
\citeres{Dicus:1988yh,Kniehl:1990iv,Brein:2003wg}. Due to the two
initial state gluons, its numerical contribution to the total cross
section can be significantly larger than the one of
$\sigma_{\text{II}}^{\zH}$, although both are formally of the same order
of perturbation theory. Its size depends on the center-of-mass energy
though: while it was negligibly small for the Tevatron, for example, it
amounts to about 4\% (6\%) at the \lhc{} for 8\,TeV (14\,TeV). Since its
impact on the theoretical uncertainty of the total cross section was
quite significant, \citere{Altenkamp:2012sx} evaluated it at the next
order in perturbation theory, i.e.\ $\order{\alpha_s^3}$, which is
formally an {\abbrev N$^3$LO} contribution to the Higgs Strahlung
process. These \qcd{} corrections where found to be of the order of
100\%, similar to what is observed in the gluon fusion process $gg\to
H$~\cite{Dawson:1990zj,Djouadi:1991tka,Graudenz:1992pv,
  Spira:1995rr,Harlander:2002wh,
  Anastasiou:2002yz,Ravindran:2003um}. While the absolute
renormalization scale dependence is almost the same at
$\order{\alpha_s^2}$ and $\order{\alpha_s^3}$, the relative variation is
smaller at $\order{\alpha_s^3}$ by a factor of two due to the large
K-factor.


\subsection{2-Higgs-Doublet Model}
\label{sec:2HDMtheory}

In the \sm{}, contributions proportional to the bottom Yukawa coupling
are typically strongly suppressed with respect to the corresponding
top-quark induced terms. This is different in extended Higgs sectors
like, for example, the \thdm{}. An introduction to the \thdm{} including
a presentation of all relevant couplings in the neutral Higgs sector is
given in Appendix~\ref{app:2HDM}. In this paper, we consider the production of
any of the three neutral Higgs bosons of the \thdm{}, generically
denoting them as $\phi\in\{h,H^0,A\}$, in association with a weak gauge
boson $V\in\{W,Z\}$.  The main differences to the \sm{} are as follows:
\begin{itemize}
\item The couplings $g_{VV}^{\phi}$ are different from the \sm{} $g_{VV}^{H}$ coupling.
\item The bottom Yukawa coupling can be significantly enhanced and thus
  cannot be neglected in general.
\item In addition to an off-shell gauge boson, also one of the three
  neutral Higgs bosons can occur as internal particle.
\end{itemize}

The implications of these changes are as follows. The \dy{} contribution
$\sigma^{\vphi}_{\dy}$ is obtained from the \sm{} expression simply by
reweighting the Higgs coupling to the gauge bosons:
\begin{equation}
\begin{split}
\sigma^{\vphi}_{\dy} = \left(g_{VV}^{\phi}\right)^2\cdot\sigma^{\vH}_\dy\,.
\end{split}
\end{equation}


\begin{figure}[ht]
\begin{center}
\begin{tabular}{cc}
\includegraphics[height=0.15\textwidth]{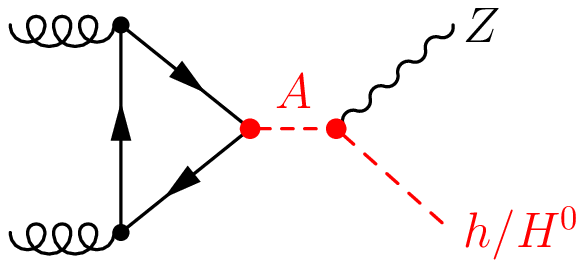}\hspace{1cm}
&\includegraphics[height=0.15\textwidth]{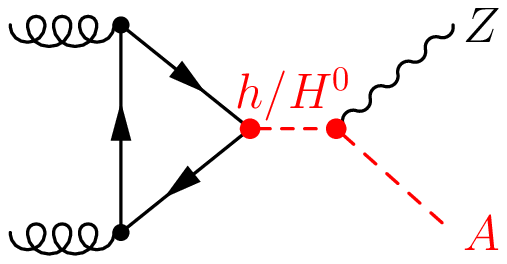}
\\
(a) & (b)
\end{tabular}
\end{center}
\vspace{-0.6cm}
\caption{Additional contributions to gluon-initiated Higgs Strahlung
in case of the \thdm{}.
}
\label{fig:ggzhAhZ}
\end{figure}


The \thdm{} version of the gluon-initiated process, $\sigma_{\ggzphi}$,
on the other hand, depends on $g_{ZZ}^\phi$ as well as on the relative
Yukawa couplings $g_b^\phi$ and $g_t^\phi$. In addition, instead of the
intermediate $Z$ boson in \fig{fig:vhinSM}\,(b), a (virtual) \cp{}-odd/even Higgs boson $\phi'$ can be produced, which decays into a
real \cp{}-even/odd Higgs boson and a $Z$ boson, see
\fig{fig:ggzhAhZ}~\cite{Kao:1991xg,Yin:2002sq,Kao:2003jw,Yang:2003kr,
  Li:2005qna,Kniehl:2011aa}. As we will see, this contribution is
absolutely essential for the correct description of the Higgs Strahlung
process in the \thdm{}.  This is particularly true if the mass of the
intermediate particle $\phi'$ is larger than the threshold, $m_{\phi'} >
m_Z+m_\phi$, in which case the propagator becomes resonant at the
partonic cms energy $\sqrt{\hat s}=m_{\phi'}$.  We regulate the
associated divergence by replacing the propagator as
\begin{equation}
\begin{split}
\frac{1}{\hat{s}-m_{\phi'}^2} \to \frac{1}{\hat{s}-m_{\phi'}^2 +
  im_{\phi'}\Gamma_{\phi'}} \,,
\label{eq:bw}
\end{split}
\end{equation}

where $\Gamma_{\phi'}$ is the total width of $\phi'$. Given a specific
\thdm{}, $\Gamma_{\phi'}$ is calculable; we use the value provided by
     {\tt 2HDMC}\,~\cite{Eriksson:2009ws,Eriksson:2010zzb}, which can be
     linked to the new version of \vhnnlo{}.

Since the $\order{\alpha_s^3}$ effects for $\sigma_{\ggzphi}$ are only
known for vanishing bottom Yukawa coupling~\cite{Altenkamp:2012sx}, we
will disregard these corrections in the \thdm{} analysis and take into
account only the $\order{\alpha_s^2}$
results~\cite{Dicus:1988yh,Kniehl:1990iv,Brein:2003wg}.

For the \thdm{} generalizations $\sigma^{\vphi}_\text{I}$ and
$\sigma^{\zphi}_\text{II}$ of $\sigma^{\vH}_\text{I}$ and
$\sigma^{\zH}_\text{II}$, also Feynman diagrams with closed bottom
instead of top quark loops should be taken into account. Due to the fact
that one cannot evaluate them in an effective theory analogous to the
heavy-top limit, these contributions are extremely difficult to
calculate and thus currently unavailable. Considering the smallness of
these terms in the \sm{}, however, it should be fair to neglect these
terms altogether. Unless stated otherwise, we therefore set
$\sigma^{\vphi}_\text{I}=\sigma^{\zphi}_\text{II}=0$ in what follows.


\begin{figure}[ht]
\begin{center}
\begin{tabular}{cccc}
\includegraphics[height=0.15\textwidth]{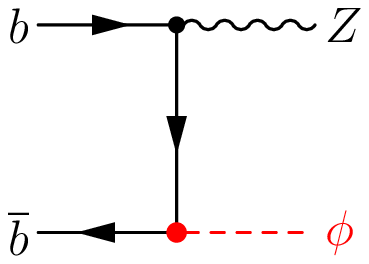}\hspace{0.4cm}
&\includegraphics[height=0.15\textwidth]{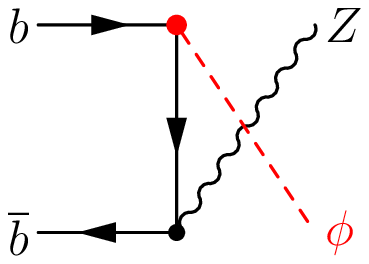}\hspace{0.4cm}
&\includegraphics[height=0.15\textwidth]{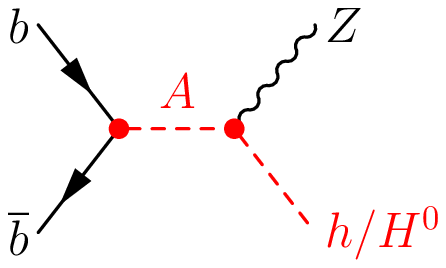}\hspace{0.4cm}
&\includegraphics[height=0.15\textwidth]{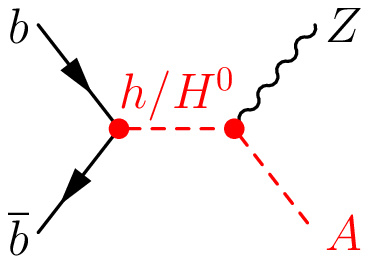}
\\
(a) & (b) & (c) & (d)
\end{tabular}
\end{center}
\vspace{-0.6cm}
\caption{$\bbzphi$ contributions to Higgs Strahlung in the \thdm{}.
}
\label{fig:bbzh}
\end{figure}


Finally, an enhanced bottom Yukawa coupling induces a new contribution
$\sigma_{\bbzphi}\sim (y_b^\phi)^2$ with respect to the \sm{}, where the
final state $Z\phi$ is produced in association with bottom
quarks~\cite{Yin:2002sq,Kao:2004vp,Yang:2003kr,Li:2005qna,Kniehl:2011aa}.
If the latter are not tagged, the cross section can be calculated
in the so-called five-flavour scheme where the perturbative series
is re-arranged to resum logarithmic terms that arise from the
collinear region of the final state bottom quark momenta
(see, e.g., \citere{bbh-santander} and references therein).
The bottom quarks then appear as initial
states, $b\bar b\to \zphi$, with proper parton density functions.  The
relevant Feynman diagrams at leading order are depicted in
\fig{fig:bbzh}. Similar to the process $\ggzphi$, they partly involve
intermediate virtual Higgs bosons $\phi'$, which we treat in the very same
way as described above (see \eqn{eq:bw}).

For the purpose of the current paper, the \lo{} prediction of the
$\bbzphi$ process shall be sufficient. We will see, however, that this
contribution can be quite significant and even numerically dominant, in
particular if the intermediate Higgs boson becomes resonant. \nlo{}
corrections to this process have been calculated in
\citere{Li:2005qna}. They should be included in the analysis once
sufficient experimental data are available.

Note that $\sigma_{\bbzphi}$ exclusively refers to that part of the
partonic process $b\bar b\to \zphi$ which involves a bottom Yukawa
coupling. The same process can also be mediated for $y_b=0$, namely as
part of $\sigma_{\dy}^{\zphi}$, which we evaluate with five-flavor
\pdf{}s. Assuming massless bottom quarks, there is no interference
between these \dy{}- and the $\bbzphi$-diagrams.

To summarize, we write the cross section for Higgs Strahlung in the
\thdm{} as
\begin{equation}
\begin{split}
\sigma^{\vphi} = \sigma^{\vphi}_{\dy}
+\delta_{V\!Z}\cdot\left(\sigma_{\ggzphi{}}
+ \sigma_{\bbzphi}\right)\,.
\label{eq:sigmapartsthdm}
\end{split}
\end{equation}


\subsection{Implementation in \vhnnlo}\label{sec:vhnnlo}

For the numerical evaluation of the individual contributions, we use the
code \vhnnlo{}~\cite{Brein:2012ne} which we extended to the
\thdm{}. It turns out convenient to link \vhnnlo{} to {\tt
  2HDMC}\,~\cite{Eriksson:2009ws,Eriksson:2010zzb} to allow for
different parametrizations of the \thdm{}, consistency checks of the
input parameters, and additional information about the decays of the
involved scalar particles.  The latest version of \vhnnlo{} includes:
\begin{itemize}
\item The Drell-Yan-like terms $\sigma_{\dy}^{\vphi}$ through \nnlo{} by
integrating the results of {\tt zwprod}~\cite{Hamberg:1991np}
and re-weighting with the proper \thdm{} coupling.
\item The gluon-initiated terms $\ggzphi$ at $\order{\alpha_s^2}$ for
  finite top and bottom Yukawa coupling, including the full top- and
  bottom-quark mass dependence, as well as terms with internal Higgs
  bosons, see \fig{fig:ggzhAhZ}. For the top- and bottom-quark masses,
  the on-shell values are used throughout the calculation of the
  $\ggzphi$ contribution.  The corresponding part of the original
  \vhnnlo{} has been fully replaced by an updated code which provides
  higher flexibility concerning the choice of physical parameters. It
  was generated with the help of {\tt
    FeynArts/FormCalc}~\cite{Hahn:2000kx,Hahn:1998yk} and requires
  \vhnnlo{} to be linked to the {\tt LoopTools}~\cite{Hahn:1998yk} and
         {\tt CUBA}~\cite{Hahn:2004fe} libraries.  Regarding the
         resonant exchange of a Higgs boson $\phi'\in\{h,H^0,A\}$ (see
         \fig{fig:ggzhAhZ}), we introduce the Breit-Wigner function as
         given in \eqn{eq:bw} and insert the numerical value for the
         total width of $\phi'$ provided by {\tt
           2HDMC}~\cite{Eriksson:2009ws,Eriksson:2010zzb}. The numerical
         integration of the Breit-Wigner peak is done by suitable Monte
         Carlo sampling.
\item The bottom-quark initiated terms $\bbzphi$ at \lo{}.  The bottom
  mass entering the Yukawa coupling is taken to be the $\msbar$ mass at
  the energy scale $\sqrt{\hat{s}}$ of the incoming partons.
  The resonant contribution from internal (pseudo)scalars 
  (see \figs{fig:bbzh}\,(c,d)) is treated in the very same way as for
  the $\ggzphi$ terms, see above.
\end{itemize}
The only genuinely new terms with respect to the previous version of
\vhnnlo{} are $\bbzphi$ and the $s$-channel contributions to
$\ggzphi$. Corresponding results have been reported on in the literature
before~\cite{Kao:1991xg,Yin:2002sq,Kao:2003jw,Kao:2004vp,Yang:2003kr,
  Li:2005qna,Kniehl:2011aa}. We re-calculated them with the help of {\tt
  FeynArts/FormCalc}~\cite{Hahn:2000kx,Hahn:1998yk}, employing an
adapted version of the \thdm{} model file. Our analytic $\bbzphi$ result
agrees with the one of \citere{Kniehl:2011aa}.  Numerical comparison of
our results with the literature is impeded by their rather crucial
dependence on the precise value of the parameters of the Higgs
sector. Since most of the previous analysis focus on the \mssm{}, such
comparisons could only be done at a rather qualitative level. 

For $\bbzphi$ ($\phi\in\{h,H^0,A\}$) and $\ggzA$ production, we find
satisfactory numerical agreement with \citere{Kniehl:2011aa}, but we
fail to reproduce the numbers for $gg\to\zh$ and $gg\to\zHo$. However,
while our results for $\zh$ production obey the required decoupling
limit by approaching the \sm{} result for large $M_A$, for example, we
cannot verify this behavior in \citere{Kniehl:2011aa}. The source of
this difference is unresolved.\footnote{Thanks to B.\,Kniehl for
  discussions on this issue.}

In case of the \sm{}, higher order terms as presented in
\sct{sec:theorySM} are available and recommended to be included:
\begin{itemize}
\item Electro-weak contributions by assuming full factorization with the
  Drell-Yan-like terms, i.e. by replacing~\cite{Brein:2004ue}
  \begin{equation}
    \begin{split}
      \sigma_{\dy}^{\vH} \to
      (1+\delta_{\ew}^{\vH})\sigma_{\dy}^{\vH}\,.
    \end{split}
  \end{equation}
  The correction factor $\delta_{\ew}^{\vH}$ is implemented by
  interpolating a table of numerical results obtained from
  \citeres{Ciccolini:2003jy,Dittmaier:2011ti}.
\item The terms $\sigma_{\text{I}}^{\vH}$ and $\sigma_{\text{II}}^{\zH}$,
  by implementing the results of \citere{Brein:2011vx}.
\item The gluon-initiated terms $\ggzH$ at $\order{\alpha_s^3}$
  with vanishing bottom Yukawa coupling, by implementing the
  perturbative correction factor in the heavy-top
  approximation~\cite{Altenkamp:2012sx}.
\end{itemize}


\subsection{Input parameters}\label{sec:input}

Throughout this paper, we use the following numerical values for the
\sm{} parameters:
\begin{align}\nonumber
&m_t^{\text{pole}} = 172.3\,\text{GeV},\quad m_b^{\text{pole}}=4.75\,\text{GeV},\quad
m_b^{\overline{\text{MS}}}(m_b)=4.16\,\text{GeV},\\
&m_W=80.398\,\text{GeV},\quad
m_Z=91.1876\,\text{GeV},\quad \Gamma_W=2.141\,\text{GeV},
\quad \Gamma_Z=2.4952\,\text{GeV},\\\nonumber
&G_F=1.16637\cdot10^{-5}\,\text{GeV}^{-2}, \quad \sin^2\theta_C=0.0508\,,
\end{align}
where $\theta_C$ is the Cabbibo mixing angle (mixing with the third
quark generation can be safely neglected); the notation for the other
quantities should be self-explanatory.  Hadronic cross sections are
calculated for a cms energy of $\sqrt{s}=14$\,TeV.  Furthermore, we use
a value of $\alpha_s(m_Z)=0.119$ as input for {\tt 2HDMC}. For the cross
section calculations however, the strong coupling constant is taken
according to the corresponding parton density functions.  As default set
we use {\abbrev MSTW2008NNLO}~\cite{Martin:2009iq}, for which
$\alpha_s(m_Z)=0.11707$.


\subsection{The $\wphi/\zphi$ ratio}
\label{sec:wphizphi}

From the previous discussion, it is clear that the ratio of the cross
sections for $\wphi$ and $\zphi$ production
\begin{equation}
\begin{split}
\rwzh{\phi} = \frac{\sigma^{\wphi}}{\sigma^{\zphi}}
\label{eq:ratio}
\end{split}
\end{equation}
is rather insensitive to radiative corrections in the \sm{}.  This
feature is shared with the Drell-Yan production of lepton pairs, of
course. However, due to the fact that $\rwzh{\phi}$ is sensitive to
Higgs couplings to fermions, it provides a useful handle on possible
deviations from the \sm{} as soon as precise measurements of the ratio
in \eqn{eq:ratio} are available. From the experimental point of view,
the similarity of the $\wphi$ and $\zphi$ production processes should be
reflected in a reduced uncertainty on the measured value of
$\rwzh{\phi}$ as well. Any dependence on the beam luminosity drops out,
for example; also $b$-tagging efficiencies are basically the same in the
numerator and the denominator, with a marginal difference from
kinematics due to $m_W\neq m_Z$.

The theoretical prediction for $\rwzh{H}$ within the \sm{} is rather
precise, since both the numerator and the denominator are known with
\nnlo{} accuracy.  Using \vhnnlo{}, we calculate $\sigma_\dy^{\vH}$
including electro-weak corrections, $\sigma_{\ggzH}$ at
$\order{\alpha_s^3}$, $\sigma^{\vH}_\text{I}$, $\sigma^{\zH}_\text{II}$
and $\sigma_{\bbzH}$, using the input parameters from \sct{sec:input}.
When varying parameters for error estimation, we take the ratio at each
point. We use the \nnlo{} \pdf{} sets from
{\abbrev MSTW}2008~\cite{Martin:2009iq}, {\abbrev
  CT}10~\cite{Lai:2010vv}, and {\abbrev NNPDF}23~\cite{Ball:2012cx},
calculate the envelope of the error bands obtained from each set, and
take its center and half width as prediction for $\rwzh{H}$ and its
$\pdf{}+\alpha_S$ uncertainty.  For the estimation of the scale
uncertainty, we vary the renormalization and factorization scale
separately in an interval $[\frac{1}{3}\mu_0,3\mu_0]$ around our default
choice $\mu_0=m_{\vH}$, the invariant mass of the $\vH$ system.  The
total uncertainty is obtained by adding both numbers linearly.  Our
results for various values of the Higgs mass are listed in
\tab{tab:ratio}.

\begin{table}
\begin{center}
\begin{tabular}{|c|c|c|c|c|}
\hline
$m_H$\,[GeV] & $\rwzh{H}^\sm$
    & $\Delta_{\text{scale}}$[\%] & $\Delta_{\pdf+\alpha_s}$[\%] &
$\Delta_\text{total}$[\%] \\\hline
\hline
100 & 1.676 & $\pm 1.9$ & $\pm 0.7$ & $\pm 2.6$ \\
110 & 1.630 & $\pm 2.3$ & $\pm 0.6$ & $\pm 2.9$ \\
120 & 1.587 & $\pm 2.4$ & $\pm 0.8$ & $\pm 3.3$ \\
122 & 1.580 & $\pm 2.6$ & $\pm 0.8$ & $\pm 3.3$ \\
124 & 1.570 & $\pm 2.7$ & $\pm 0.7$ & $\pm 3.3$ \\
125 & 1.570 & $\pm 2.6$ & $\pm 0.7$ & $\pm 3.4$ \\
126 & 1.563 & $\pm 2.7$ & $\pm 0.7$ & $\pm 3.4$ \\
128 & 1.556 & $\pm 2.8$ & $\pm 0.9$ & $\pm 3.7$ \\
130 & 1.545 & $\pm 2.8$ & $\pm 0.8$ & $\pm 3.6$ \\
140 & 1.501 & $\pm 3.1$ & $\pm 1.0$ & $\pm 4.1$ \\
150 & 1.470 & $\pm 3.4$ & $\pm 0.8$ & $\pm 4.1$ \\
160 & 1.428 & $\pm 3.6$ & $\pm 0.8$ & $\pm 4.5$ \\
170 & 1.406 & $\pm 3.7$ & $\pm 0.9$ & $\pm 4.6$ \\
180 & 1.385 & $\pm 3.9$ & $\pm 0.9$ & $\pm 4.7$ \\
190 & 1.388 & $\pm 4.0$ & $\pm 0.9$ & $\pm 5.0$ \\
200 & 1.381 & $\pm 4.1$ & $\pm 0.9$ & $\pm 5.0$ \\
210 & 1.382 & $\pm 3.9$ & $\pm 0.9$ & $\pm 4.9$ \\
220 & 1.388 & $\pm 3.8$ & $\pm 0.9$ & $\pm 4.7$ \\
230 & 1.397 & $\pm 3.6$ & $\pm 0.8$ & $\pm 4.5$ \\
240 & 1.412 & $\pm 3.3$ & $\pm 0.8$ & $\pm 4.2$ \\
250 & 1.430 & $\pm 3.1$ & $\pm 0.8$ & $\pm 3.9$ \\
260 & 1.446 & $\pm 2.9$ & $\pm 0.8$ & $\pm 3.6$ \\
270 & 1.469 & $\pm 2.6$ & $\pm 0.8$ & $\pm 3.3$ \\
280 & 1.487 & $\pm 2.4$ & $\pm 0.7$ & $\pm 3.0$ \\
290 & 1.508 & $\pm 2.1$ & $\pm 0.7$ & $\pm 2.8$ \\
300 & 1.524 & $\pm 1.9$ & $\pm 0.7$ & $\pm 2.5$ \\

\hline
\end{tabular}
\caption[]{\label{tab:ratio} Ratio $\sigma^{\wH}$ over $\sigma^{\zH}$ in the \sm{} 
for $\sqrt{s}=$14\,TeV with error estimation from scale variation and $\pdf+\alpha_S$ uncertainties. 
The numerical error induced by the Monte Carlo integration 
is expected to have an effect on the last digit only.}
\end{center}
\end{table}

It is instructive to compare the uncertainty of the ratio $\rwzh{H}$ to those of the $\vH$
cross sections. For $m_H=125$\,GeV we obtain
\begin{equation}
\begin{split}
 \sigma_\sm^{\wH} &= 1.520\, \text{pb} \pm 0.6\%^\text{scale} \pm
 2.1\%^{\pdf+\alpha_S}, \\ 
 \sigma_\sm^{\zH} &= 0.970\, \text{pb} \pm 3.2 \%^\text{scale} \pm
 1.9\%^{\pdf+\alpha_S},\\
 \rwzh{H}^\sm    &= 1.570             \pm 2.6\%^\text{scale} \pm
 0.7\%^{\pdf+\alpha_S}. 
\end{split}
\end{equation}
An uncorrelated error estimate for the ratio, i.e. adding the relative
uncertainties of $\sigma_\sm^{\wH}$ and $\sigma_\sm^{\zH}$ in
quadrature, would lead to about $5.8\%$ in total compared to $3.4\%$. 
 Thus we observe that a significant part of the uncertainties
indeed cancels in the ratio.  The residual uncertainty is dominated by
the scale uncertainty of the $\ggzH$ contribution to $\sigma^{\zH}$,
which has no counterpart in $\sigma^{\wH}$ to cancel against.

In the rest of this paper, we will study the influence of possible
non-\sm{} physics on the ratio $\rwzh{\phi}$ for the production of a
neutral Higgs boson $\phi$.  In a first step, we will simply modify
the \sm{} Yukawa couplings, without any particular
underlying model in mind. Subsequently, we will study a general \thdm{},
taking into account all the effects discussed in
\sct{sec:2HDMtheory}.

In the following discussion, we use \eqn{eq:sigmapartsthdm}
as the definition of $\sigma^{\vphi}$, without $\order{\alpha_s^3}$
corrections to $\sigma_{\ggzH}$ and electro-weak effects,
since they are only known in the \sm{}.



\section{Effect of the Yukawa couplings}
\label{sec:SMYukawas}

In this section, we will not respect any constraints on the third
generation Yukawa couplings of the \sm{}, but simply test the
sensitivity of the ratio $\rwzh{\phi}$ on these parameters. For that
purpose, we rescale the Yukawa couplings as
\begin{equation}
\begin{split}
y_t = \kappa\,y_t^\sm\qquad\mbox{and/or}\qquad y_b = y_b^\sm/\kappa\,,
\label{eq:yukvar}
\end{split}
\end{equation}
where $y_t^\sm$ and $y_b^\sm$ are the \sm{} values of the top- and the
bottom Yukawa couplings.


\begin{figure}[ht]
\begin{center}
\begin{tabular}{ccc}
\includegraphics[width=0.3\textwidth]{%
  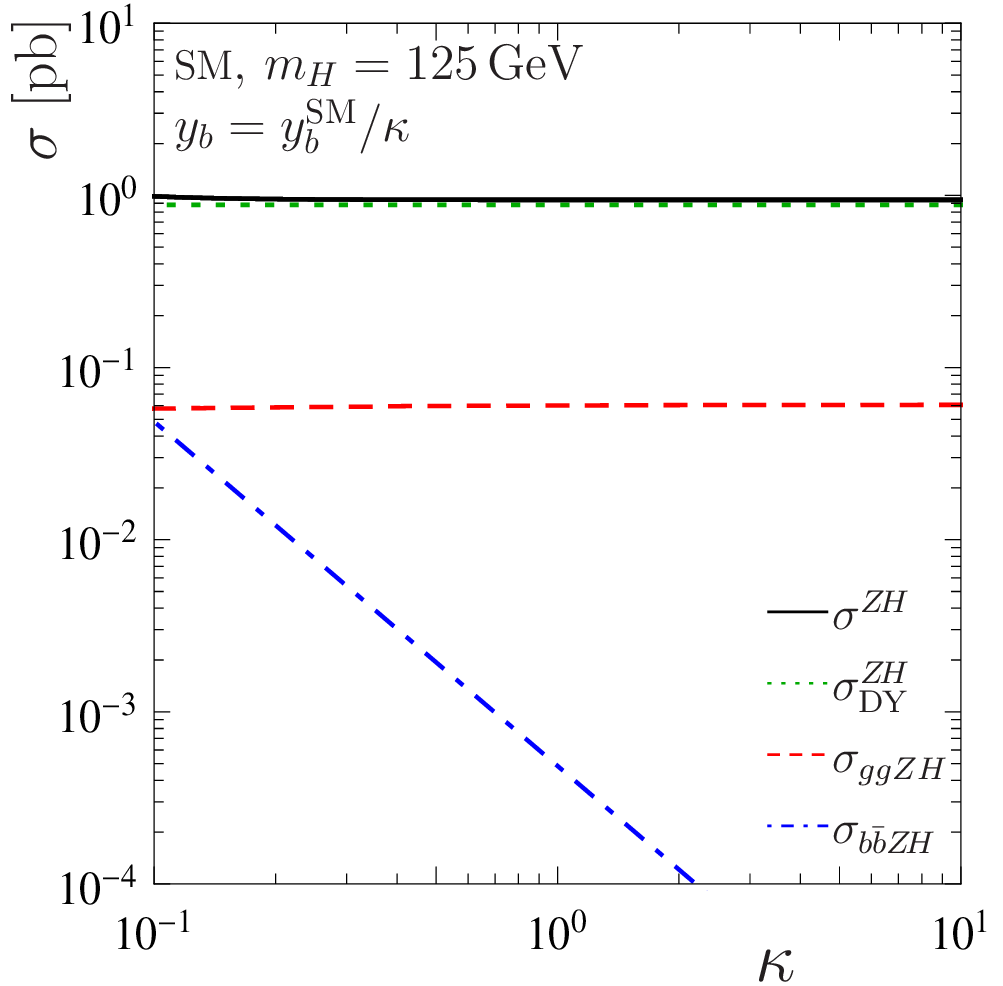} &
\includegraphics[width=0.3\textwidth]{%
  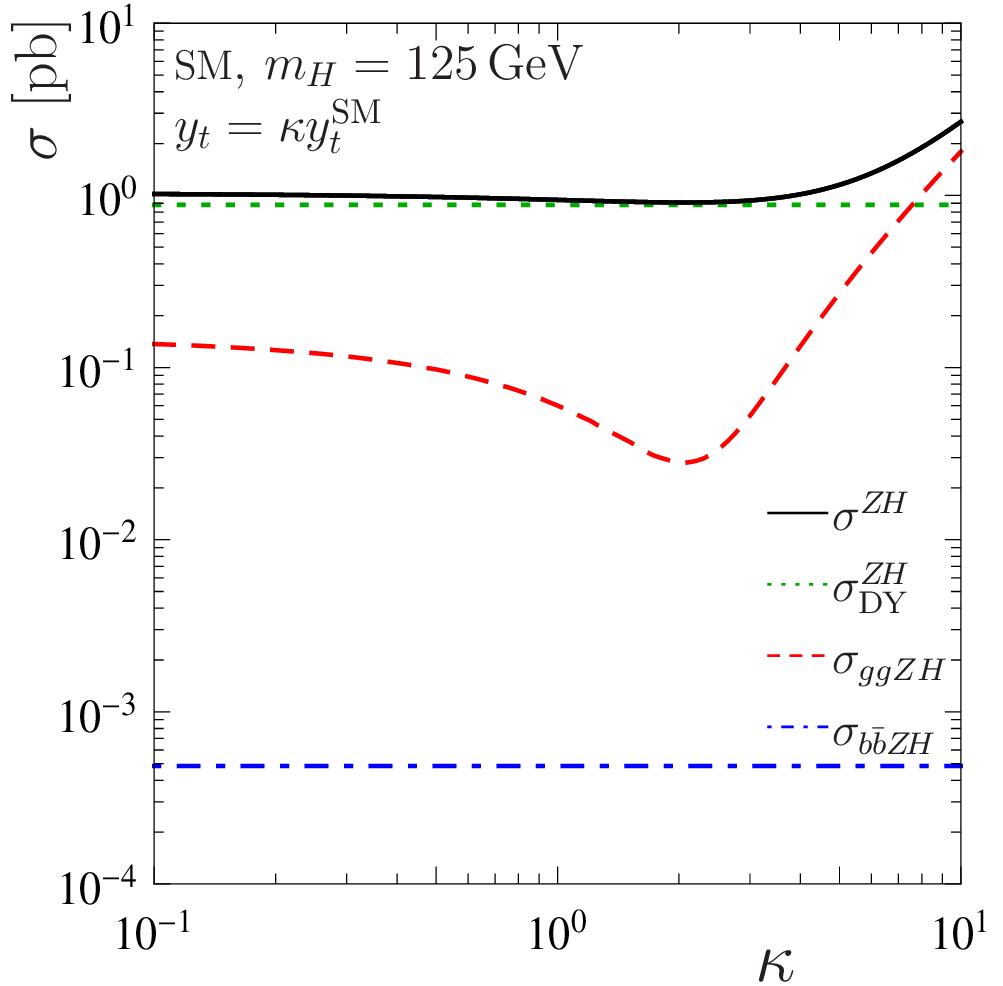} &
\includegraphics[width=0.3\textwidth]{%
  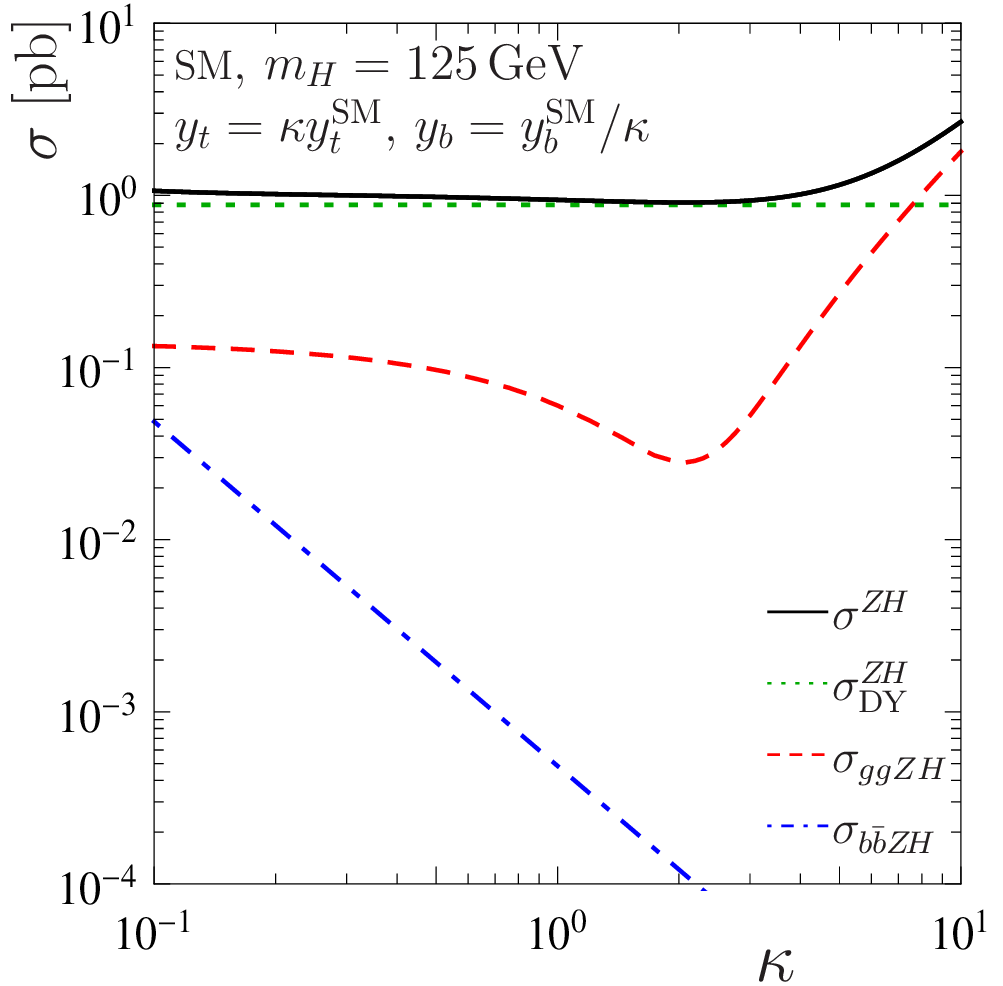} \\[-0.3cm]
 (a) & (b) & (c)\\
\includegraphics[width=0.3\textwidth]{%
  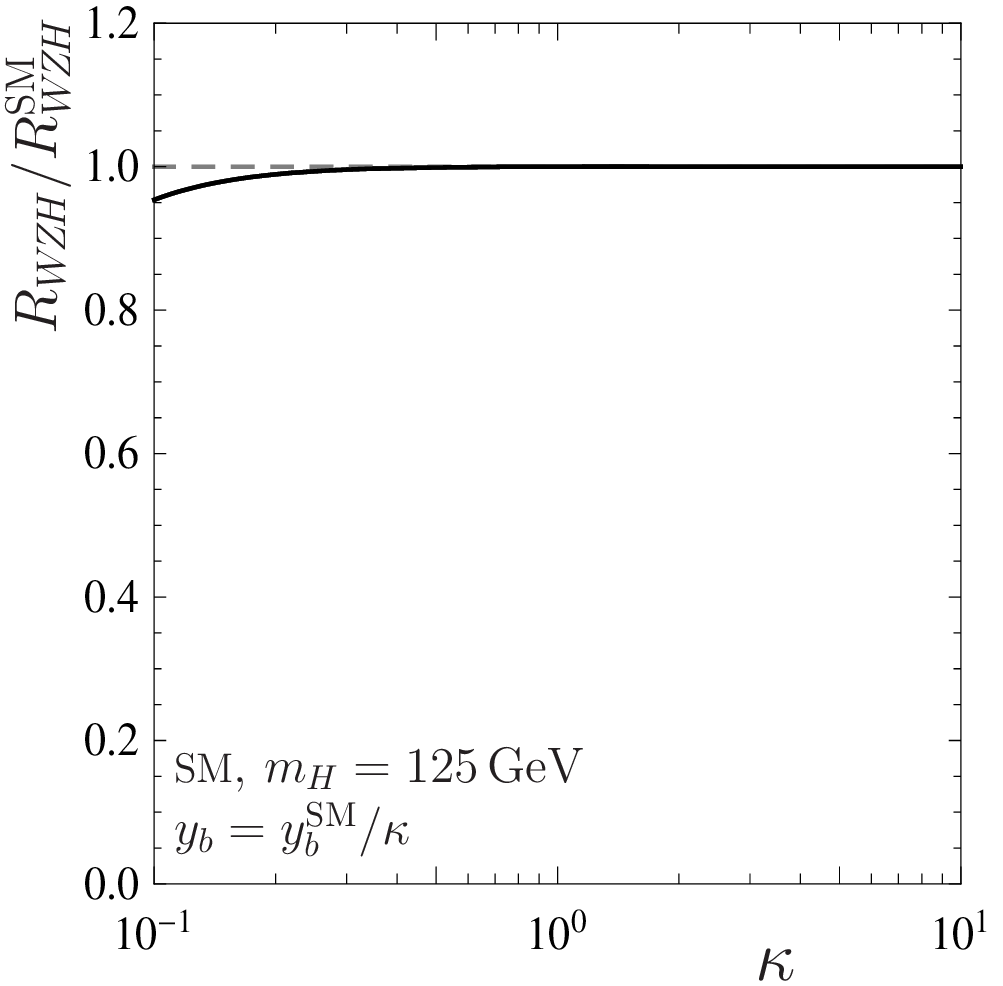} &
\includegraphics[width=0.3\textwidth]{%
  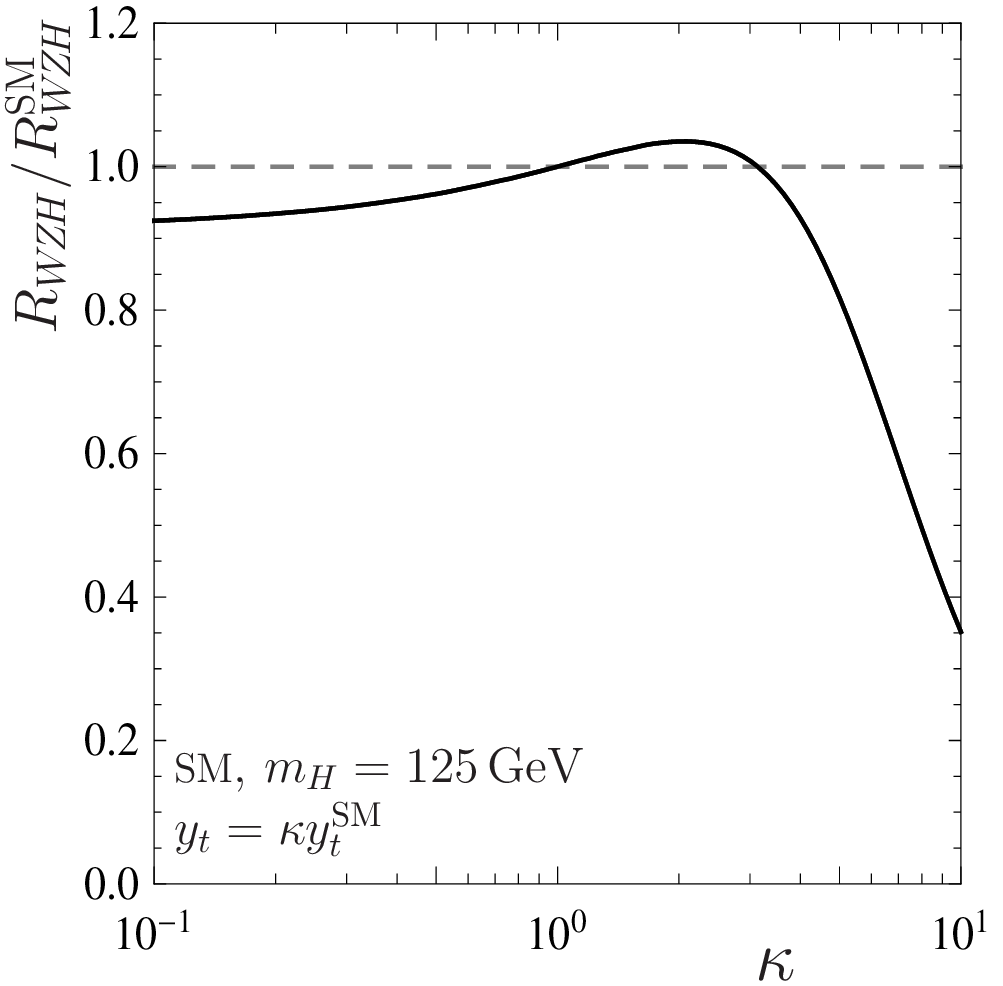} &
\includegraphics[width=0.3\textwidth]{%
  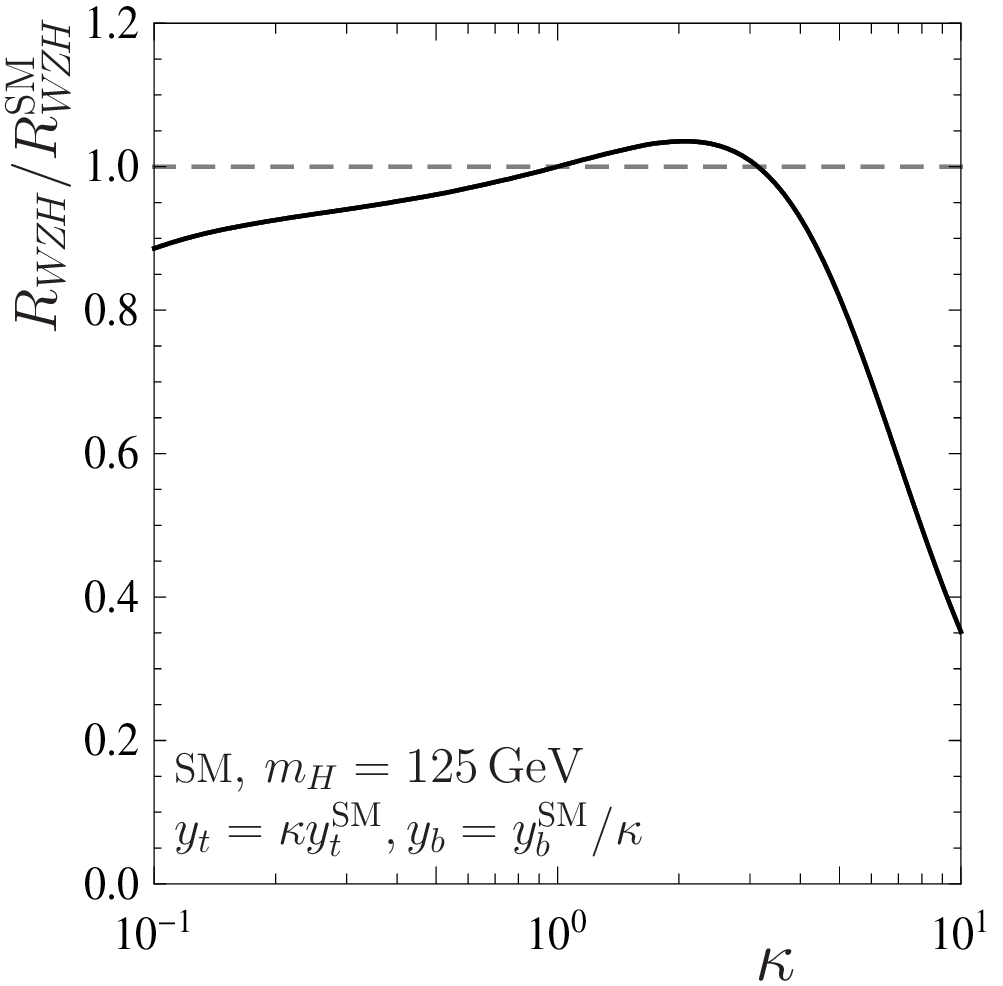} \\[-0.3cm]
(d) & (e) & (f)
\end{tabular}
\end{center}
\vspace{-0.6cm}
\caption[]{\label{fig:SMYuk} (a--c) $\sigma(pp\rightarrow \zH)$ (black/solid),
  $\sigma^{\zH}_{\dy}$ (green/dotted),
  $\sigma_{\ggzH{}}$ (red/dashed) and $\sigma_{\bbzH}$ (blue/dash-dotted)
  in pb for $\sqrt{s}=14$\,TeV and $m_H=125$\,GeV as a function of
  $\kappa$, where (a) $y_t= y_t^\sm$ and $y_b = y_b^\sm{}/\kappa$, (b)
  $y_t=\kappa y_t^\sm$ and $y_b = y_b^\sm{}$, (c) $y_t=\kappa y_t^\sm$
  and $y_b = y_b^\sm{}/\kappa$; (e-f) the ratio $\rwzh{H}/\rwzh{H}^\sm$ for cases
  (a-c).}
\end{figure}


\figs{fig:SMYuk}\,(a-c) show the various contributions to the total
$\zH$ production cross section as discussed above.  For $\kappa=1$, all
figures show the identical \sm{} cross section contributions.  In
\fig{fig:SMYuk}\,(a), the top Yukawa coupling is fixed to
$y_t=y_t^\sm{}$, and the bottom Yukawa coupling is varied according to
\eqn{eq:yukvar}; in \fig{fig:SMYuk}\,(b), $y_b=y_b^\sm$ and $y_t$
varies; and in \fig{fig:SMYuk}\,(c), both couplings vary according to
\eqn{eq:yukvar}.  Obviously, $\sigma_{\bbzH}$ (blue/dash-dotted) is
proportional to $y_b^2$; even at large $y_b$ (small $\kappa$,
figure~(a,c)), it is down by almost a factor of 20 relative to the \dy{}
contribution (green/dotted).

The dependence of the $\ggzH$ contribution (red/dashed) on the bottom
Yukawa coupling is very small.  Its $y_t$ dependence is non-trivial,
though, due to interference terms. Over a large range of $y_t$, the
dominant contribution to $\ggzH$ originates from terms
$gg\to Z^\ast\to \zH$ that do not depend on any Yukawa coupling (see
\fig{fig:vhinSM}\,(b)).  Once $y_t$ is larger than about $3\,y_t^\sm$,
the $\ggzH$ component rises steeply.  It surpasses the \dy{}
contribution at about $y_t=7\,y_t^\sm$, exceeding it by more than a
factor of two at $y_t=10\,y_t^\sm$.

The resulting $\wH/\zH$ ratios $\rwzh{H}$, normalized to the \sm{} ratio
$\rwzh{H}^{\sm}$, are shown as the solid line in
\figs{fig:SMYuk}\,(d-f). Note that for consistency also the \sm{} result
has been obtained according to \eqn{eq:sigmapartsthdm} here, with
electro-weak corrections and $\order{\alpha_s^3}$ effects to
$\sigma_{\ggzH}$ neglected.
For $\kappa\lesssim 4$, the deviation from the \sm{} prediction (dashed)
does not significantly exceed 10\% within the range of $\kappa$
considered here. An increased top Yukawa coupling, however, leads to a
sharp decrease of $\rwzh{H}$: for $\kappa=10$, the value of this ratio
is almost three times smaller than its \sm{} prediction.

To conclude this section, let us remark that, apart from the magnitude
of the top Yukawa coupling, an anomalous sign of the top Yukawa coupling
would be an undoubtful indicator of New Physics. Its measurement clearly
requires interference effects; a determination through associated Higgs
and single-top production has been suggested in
Ref.\,\cite{Biswas:2012bd,Biswas:2013xva}.

The $gg\to \zH$ process offers another, albeit more indirect way to
measure the sign of the top Yukawa coupling due to the interference
between box- and triangle diagrams (see, e.g., \figs{fig:vhinSM}\,(b)
and (c), respectively)\,\cite{Englert:2013vua}. Since this interference
is destructive in the \sm{}, the ratio $\rwzh{H}$ decreases when
changing the top Yukawa coupling to its negative.  For example, at
$m_H=125$\,GeV and $\sqrt{s}=14$\,TeV, the decrease is about 20\% when
using the \lo{} result for $gg\to \zH$; we estimate that this decrease
can be as large as 30\% when higher orders are included. In the boosted
regime, this effect should be even more prominent due to the enhanced
importance of the $gg\to \zH$ process (see Section\,\ref{sec:boost}).

While the impact of a sign change in the bottom Yukawa coupling relative
to its \sm{} value is rather small, it might be worth investigating this
issue as well in extended theories. A more detailed analysis is beyond
the scope of this paper though.


\section{2-Higgs-Doublet Model}
\label{sec:2HDMdiscussion}


As discussed in \sct{sec:2HDMtheory}, the Higgs Strahlung cross section
in the \thdm{} is not only affected by the modification of the
couplings, but also by the contribution from additional Higgs bosons.
In this section we will therefore extend the discussion of the previous
section to the full \thdm{}. We start with a comment on our choice of
the \thdm{} parameters,
followed by a short note about the influence of internal scalars to
Higgs Strahlung at the partonic level of the cross section.  Then we
focus on the production of the light Higgs and subsequently extend our
discussion to the heavy and the pseudoscalar Higgs boson. In each case
our main concern is the ratio $\rwzh{\phi}$.

As explained before, $\sigma^{\vphi}_{\text{I}}$ and
$\sigma^{\zphi}_{\text{II}}$ are not taken into account in the
subsequent discussion of production cross sections in the \thdm{}.  The
contribution $\sigma_{\ggzphi}$ is calculated at $\order{\alpha_s^2}$,
$\sigma_{\dy}^{\vphi}$ without electro-weak corrections.


\subsection{Choice of \thdm{} parameters}

Apart from the \sm{} input, the Higgs Strahlung cross section depends on
the \thdm{} Yukawa type (see Appendix~\ref{app:2HDM}) as well as the parameters
$\tan\beta$, $\sin(\beta-\alpha)$, and the masses of the neutral Higgs
bosons. If an intermediate Higgs state (see, e.g., \fig{fig:ggzhAhZ})
becomes resonant, the total decay width of that state enters the
calculation, see \eqn{eq:bw}. This introduces a dependence on the
parameter $m_{12}$ which, however, is weak. We therefore set $m_{12}=0$
in all our numerical examples.  Note that according to stability,
perturbativity, and unitarity requirements, this choice is actually
allowed only as long as $\tan\beta\lesssim 1$.  Different choices of
$m_{12}$ can considerably loosen this constraint though, without
affecting our numerical results very much.

Identifying the observed Higgs signal at the \lhc{} with the light Higgs
of the \thdm{} forces the value of $\sin(\beta-\alpha)$ to be close to
$1$.  In this paper, we largely disregard such constraints and always
scan the full range of $\sin(\beta-\alpha)$.


\subsection{$\ggzphi$ and $\bbzphi$ contributions 
  involving internal (pseudo)scalars} 

As pointed out in \sct{sec:2HDMtheory}, the cross section $pp\rightarrow
\phi Z$ in the \thdm{} includes $\ggzphi$ and $\bbzphi$ contributions
that involve an internal scalar different from the final state Higgs
particle $\phi$.  Before moving on to the full results, let us look at
this contribution in a bit more detail.

\begin{figure}[ht]
\begin{center}
\begin{tabular}{ccc}
\includegraphics[width=0.3\textwidth]{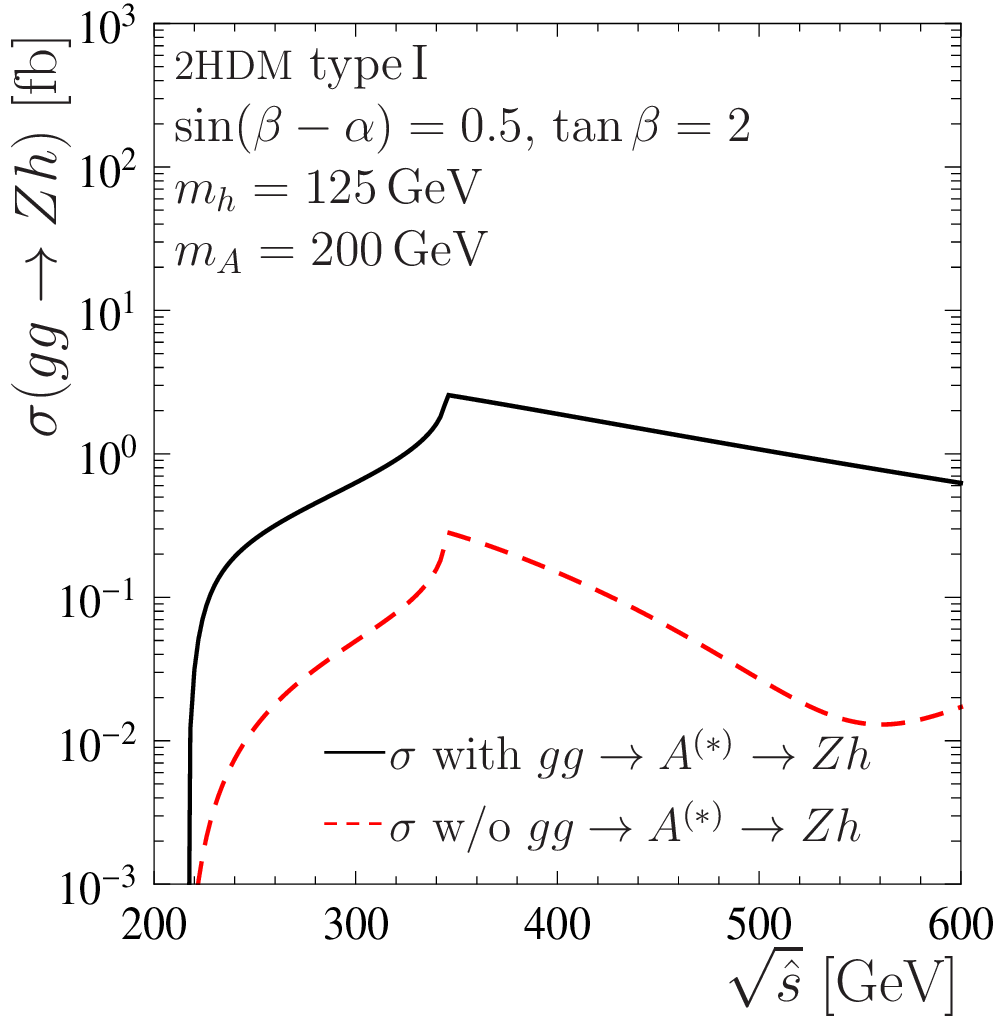}
&\includegraphics[width=0.3\textwidth]{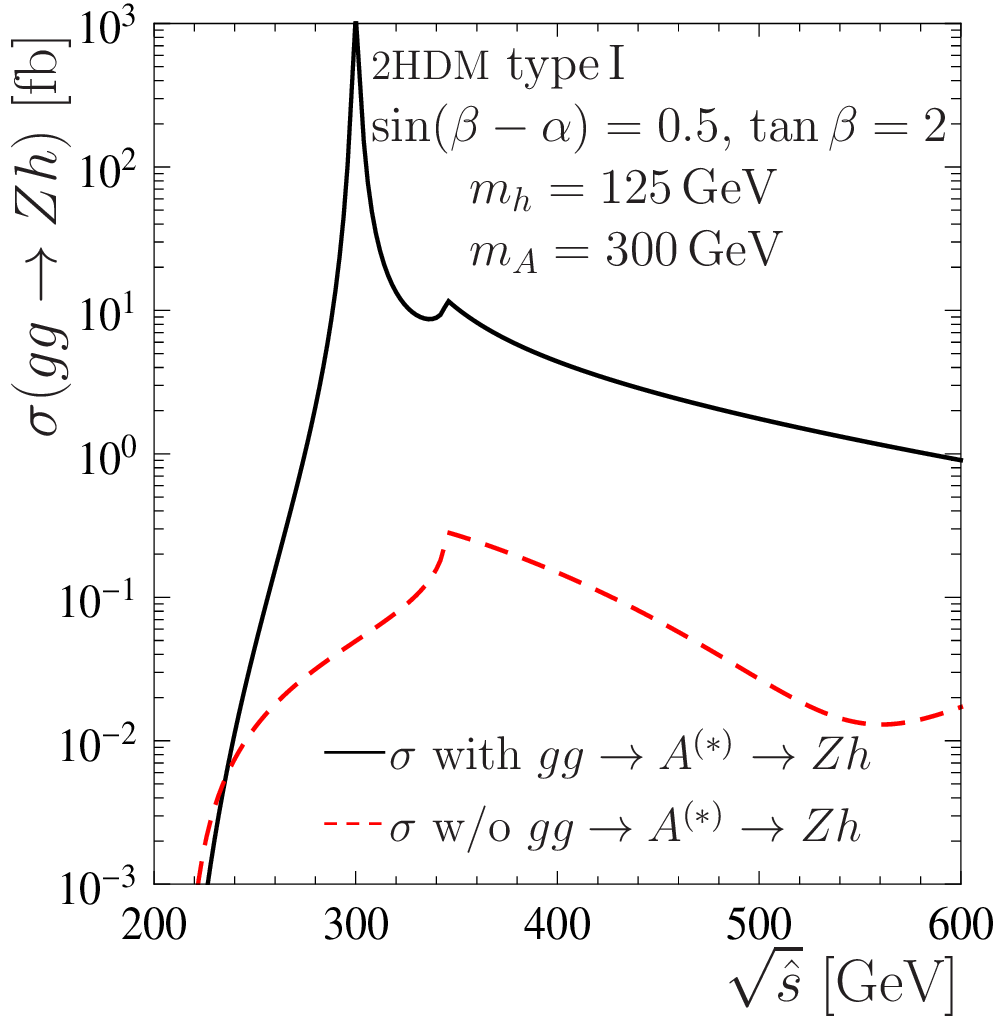}
&\includegraphics[width=0.3\textwidth]{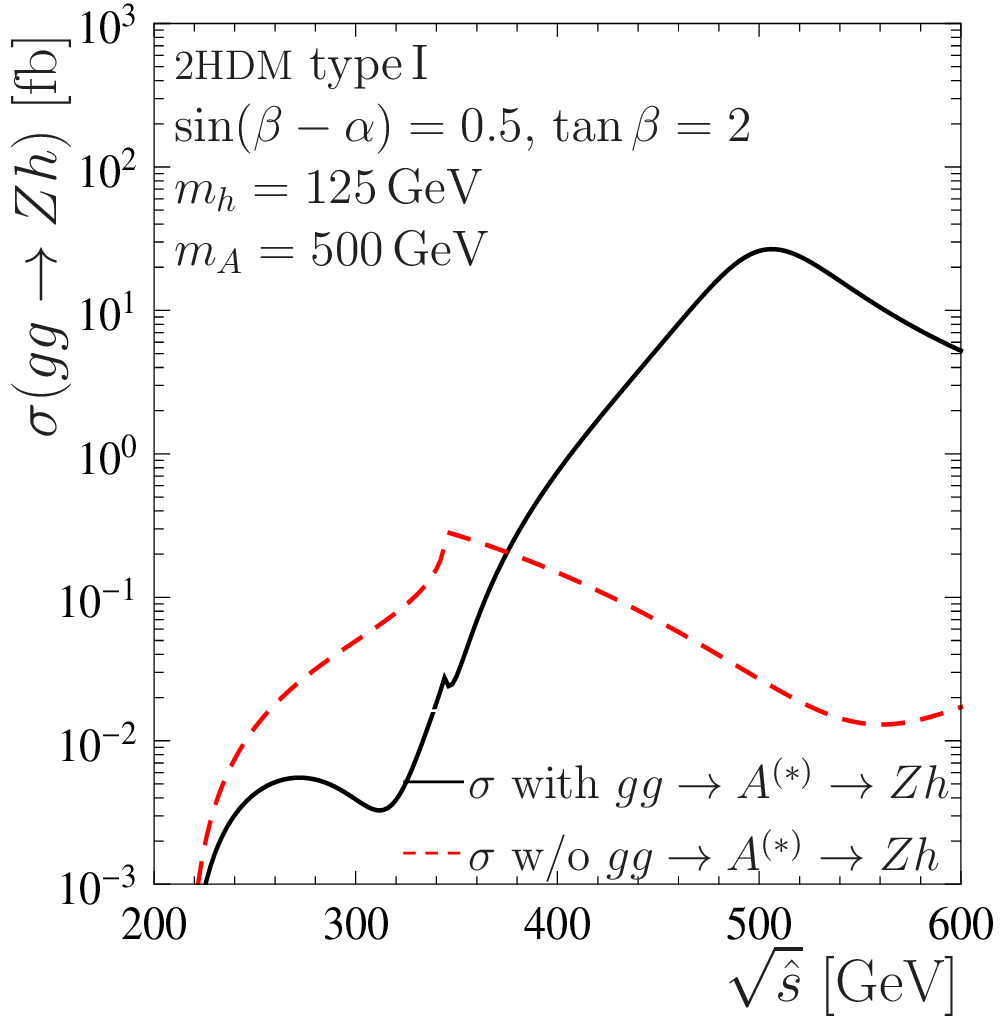}
\\[-0.3cm]
(a) & (b) & (c)
\end{tabular}
\end{center}
\vspace{-0.6cm}
\caption{$\sigma(gg\rightarrow \zh)$ in fb at the partonic level as a
  function of $\sqrt{\hat{s}}$ in GeV for $m_h=125$\,GeV,
  $\sin(\beta-\alpha)=0.5$ and $\tan\beta=2$.  The red/dashed line
  corresponds to $\sigma$ without internal pseudoscalar. The black lines
  show $\sigma$ with the contribution involving an internal pseudoscalar
  Higgs for (a) $m_A=200$\,GeV, (b) $m_A=300$\,GeV and (c)
  $m_A=500$\,GeV.  }
\label{fig:resonance}
\end{figure}

Consider the production of the light Higgs $h$ with mass $m_h=125$\,GeV
for $\sin(\beta-\alpha)=0.5$ and $\tan\beta=2$. \fig{fig:resonance}
shows the partonic cross section $gg\rightarrow Zh$ as a function of the
partonic cms energy $\sqrt{\hat s}$, with (black/solid) and without
(red/dashed) the pseudoscalar $A$ as internal particle, for three
different values of its mass (a-c). Below the kinematical threshold
$\sqrt{\hat{s}}<m_h+m_Z\approx 216$\,GeV, the $\ggzh$ contribution
vanishes.  In \fig{fig:resonance}\,(a), the mass of the pseudoscalar
Higgs is $m_A=200$\,GeV; it is therefore only produced
off-shell.
Already then, the numerical contribution to the partonic cross section
is quite significant.  In \fig{fig:resonance}\,(b) and (c), on the other
hand, the pseudoscalar Higgs is assumed heavier and therefore becomes
resonant. For $m_A=300$\,GeV (\fig{fig:resonance}\,(b)), its total width
is $\Gamma_A=2.82$\,GeV which leads to a sharp peak in the partonic
cross section. For $m_A=500$\,GeV (\fig{fig:resonance}\,(c)), on the
other hand, the corresponding width becomes $\Gamma_A=75.0$\,GeV and the
peak is much broader.  Independent of $m_A$, the heavy and charged Higgs
masses are chosen to be $m_{H^0}=m_{H^\pm}=200$\,GeV.

All cases of \fig{fig:resonance} show the significance of the internal
scalar to the $\ggzh$ contribution; a similar discussion applies to
$\ggzHo$ and $\ggzA$ production, as well as to the $\bbzphi$ mode. The
numerical effects of these terms on the hadronic cross sections and in
particular on the ratio $\rwzh{\phi}$ will be discussed in the
subsequent section.


\subsection{Light Higgs}\label{sec:light}


\begin{figure}[ht]
\begin{center}
\begin{tabular}{ccc}
\includegraphics[width=0.3\textwidth]{%
  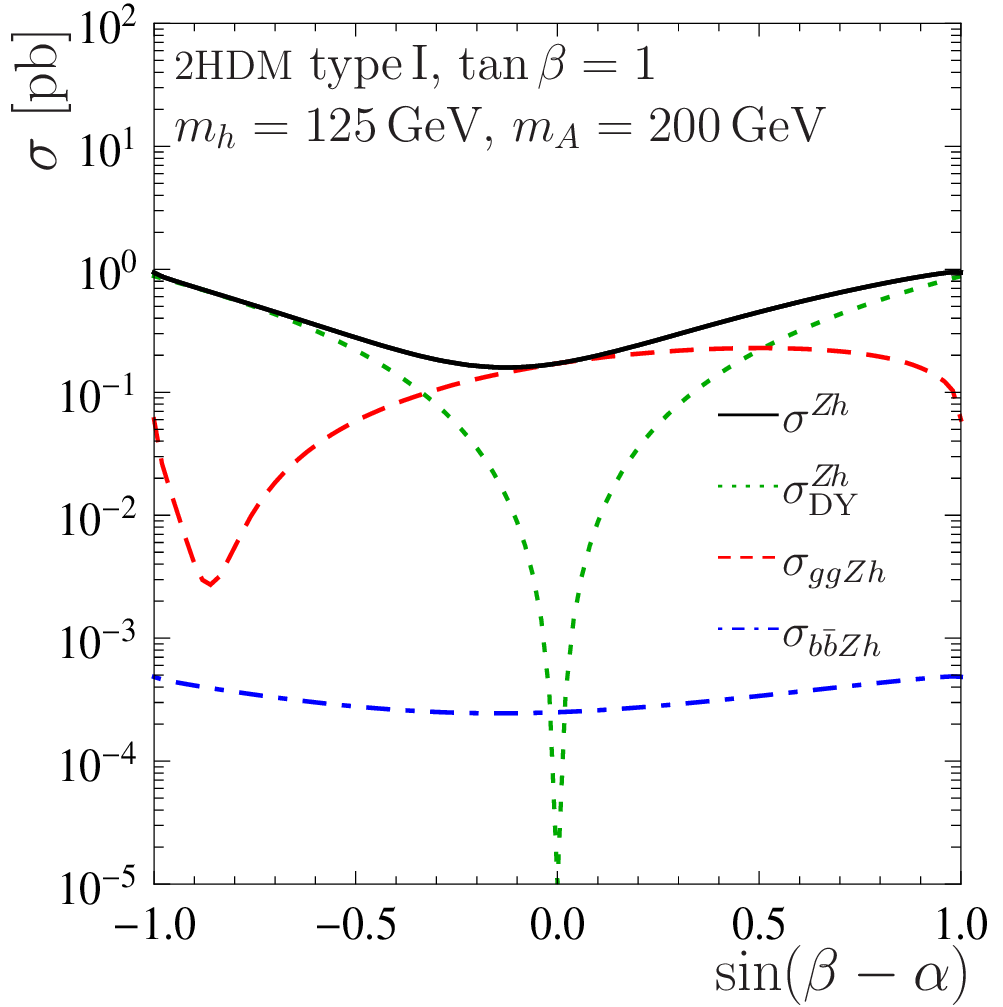} &
\includegraphics[width=0.3\textwidth]{%
  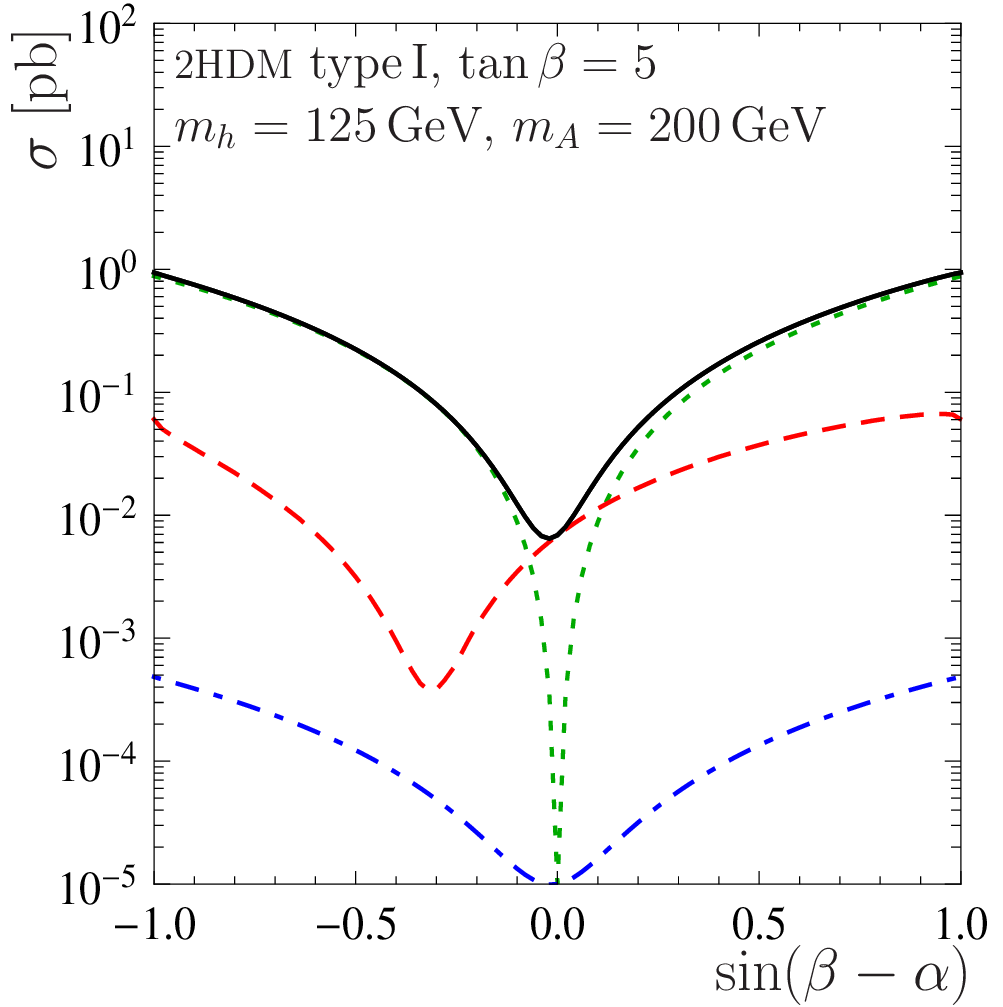} &
\includegraphics[width=0.3\textwidth]{%
  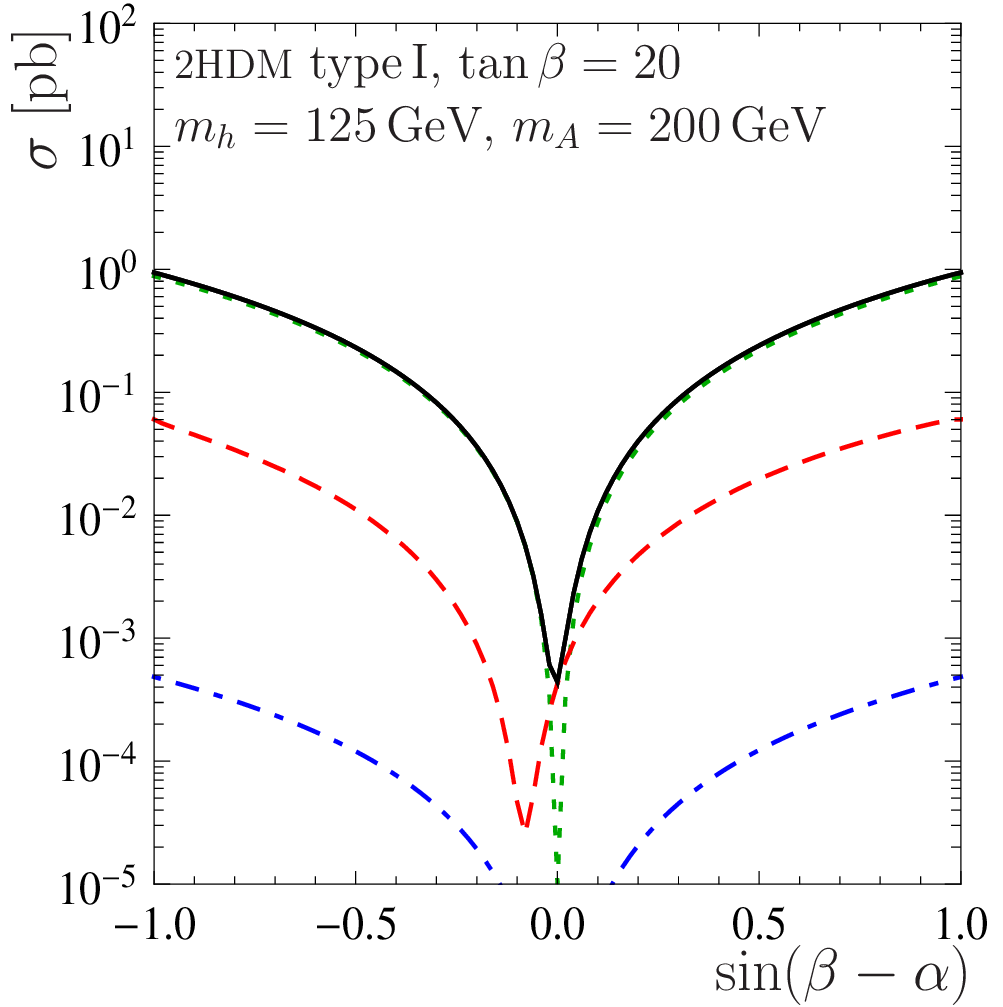} \\[-0.3cm]
(a) & (b) & (c)\\
\includegraphics[width=0.3\textwidth]{%
  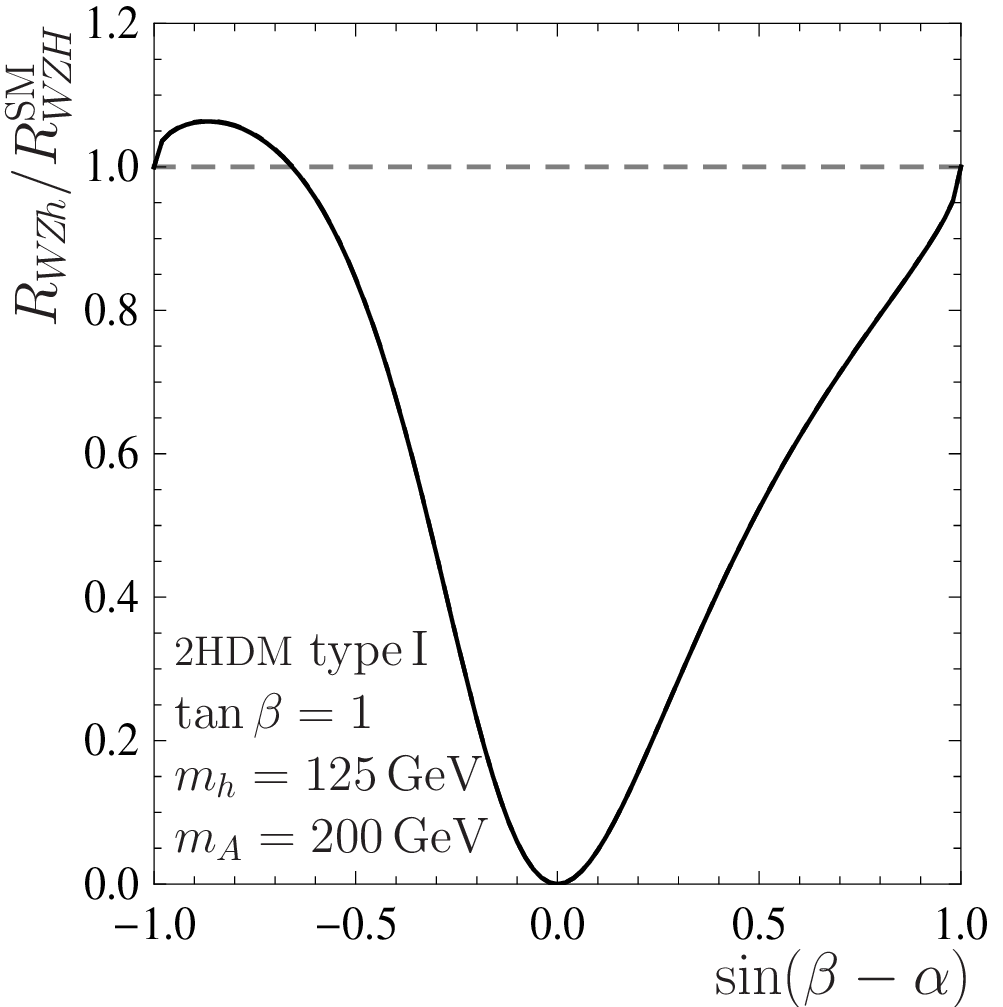} &
\includegraphics[width=0.3\textwidth]{%
  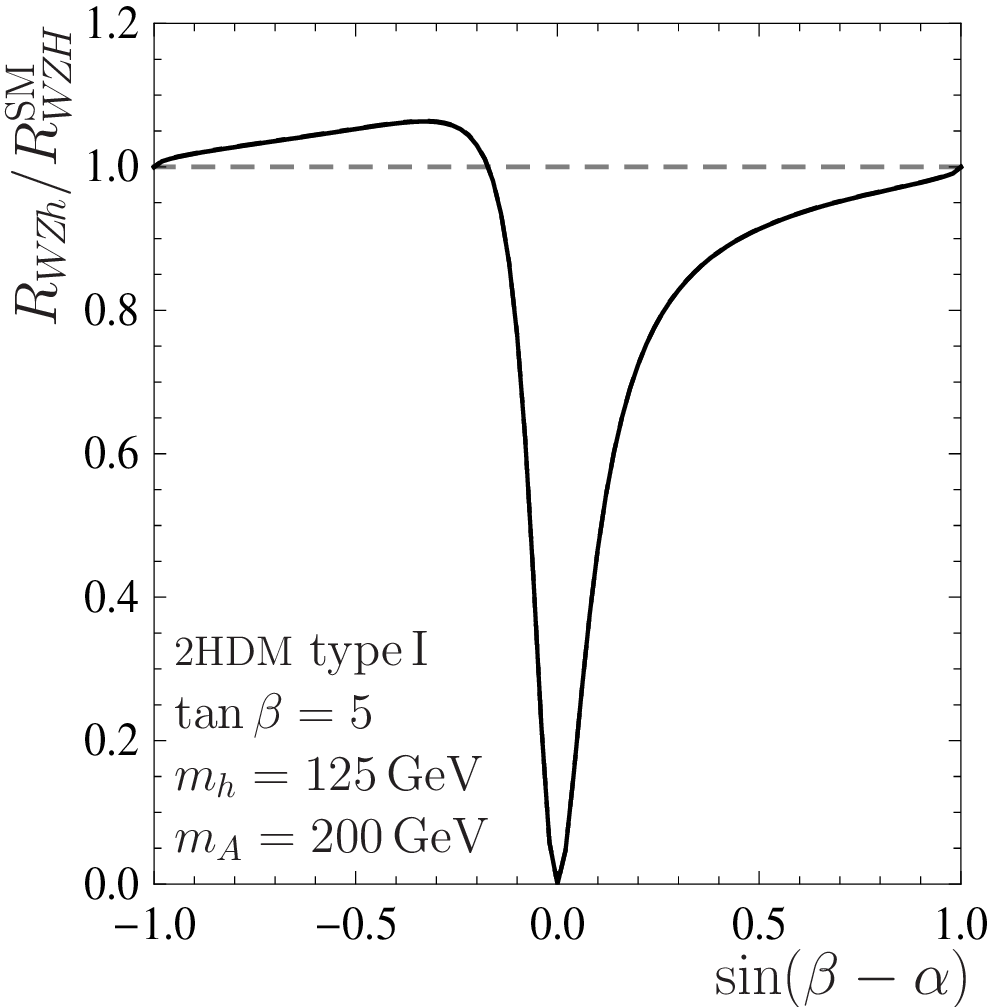}&
\includegraphics[width=0.3\textwidth]{%
  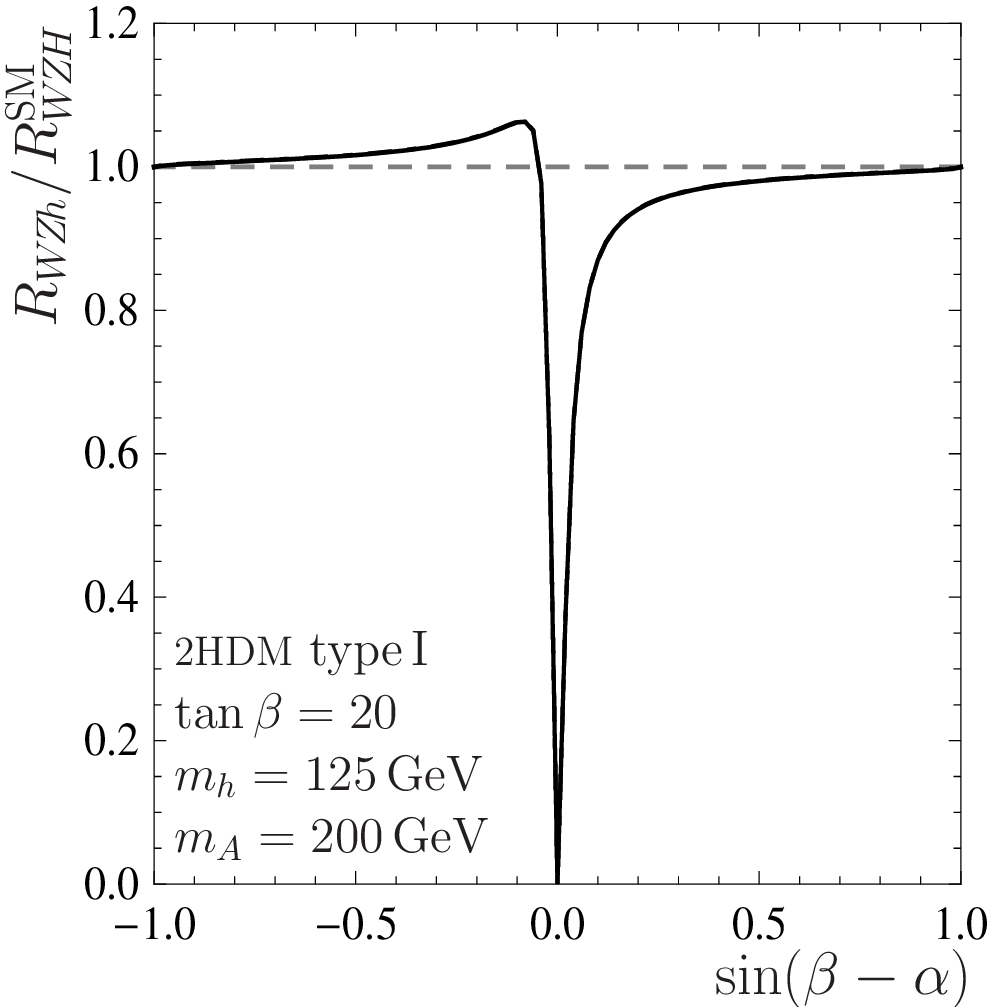}\\[-0.3cm]
(d) & (e) & (f)
\end{tabular}
\end{center}
\vspace{-0.6cm}
\caption{(a-c) $\sigma(pp\to \zh)$ (black/solid), $\sigma_{\dy}^{\zh}$
  (green/dotted), $\sigma_{{\ggzh}}$ (red/dashed) and $\sigma_{\bbzh}$
  (blue/dash-dotted) in pb for $\sqrt{s}=14$\,TeV and $m_h=125$\,GeV as
  a function of $\sin(\beta-\alpha)$ for a type\,I \thdm{} with
  $m_A=m_{H^0}=m_{H^\pm}=200$\,GeV using (a) $\tan\beta=1$, (b)
  $\tan\beta=5$ and (c) $\tan\beta=20$; (d-e) the ratio
  $\rwzh{h}/\rwzh{H}^\sm$ for the cases (a-c).  }
\label{fig:lighthiggsmA200_2T1}
\end{figure}


The black solid line in \figs{fig:lighthiggsmA200_2T1} (a-c) shows the
hadronic cross section $\sigma(pp\to \zh)$ in pb for $\sqrt{s}=14$\,TeV
and $m_h=125$\,GeV in a type\,I \thdm{} as a function of
$\sin(\beta-\alpha)$ for $\tan\beta=1$, $5$, and $20$, respectively. The
pseudoscalar mass is taken to be $m_A=200$\,GeV; the other \thdm{}
parameters are irrelevant for this example.  The remaining lines in
these figures correspond to different contributions entering the total
cross section.

The \dy{} terms (green/dotted) are proportional to the square of
$g_{VV}^{h}=\sin(\beta-\alpha)$ and do not depend on any other
parameters. They are therefore identical in all three cases, vanish at
$\sin(\beta-\alpha)=0$, and are symmetric around this point.

The $\ggzh$ terms are displayed as the red/dashed line; their amplitude
involves contributions proportional to $g_{VV}^h$ and contributions
proportional to the Yukawa couplings $g_b^h$ and $g_t^h$.  Additionally,
the process $gg\rightarrow A^*\rightarrow Zh$ is relevant, whose
amplitude is proportional to $g_{AZ}^{h}=\cos(\beta-\alpha)$ and $g_b^A$
or $g_t^A$.  In case of the type\,I \thdm{}, both $g_b^h$ and $g_t^h$
(and $g_b^A$ and $g_t^A$) decrease with increasing $\tan\beta$, which
implies that for larger values of $\tan\beta$ the overall cross section
$\sigma_{{\ggzh}}$ is dominated by $g_{VV}^h$. For smaller values of
$\tan\beta$, the $\ggzh$ terms can become dominant even for rather large
values of $|\sin(\beta-\alpha)|$.

The $\bbzh$ contribution (blue/dash-dotted) is of no relevance in the
type\,I \thdm{}. Note that for $\sin(\beta-\alpha)=\pm 1$, all
contributions assume their \sm{} values as then
$\cos\alpha/\sin\beta = \pm 1$ and therefore $y_{t,b} = \pm y_{t,b}^\sm$
(see Appendix~\ref{app:2HDM}).

\figs{fig:lighthiggsmA200_2T1}\,(d-f) show the corresponding ratios
$\rwzh{h}$ with respect to the \sm{} ratio $\rwzh{H}^\sm$
for the three different values of $\tan\beta$ as the solid
line by taking into account all available contributions. At
$\tan\beta=1$, the ratio exhibits a remarkably strong dependence on
$\sin(\beta-\alpha)$, also close to the edges. Towards larger values of
$\tan\beta$, the curves become flatter for $\sin(\beta-\alpha)=\pm 1$
since the $\zh$-specific contributions become more and more
suppressed. At $\sin(\beta-\alpha)=0$, $\wh$ production is identical to
zero in our approximation (recall that we neglect
$\sigma_\text{I}^{\vh}$ as well as
$\sigma_\text{II}^{\zh}$). \figs{fig:lighthiggsmA200_2T1}\,(d-f) also
include a dash-dotted curve which, for the parameters of this example, is
almost indistinguishable from the solid curve. It shows the ratio
$\rwzh{h}$ {\it without} the $\sigma_{\bbzh}$ contribution; this might
be a useful quantity if the $\bbzphi$ process can be excluded
efficiently by applying a $b$-tagging veto. In the current example, this
is irrelevant, of course, but we will use the same notation in other
examples, where the $\bbzphi$ process is numerically much more
important.


\begin{figure}[ht]
\begin{center}
\begin{tabular}{ccc}
\includegraphics[width=0.3\textwidth]{%
  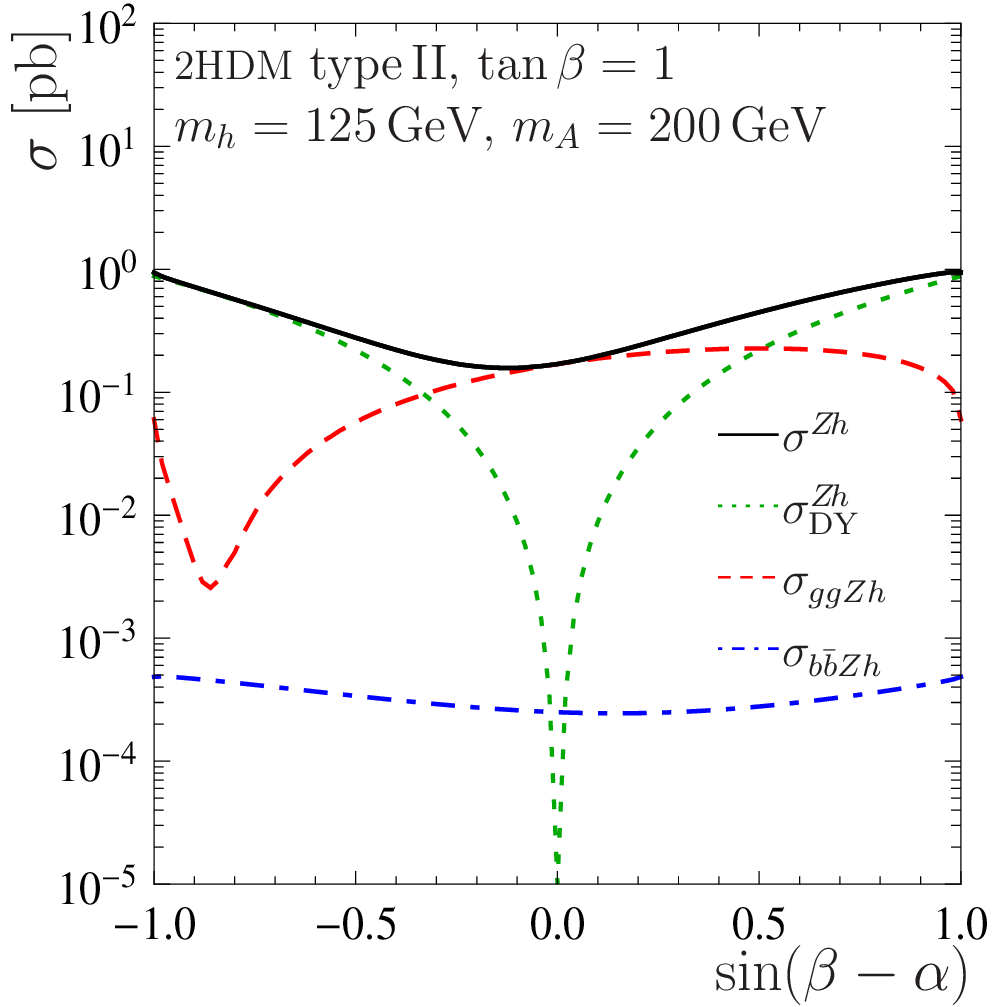} &
\includegraphics[width=0.3\textwidth]{%
  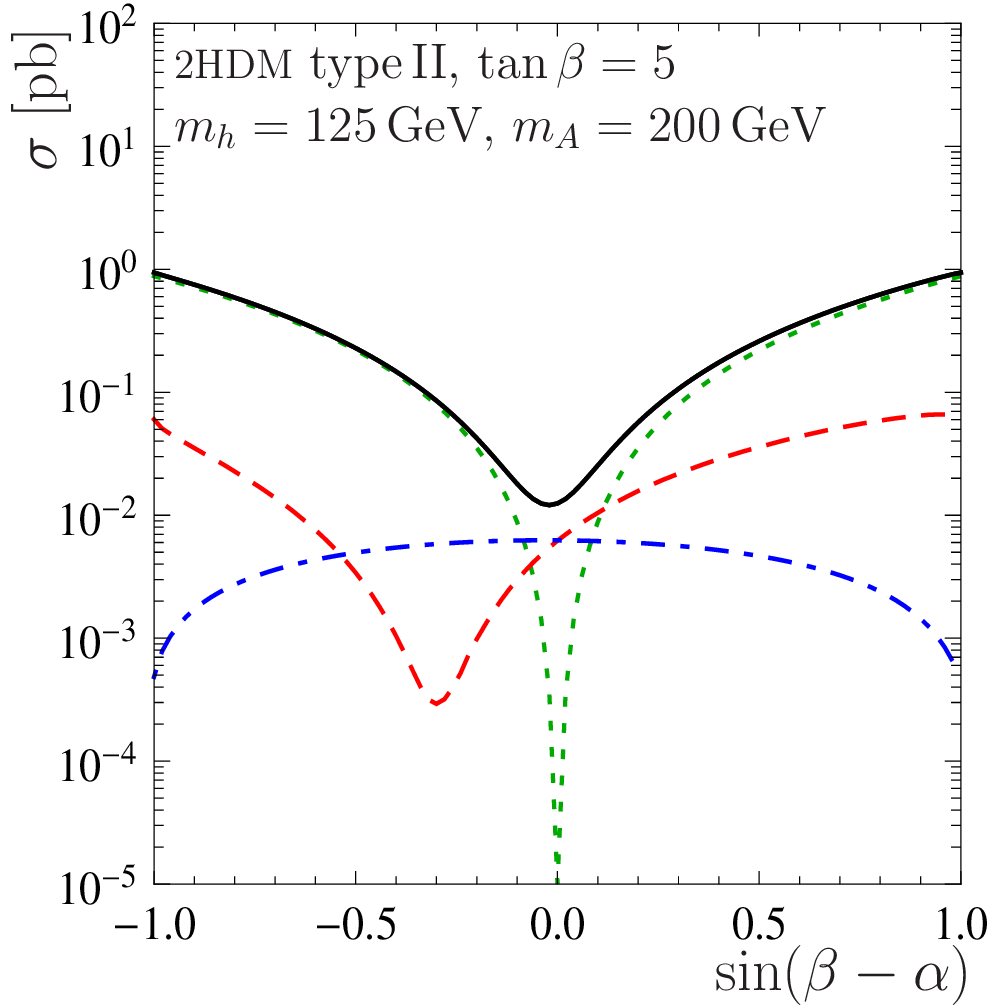} &
\includegraphics[width=0.3\textwidth]{%
  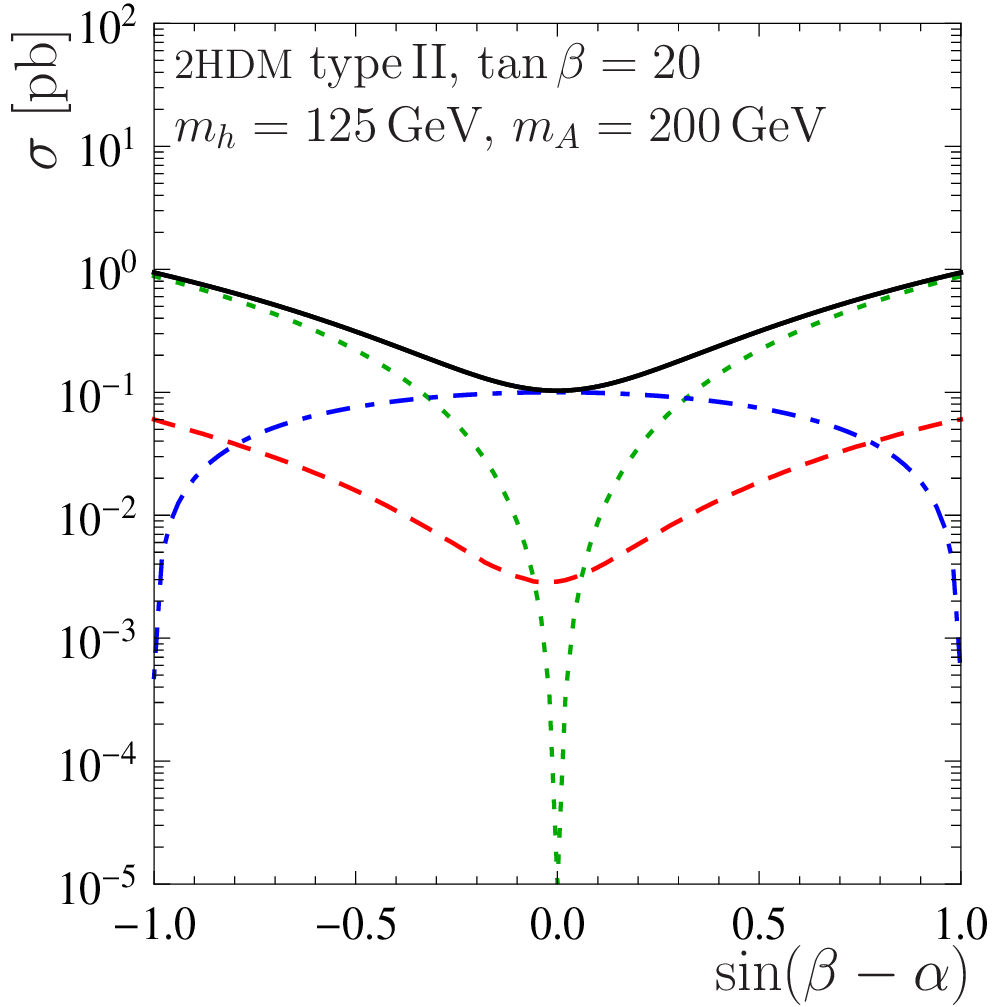} \\[-0.3cm]
(a) & (b) & (c)\\
\includegraphics[width=0.3\textwidth]{%
  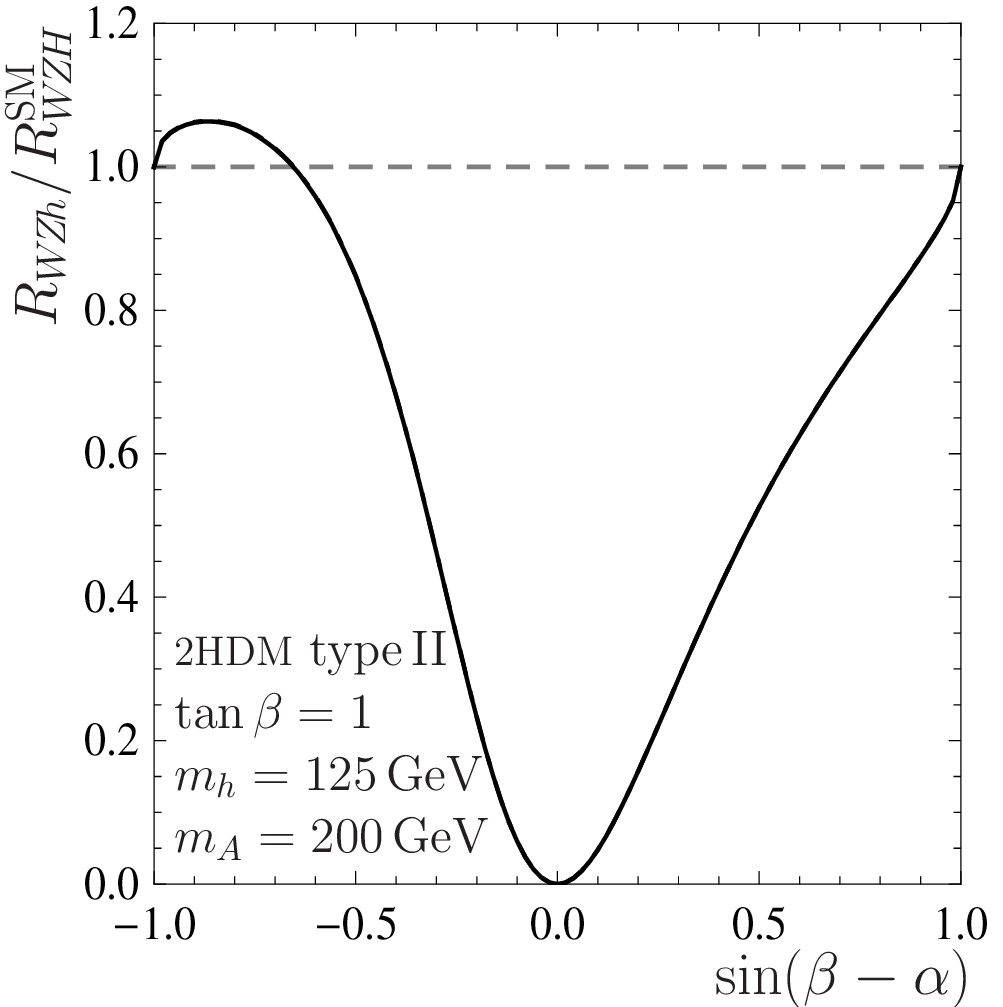} &
\includegraphics[width=0.3\textwidth]{%
  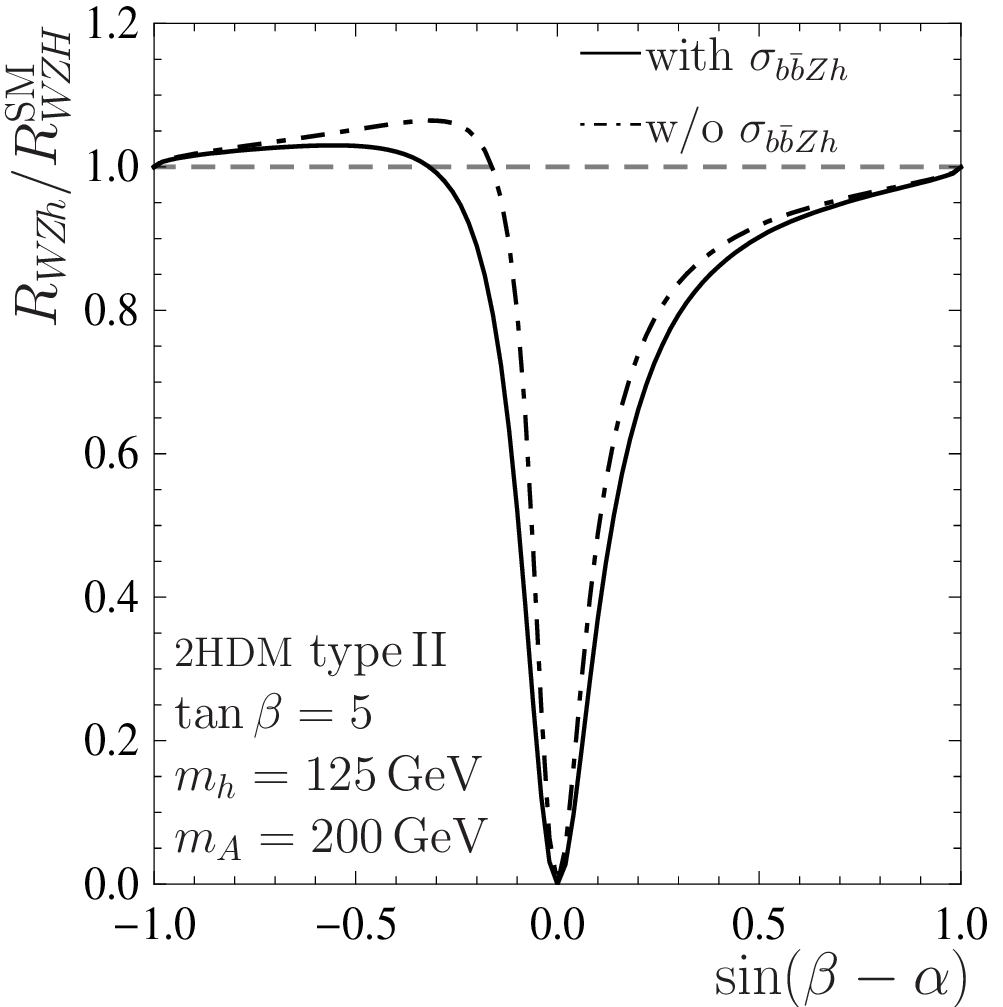}&
\includegraphics[width=0.3\textwidth]{%
  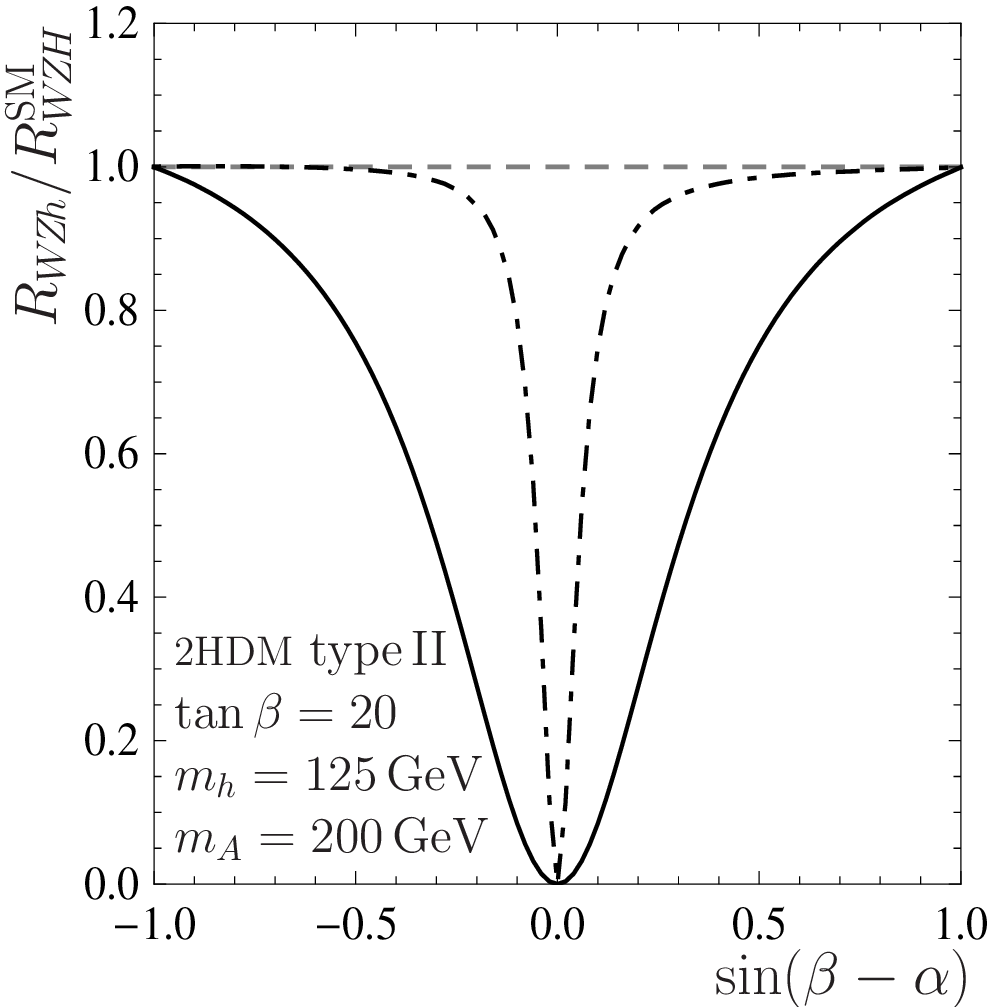}\\[-0.3cm]
(d) & (e) & (f)
\end{tabular}
\end{center}
\vspace{-0.6cm}
\caption{(a-c) $\sigma(pp\to \zh)$ (black/solid), $\sigma_{\dy}^{\zh}$
  (green/dotted), $\sigma_{{\ggzh}}$ (red/dashed) and $\sigma_{\bbzh}$
  (blue/dash-dotted) in pb for $\sqrt{s}=14$\,TeV and $m_h=125$\,GeV as
  a function of $\sin(\beta-\alpha)$ for type\,II \thdm{} with
  $m_A=m_{H^0}=m_{H^\pm}=200$\,GeV using (a) $\tan\beta=1$, (b)
  $\tan\beta=5$ and (c) $\tan\beta=20$; (d-e) the ratio
  $\rwzh{h}/\rwzh{H}^\sm$ for the cases (a-c), respectively with
  $\sigma_{\bbzh}$ (solid) and without (dash-dotted).  }
\label{fig:lighthiggsmA200_2T2}
\end{figure}


In \fig{fig:lighthiggsmA200_2T2} we show the corresponding results for a
type\,II \thdm{}.  The \dy{} contributions are identical to type\,I, see
\fig{fig:lighthiggsmA200_2T1}.  For $\tan\beta=1$, also the $\ggzh$
curve is quite similar to the type\,I case because the $b$-loop
contribution is very small in both cases, and the top Yukawa coupling is
identical. Since $\sigma_{\bbzh}$ is again very small, also the total
cross section --- and therefore also the ratio $\rwzh{h}$, see
\fig{fig:lighthiggsmA200_2T2}\,(d) --- is almost the same as for
type\,I.
In the limit $\sin(\beta-\alpha)=\pm 1$, again all curves assume their
\sm{} value for the same reason as discussed above.

With increasing $\tan\beta$, the bottom Yukawa coupling becomes more and
more important. For the $\ggzh$ contribution, this results in a
complicated interplay with the top Yukawa and the $gg\to \{Z,A\}^\ast\to
\zh$ terms, which first reduces the $\ggzh$ fraction due to the
destructive top-bottom interference, before the bottom effects start to
dominate, leading to an increase.  The $\bbzh$ contribution, on the
other hand, increases drastically with $\tan\beta$ and even starts to
exceed the other contributions in a wider and wider range of
$\sin(\beta-\alpha)$.  The importance of this channel is reflected also
in the ratio plots \fig{fig:lighthiggsmA200_2T2}\,(d-f): without it
(dash-dotted line), the $\sin(\beta-\alpha)$ dependence towards $\pm 1$ for
$\tan\beta >1$ is much flatter than when it is taken into account.

These examples already show that a simple reweighting of the \sm{} cross
section by $(g_{VV}^h)^2$ (which would resemble the behaviour of the
\dy{} contributions in the shown figures) is not appropriate in
general. Even in regions where $g_{VV}^h$ is close (but not equal) to
$\pm 1$, the Yukawa contributions to the amplitude proportional to
$g_t^h$ and $g_b^h$ (and $g_t^A$ and $g_b^A$) can become non-negligible.


\begin{figure}[ht]
\begin{center}
\begin{tabular}{ccc}
\includegraphics[width=0.3\textwidth]{%
  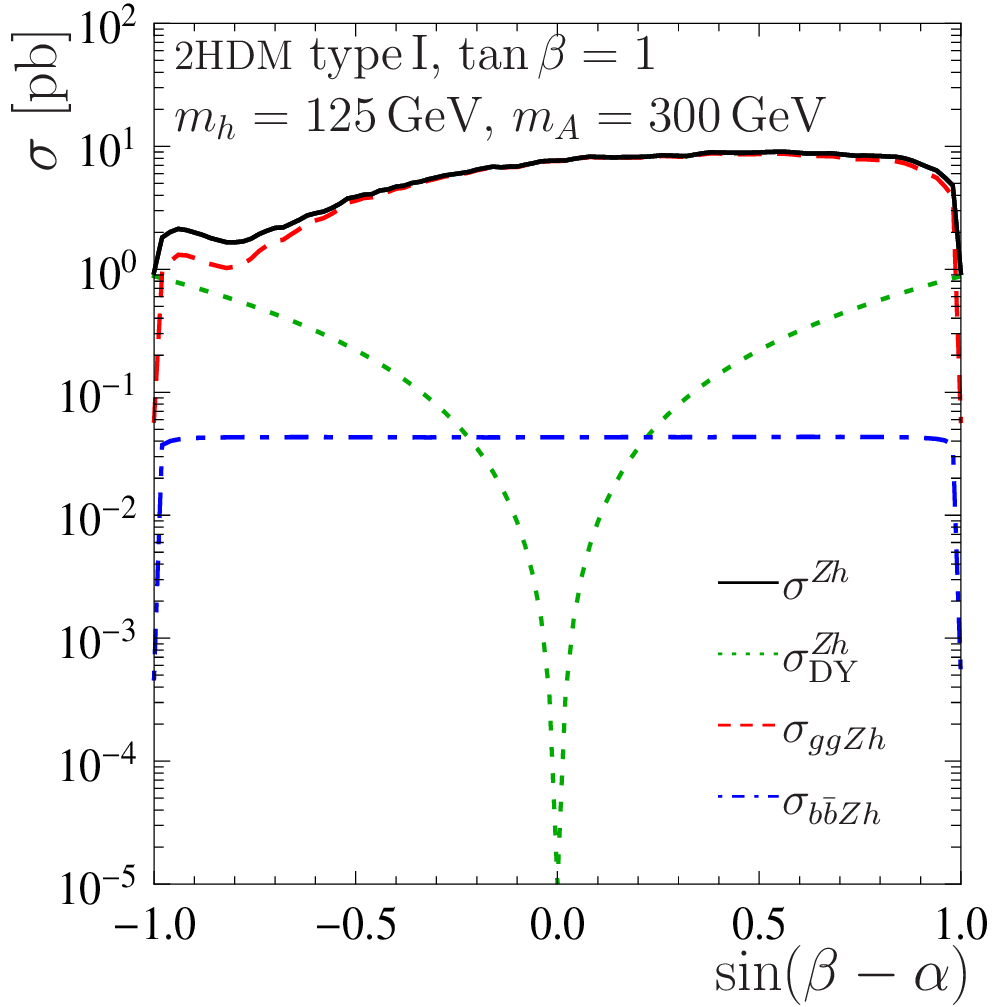} &
\includegraphics[width=0.3\textwidth]{%
  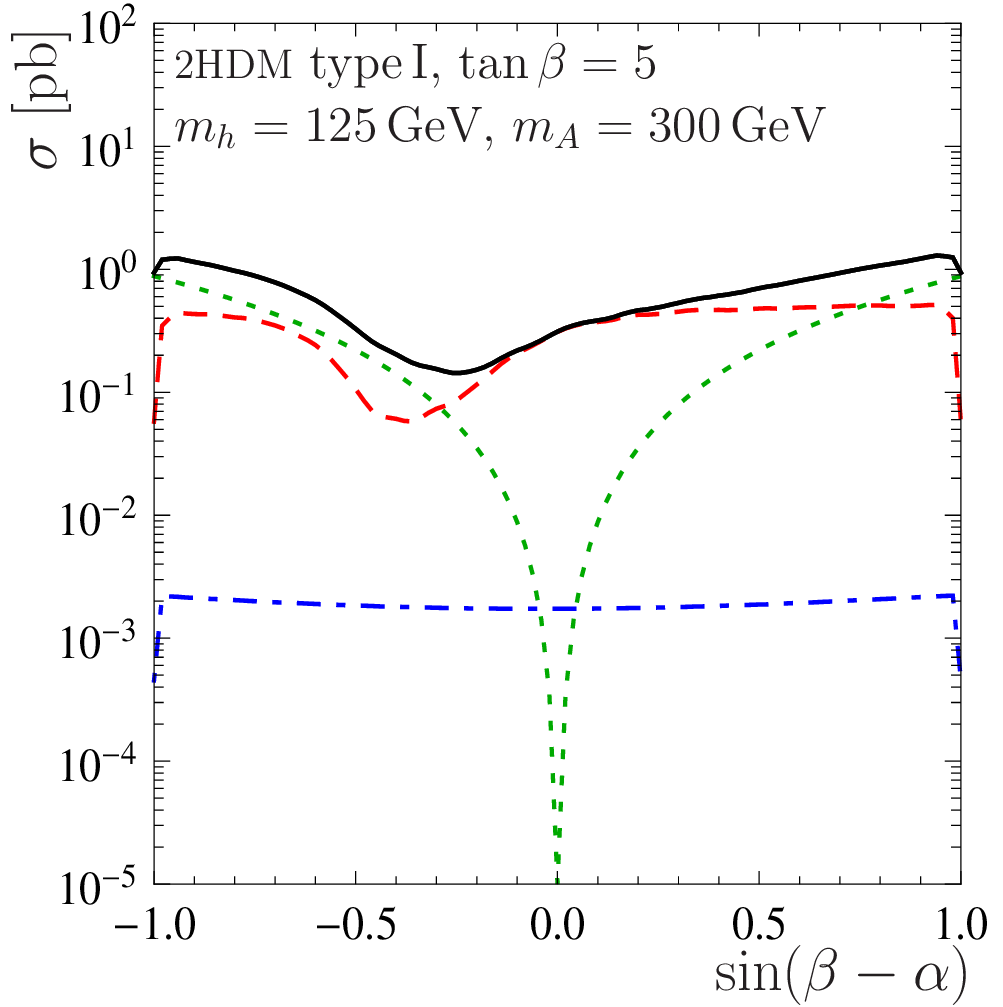} &
\includegraphics[width=0.3\textwidth]{%
  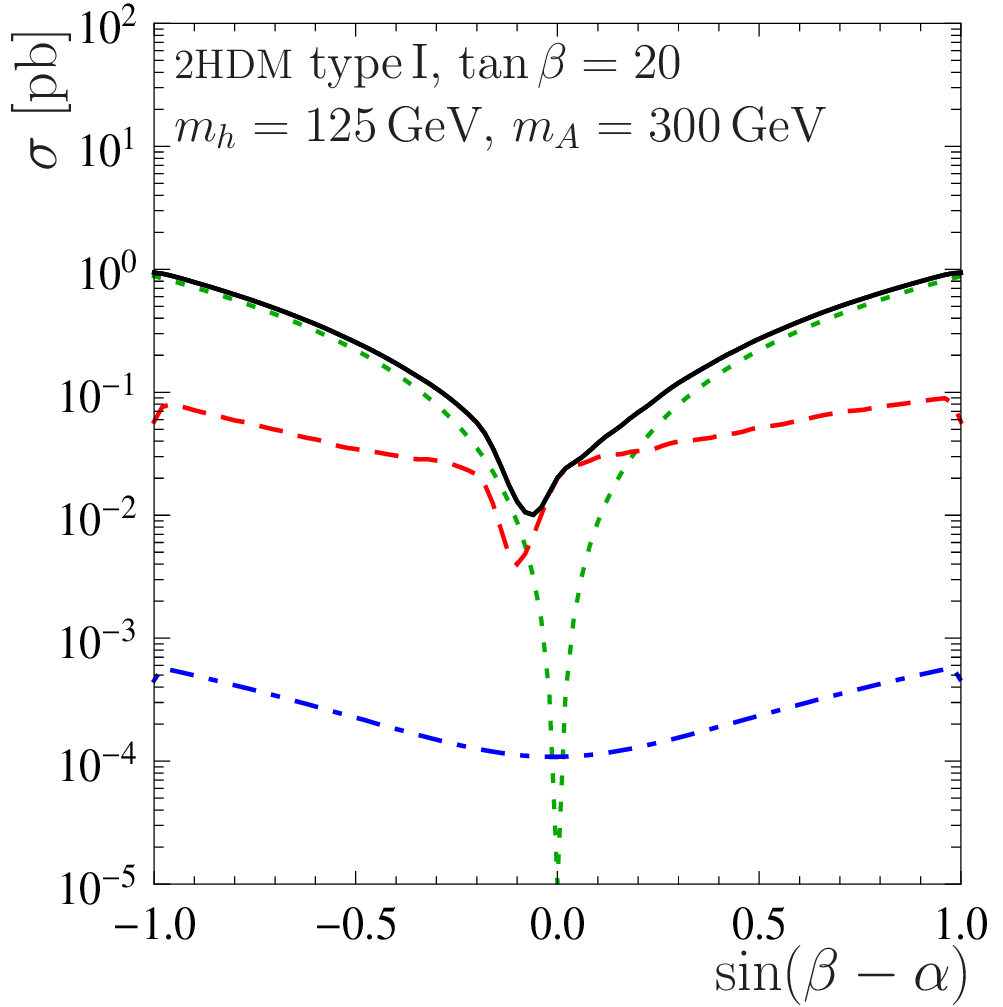} \\[-0.3cm]
(a) & (b) & (c)\\
\includegraphics[width=0.3\textwidth]{%
  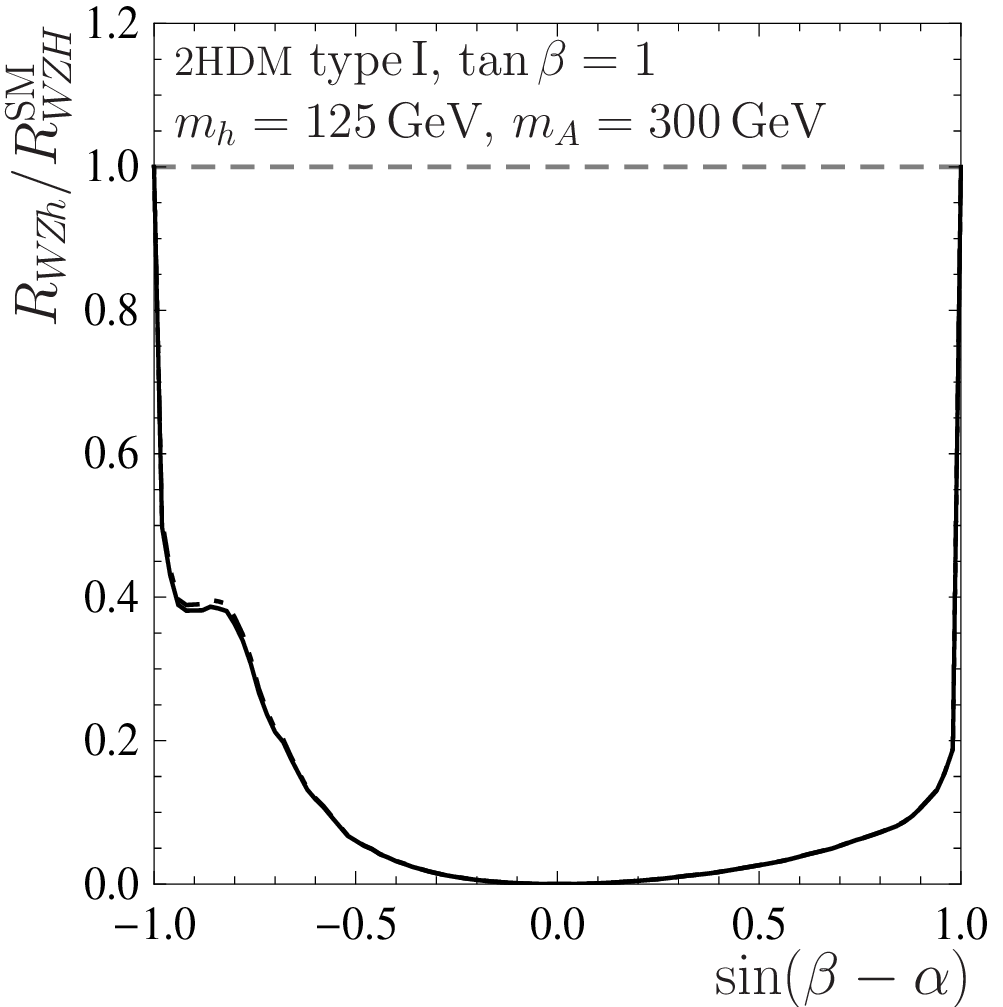} &
\includegraphics[width=0.3\textwidth]{%
  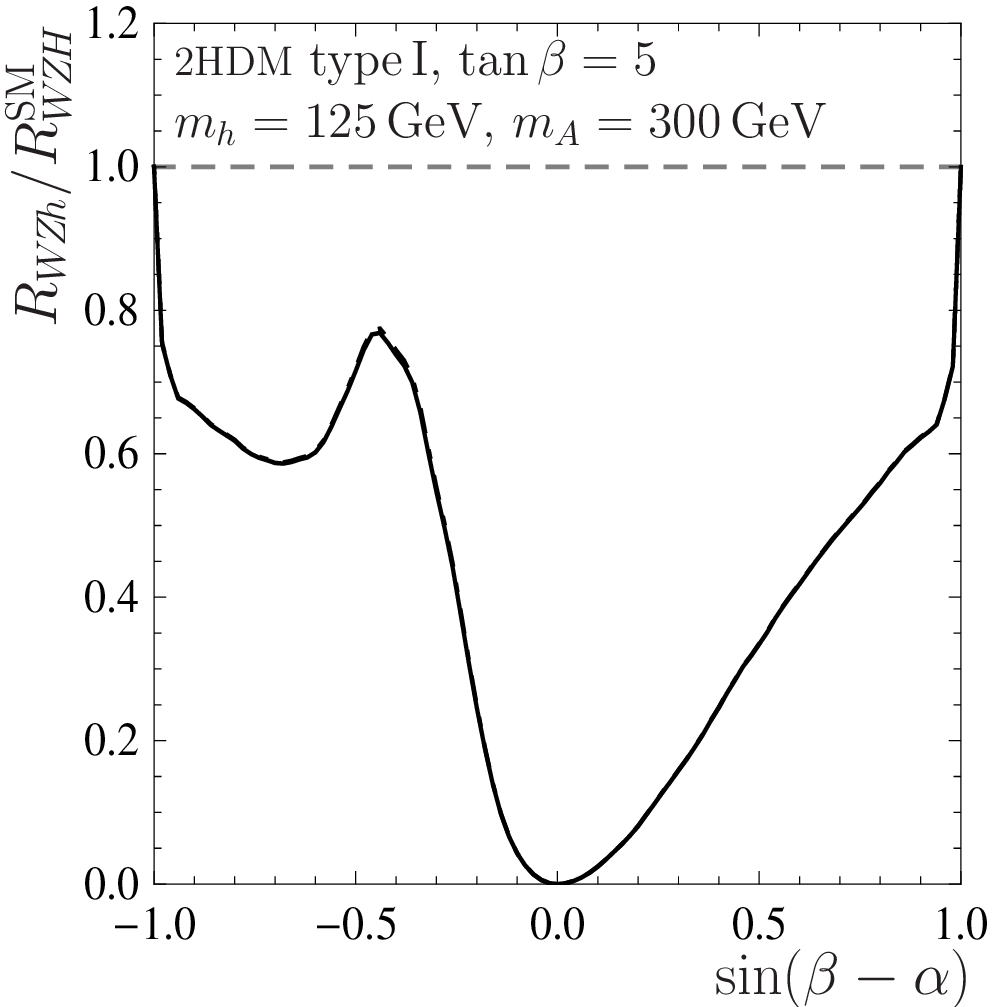}&
\includegraphics[width=0.3\textwidth]{%
  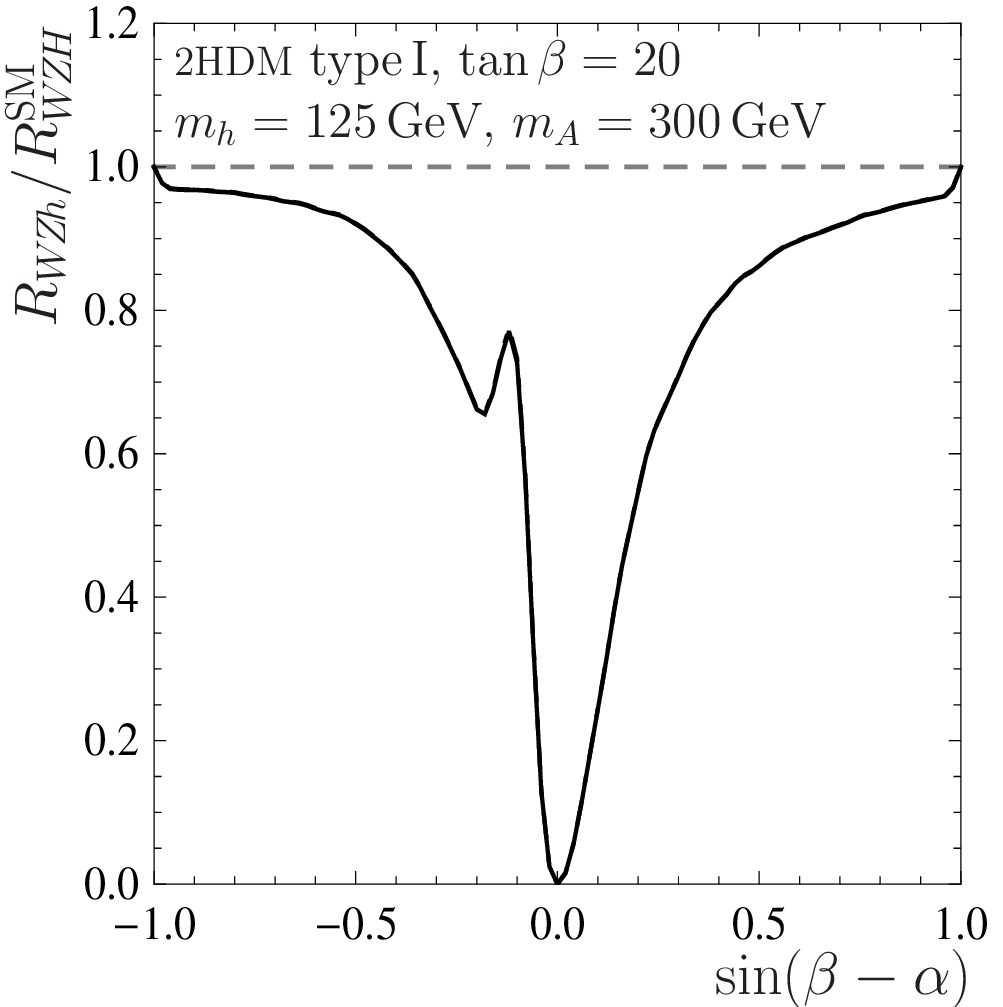}\\[-0.3cm]
(d) & (e) & (f)
\end{tabular}
\end{center}
\vspace{-0.6cm}
\caption{(a-c) $\sigma(pp\to \zh)$ (black/solid), $\sigma_{\dy}^{\zh}$
(green/dotted), $\sigma_{{\ggzh}}$ (red/dashed)
and $\sigma_{\bbzh}$ (blue/dash-dotted) in pb for $\sqrt{s}=14$\,TeV and
$m_h=125$\,GeV as a function of
$\sin(\beta-\alpha)$ for type\,I \thdm{}  with $m_A=m_{H^0}=m_{H^\pm}=300$\,GeV
using (a) $\tan\beta=1$, (b) $\tan\beta=5$ and (c) $\tan\beta=20$;
(d-e) the ratio $\rwzh{h}/\rwzh{H}^\sm$ for the cases (a-c).
}
\label{fig:lighthiggsmA300_2T1}
\end{figure}


If the mass of the pseudoscalar Higgs exceeds the kinematical threshold,
$m_A>m_h+m_Z$, it can be produced on-shell via $gg\rightarrow
A\rightarrow Zh$ or $b\overline{b}\rightarrow A\rightarrow Zh$, leading
to a substantial increase of these contributions to the cross sections
under consideration, see \fig{fig:resonance}.  The parameters for
\fig{fig:lighthiggsmA300_2T1} are identical to those of
\fig{fig:lighthiggsmA200_2T1}, except that the pseudoscalar mass $m_A$
is increased from $200$\,GeV to $300$\,GeV. Since now we have to take into
account the total width of the pseudoscalar Higgs, which in turn depends
on $m_{12}$ and the heavy and charged Higgs masses $m_{H^0}$
and $m_{H^\pm}$, we have to
specify these parameters as well. Unless stated otherwise, we will set
$m_{12}=0$ and $m_{H^0}=m_{H^\pm}=m_A$ in what follows. 

The \dy{} contributions are obviously not affected by the change in
$m_A$.  For small $\tan\beta$ and $\sin(\beta-\alpha)\neq \pm 1$, the
resonant $gg\to A\to \zh$ process dramatically increases the $\ggzh$
contribution, making it dominant by more than a factor of 10 in most of
the $\sin(\beta-\alpha)$ region.  The fact that it features a relatively
constant behaviour for most values of $\alpha$ can be understood as
follows: For small values of $\tan\beta$, the amplitude is approximately
proportional to $(g_{Z}^{Ah})^2=\cos^2(\beta-\alpha)$. Squaring the
propagator of \eqn{eq:bw} results in a Breit-Wigner function. If we
write the latter function in the narrow-width approximation
\begin{align}
 \frac{1}{(\hat{s}-m_A^2)^2+m_A^2\Gamma_A^2}\rightarrow 
 \frac{\pi}{m_A\Gamma_A}\delta(\hat{s}-m_A^2)\,,
\end{align}
the factor $\cos^2(\beta-\alpha)$ is canceled by $1/\Gamma_A$ as long
as the decay mode $A\rightarrow Zh$ is dominant.
For $|\sin(\beta-\alpha)|\rightarrow 1$ the decay mode $A\rightarrow Zh$
is suppressed. As soon as its partial width drops below the other partial widths,
mainly $\Gamma(A\rightarrow b\overline{b})$, the contribution $\sigma_{\ggzh}$
is instantly turning towards its \sm{} value.
In case $m_A>2m_t$ the decay channel $A\rightarrow t\overline{t}$
opens and lowers the pseudoscalar contributions.
For increasing
$\tan\beta$, the suppression of $g_t^A$ and $g_b^A$ reduces the impact
of the pseudoscalar as internal particle. A similar statement holds for
the contribution $\sigma_{\bbzh}$, but their
contribution is down by a factor $10^{-2}$ to $10^{-3}$.

Similarly to the case $m_A=200$\,GeV, the impact of the $\ggzh$ and
$\bbzh$ terms decreases towards larger $\tan\beta$ and the total cross
section is described by the pure \dy{} terms better and better.  For
$\tan\beta\lesssim 5$, however the ratio $\rwzh{h}$ significantly
deviates from its \sm{} value even in the regions very close to
$\sin(\beta-\alpha)=\pm 1$ (see \figs{fig:lighthiggsmA300_2T1}\,(d,e)).  In
fact, for $\tan\beta=1$, $\rwzh{h}<0.7$ (meaning
$\rwzh{h}/\rwzh{h}^\sm<0.4$) for $-0.91 < \sin(\beta-\alpha) < 0.99$.


\begin{figure}[ht]
\begin{center}
\begin{tabular}{ccc}
\includegraphics[width=0.3\textwidth]{%
  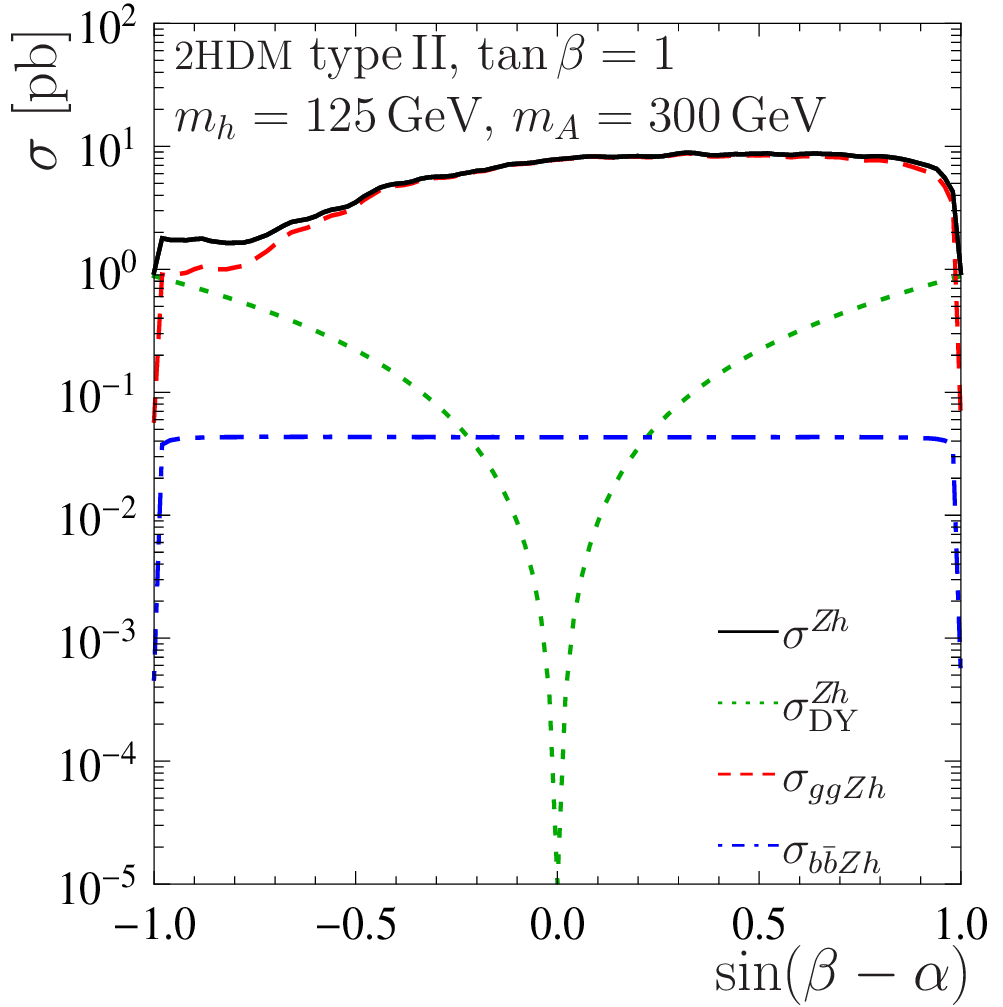} &
\includegraphics[width=0.3\textwidth]{%
  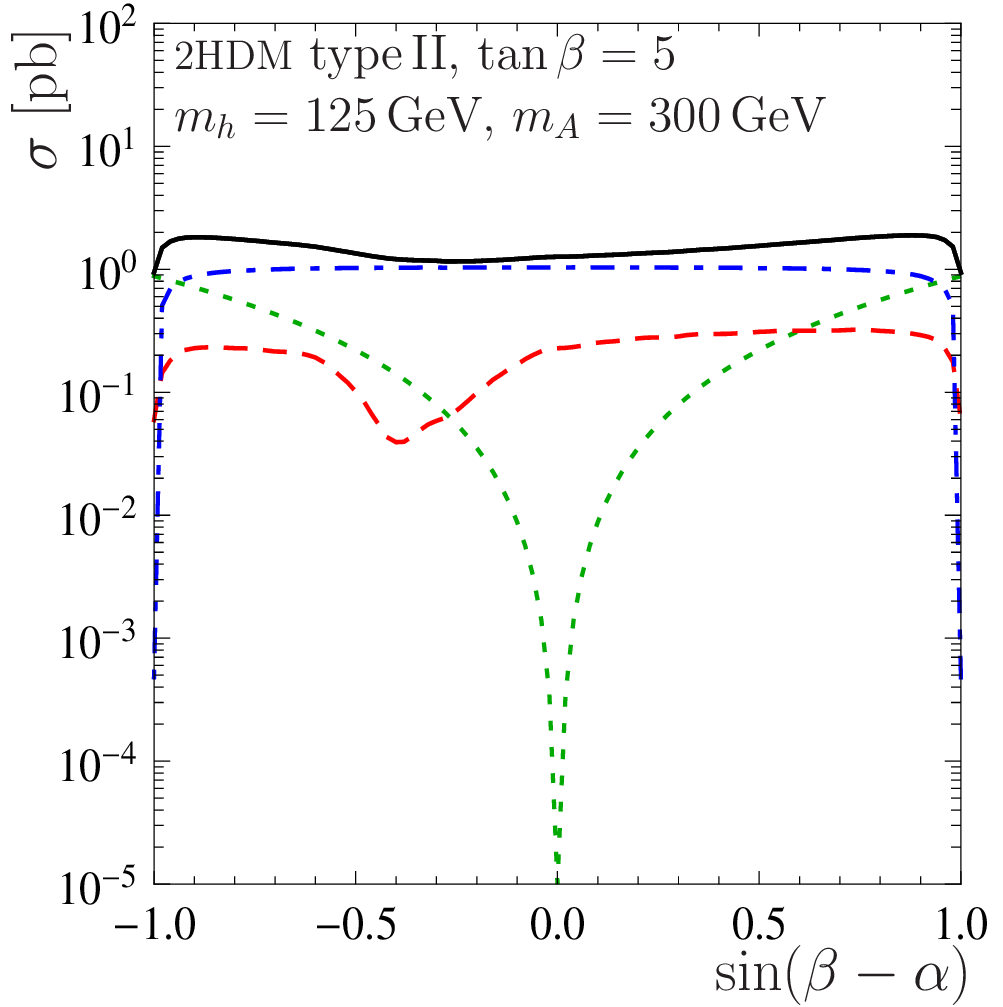} &
\includegraphics[width=0.3\textwidth]{%
  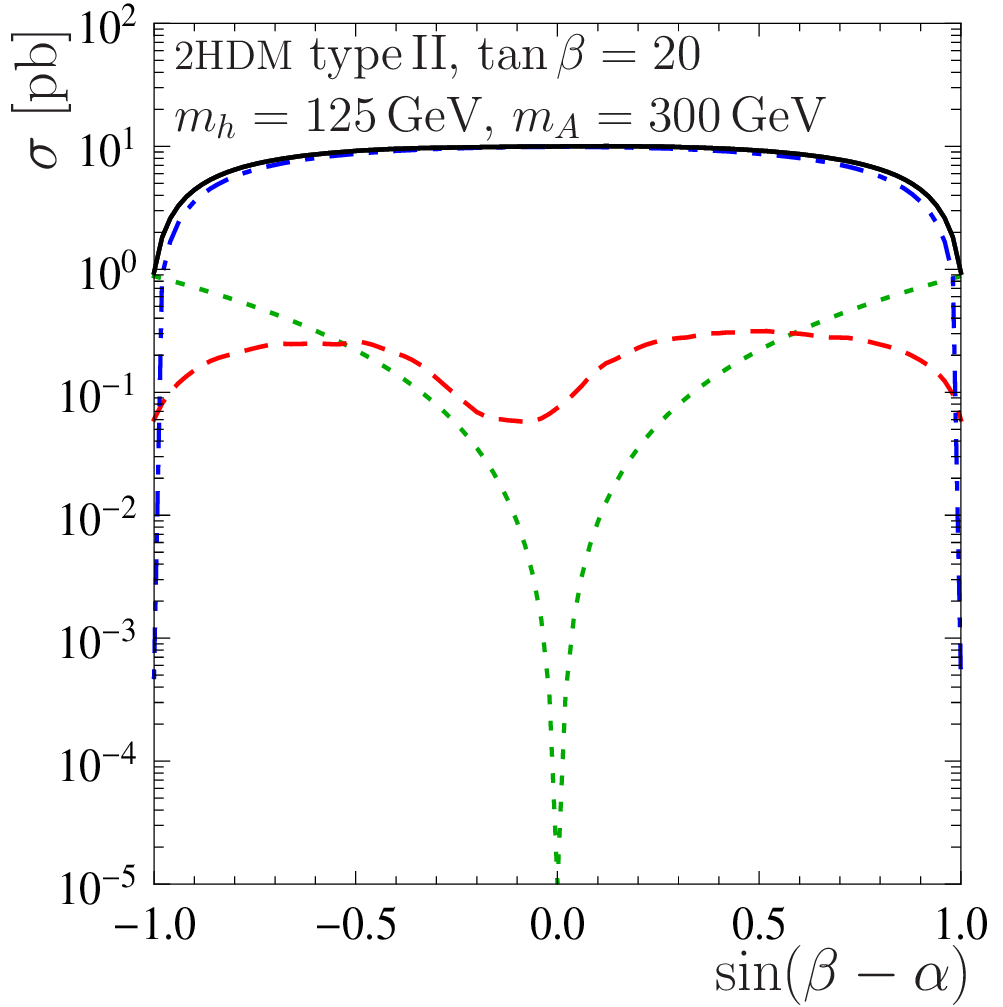} \\[-0.3cm]
(a) & (b) & (c)\\
\includegraphics[width=0.3\textwidth]{%
  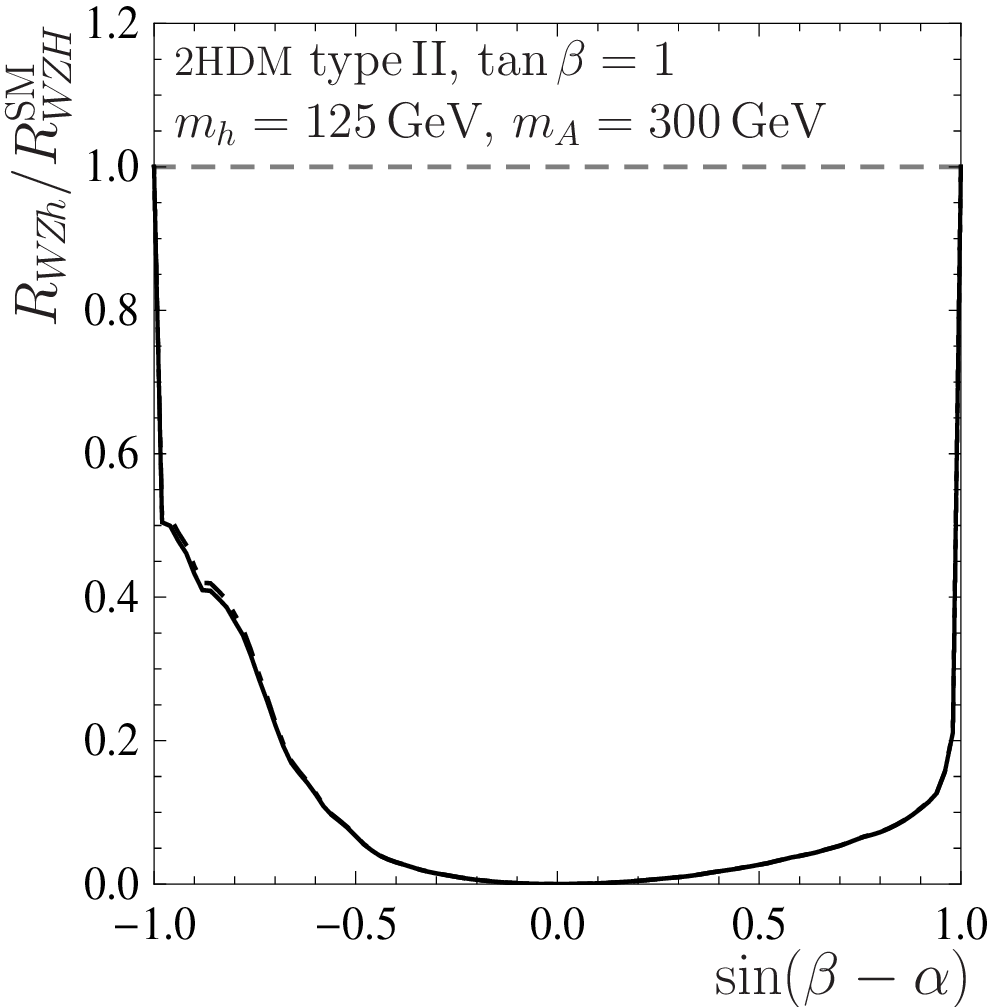} &
\includegraphics[width=0.3\textwidth]{%
  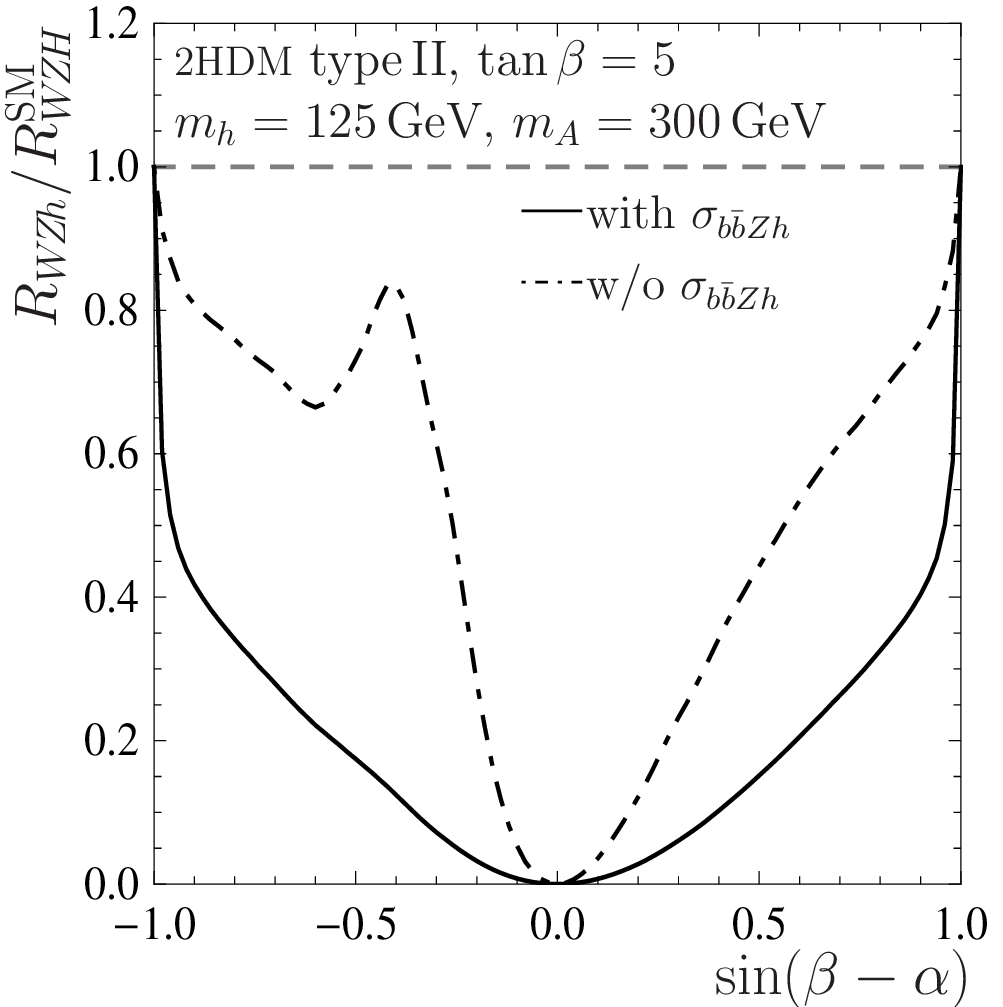}&
\includegraphics[width=0.3\textwidth]{%
  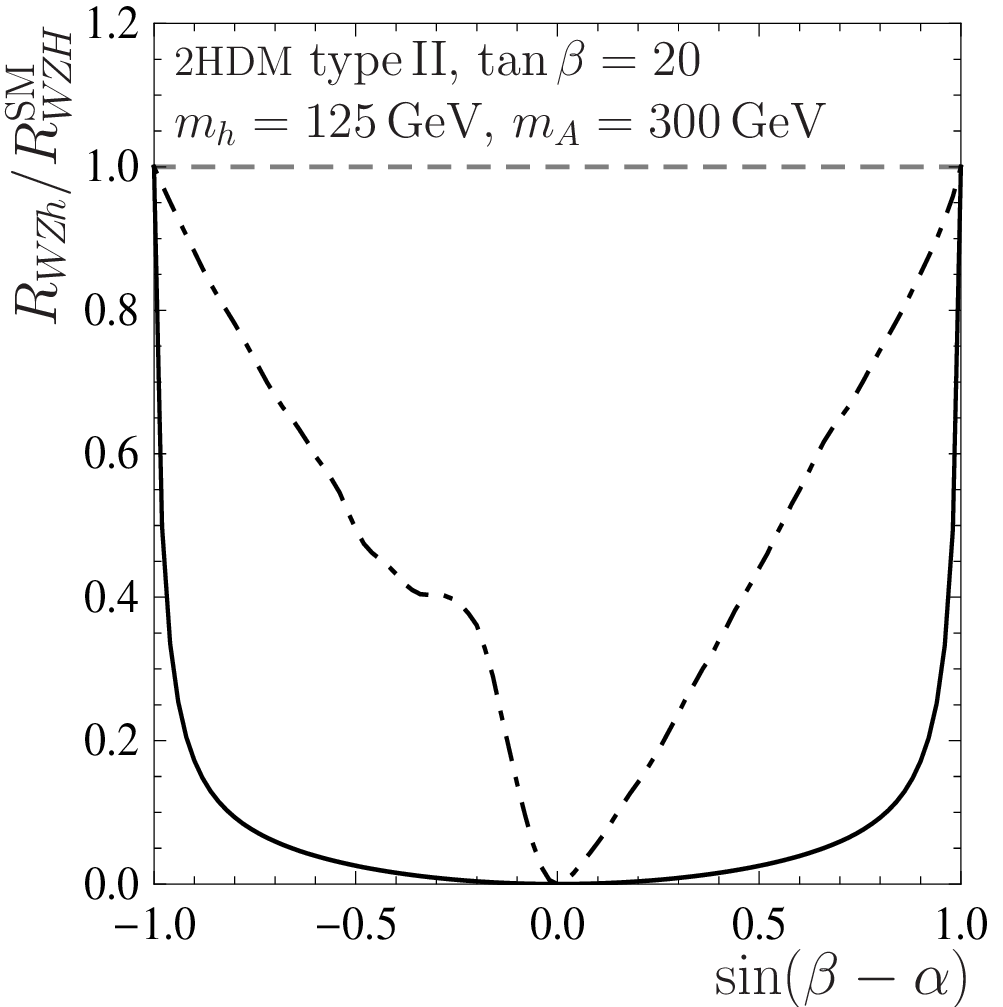}\\[-0.3cm]
(d) & (e) & (f)
\end{tabular}
\end{center}
\vspace{-0.6cm}
\caption{(a-c) $\sigma(pp\to \zh)$ (black/solid), $\sigma_{\dy}^{\zh}$
(green/dotted), $\sigma_{{\ggzh}}$ (red/dashed)
and $\sigma_{\bbzh}$ (blue/dash-dotted) in pb for $\sqrt{s}=14$\,TeV and
$m_h=125$\,GeV as a function of
$\sin(\beta-\alpha)$ for type\,II \thdm{} with $m_A=m_{H^0}=m_{H^\pm}=300$\,GeV
using (a) $\tan\beta=1$, (b) $\tan\beta=5$ and (c) $\tan\beta=20$;
(d-e) the ratio $\rwzh{h}/\rwzh{H}^\sm$ for the cases (a-c).
}
\label{fig:lighthiggsmA300_2T2}
\end{figure}


\fig{fig:lighthiggsmA300_2T2} shows the cross section
dependence on $\sin(\beta-\alpha)$ for a
type\,II \thdm{}. The parameters are the same as in
\fig{fig:lighthiggsmA200_2T2}, except that the masses
$m_A=m_{H^0}=m_{H^\pm}$ are increased from $200$\,GeV to
$300$\,GeV. Similarly to the case for $m_A=200$\,GeV, the curve for
$\tan\beta=1$ very much resembles the one for type\,I, which is again due
to the little impact of the bottom contributions, of course. Towards
larger values of $\tan\beta$, the $\bbzh$ fraction becomes dominant very
quickly. The behaviour of the ratio $\rwzh{h}$ is thus quite remarkable
for all values of $\tan\beta$, see
\figs{fig:lighthiggsmA300_2T2}\,(d-f). If the $\bbzh$ component is
disregarded (for reasons discussed above), the deviation of $\rwzh{h}$
from its \sm{} value becomes less drastic towards larger $\tan\beta$, in
particular close to $\sin(\beta-\alpha)=\pm 1$, but it still amounts to
about 20\% for $|\sin(\beta-\alpha)|<0.9$ at
$\tan\beta=5$.

The statements of this section are also valid in case of other choices
of the parameters, in particular for other choices of $m_{12}$, since
$m_{12}$ only affects the decay widths of the internal pseudoscalar, as
discussed above. If however the heavy Higgs and/or charged Higgs bosons are
light enough to open the
decay channels $A\rightarrow \zHo$ and/or $A\rightarrow W^\pm H^\mp$
in addition, the decay width of $A$ is
enlarged and thus the $\ggzh$ and $\bbzh$ cross section contributions
get reduced.  We will discuss this effect in the subsequent section for
the heavy Higgs, where $A\rightarrow Zh$, $A\rightarrow \zHo$ and
$A\rightarrow W^\pm H^\mp$ are possible.


\subsection{Heavy and pseudoscalar Higgs}


\begin{figure}[ht]
\begin{center}
\begin{tabular}{cc}
\includegraphics[width=0.3\textwidth]{%
  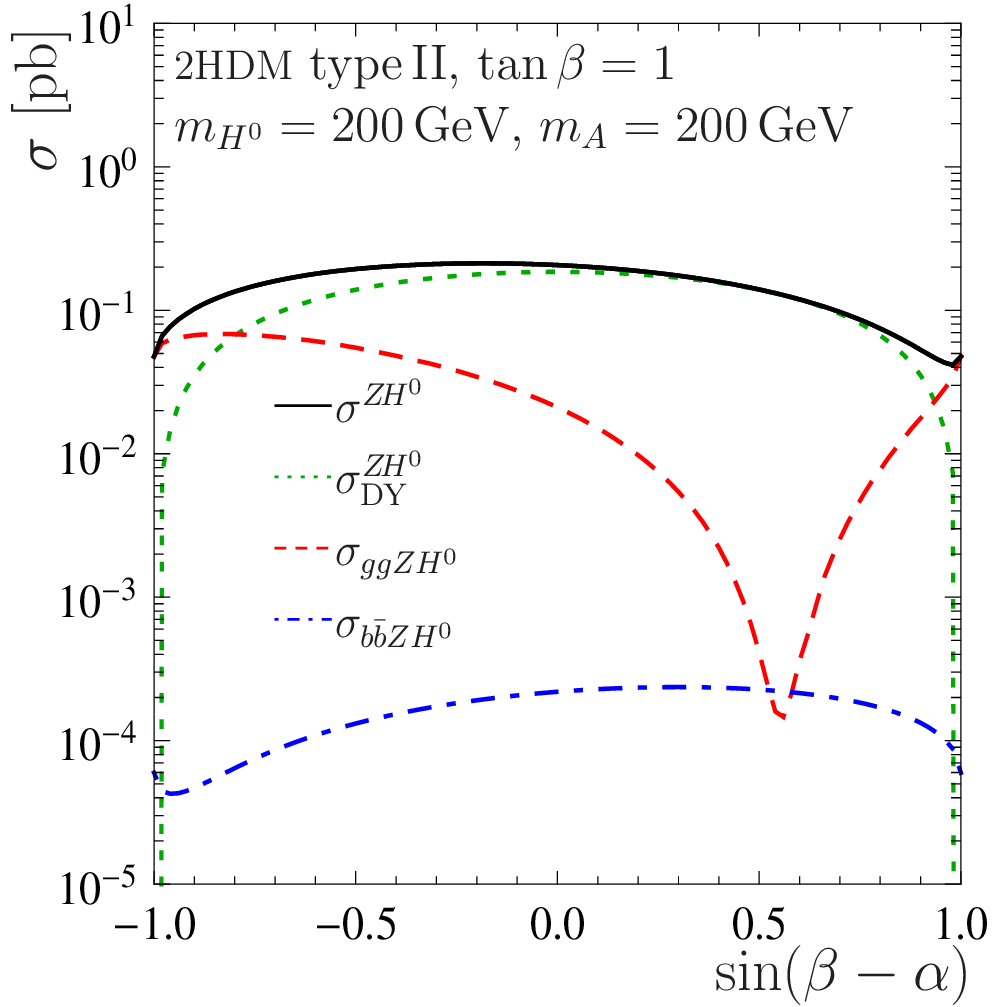} &
\includegraphics[width=0.3\textwidth]{%
  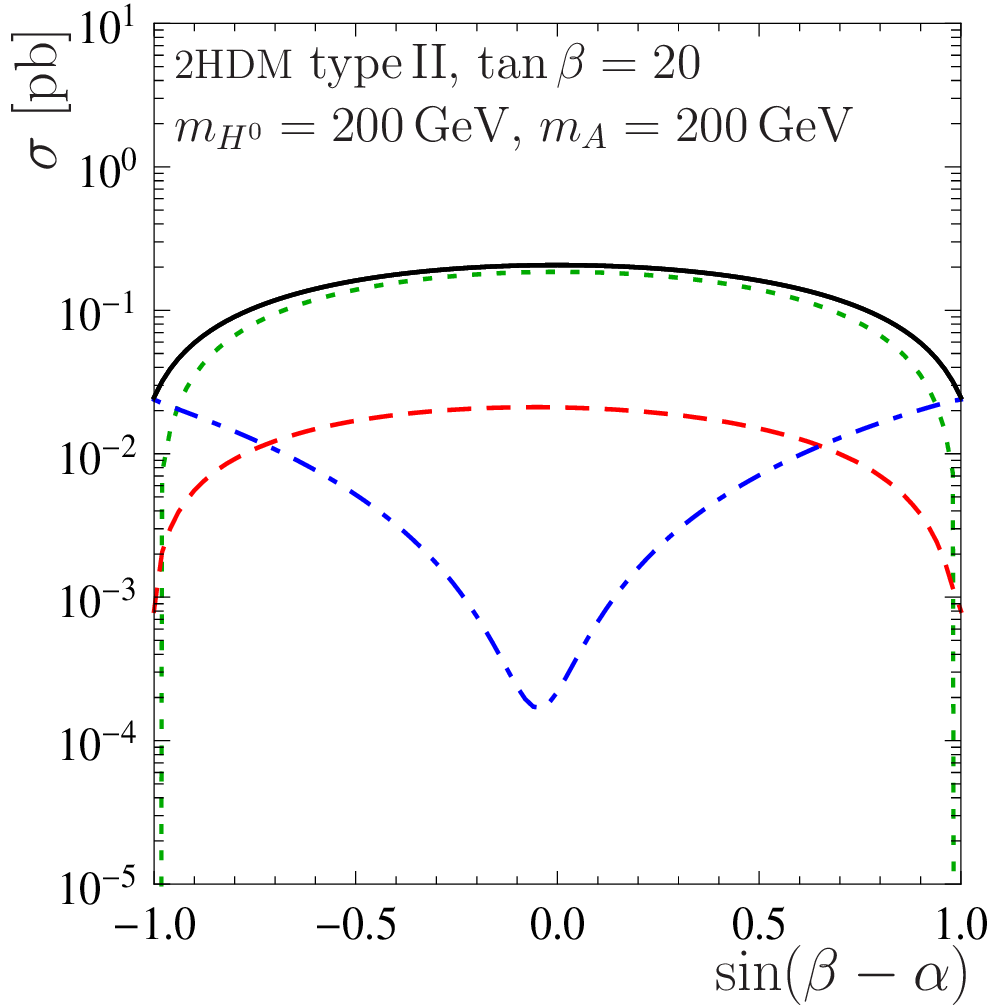} \\[-0.3cm]
(a) & (b) \\
\includegraphics[width=0.3\textwidth]{%
  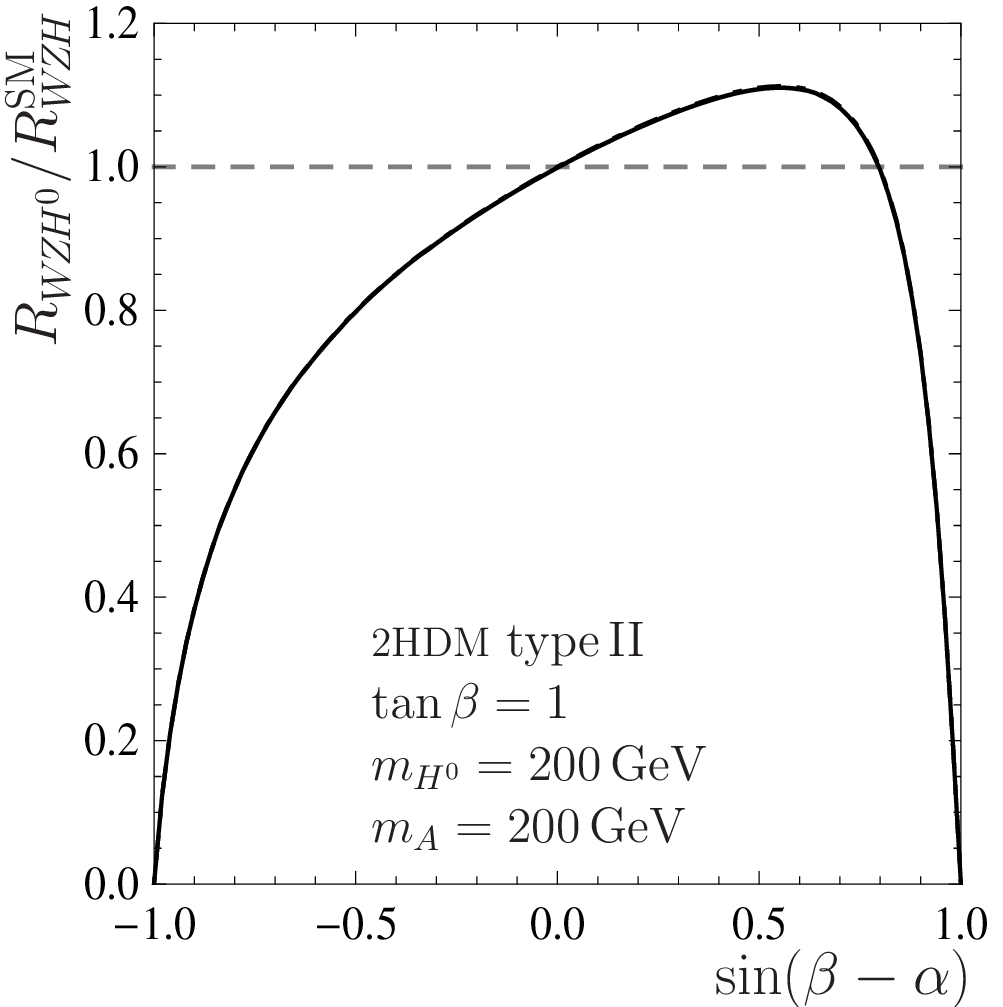} &
\includegraphics[width=0.3\textwidth]{%
  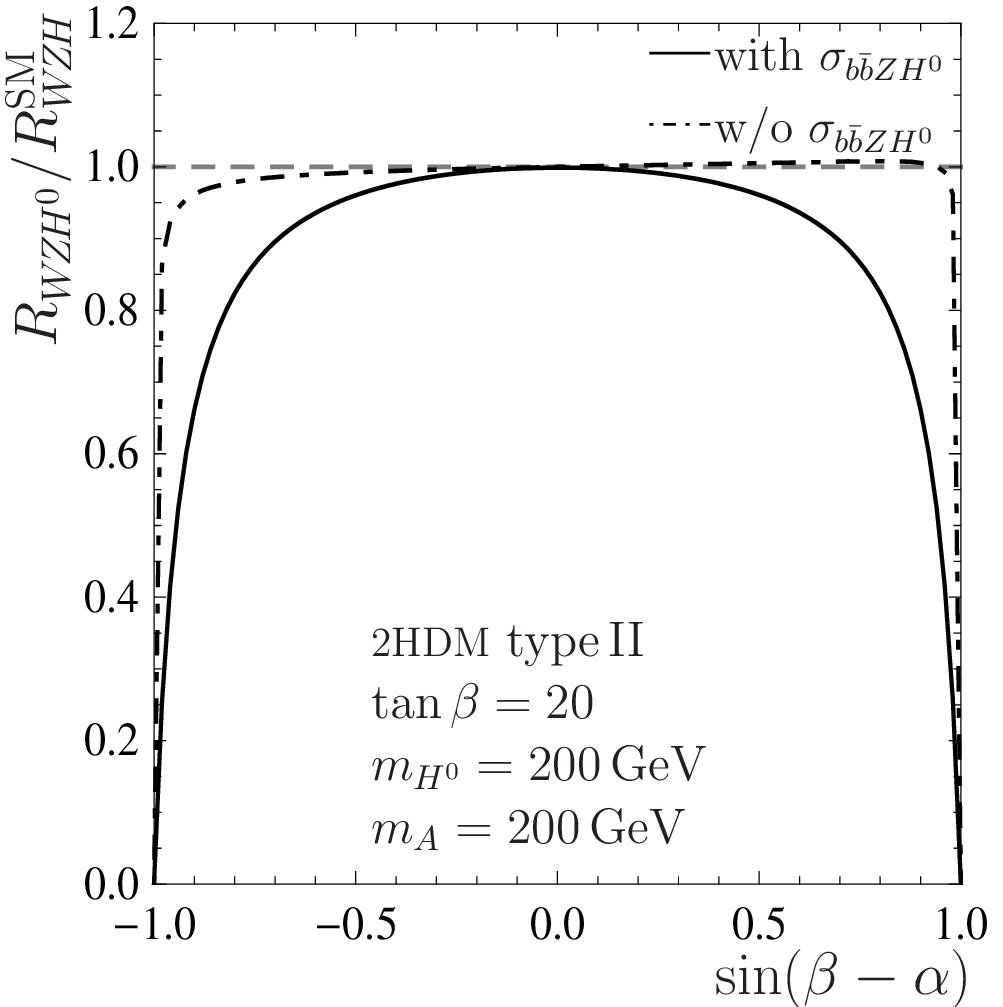} \\[-0.3cm]
(c) & (d) 
\end{tabular}
\end{center}
\vspace{-0.6cm}
\caption{(a-b) $\sigma(pp\to \zHo)$ (black/solid), $\sigma_{\dy}^{\zHo}$
  (green/dotted), $\sigma_{{\ggzHo}}$ (red/dashed) and $\sigma_{\bbzHo}$
  (blue/dash-dotted) in pb for $\sqrt{s}=14$\,TeV and $m_{H^0}=200$\,GeV
  as a function of $\sin(\beta-\alpha)$ for type\,II \thdm{} with
  $m_h=125$\,GeV, $m_A=m_{H^\pm}=200$\,GeV using (a) $\tan\beta=1$ and
  (b) $\tan\beta=20$; (c-d) the ratio $\rwzh{H^0}/\rwzh{H}^\sm$ for the
  cases (a-b), respectively with $\sigma_{\bbzHo}$ (solid) and without
  (dash-dotted).  }
\label{fig:heavyhiggsmA200_2T2}
\end{figure}



\begin{figure}[ht]
\begin{center}
\begin{tabular}{cc}
\includegraphics[width=0.3\textwidth]{%
  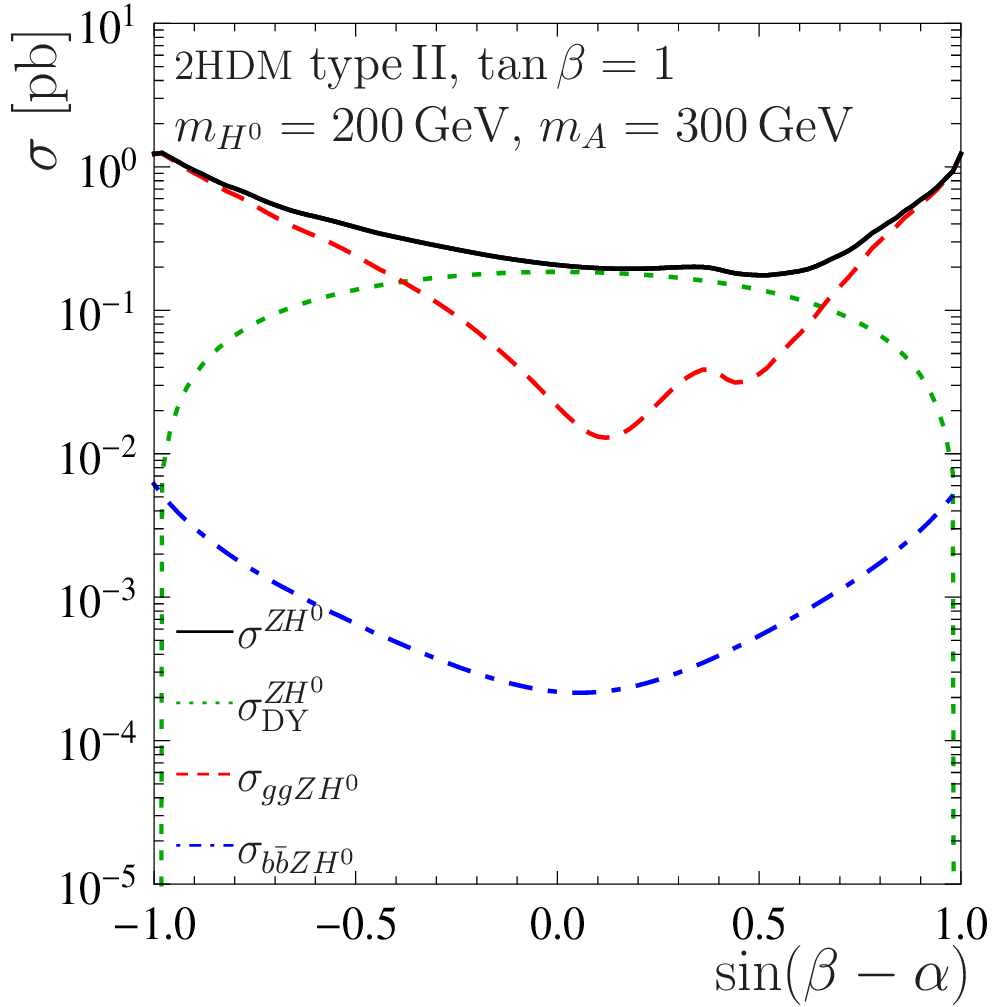} &
\includegraphics[width=0.3\textwidth]{%
  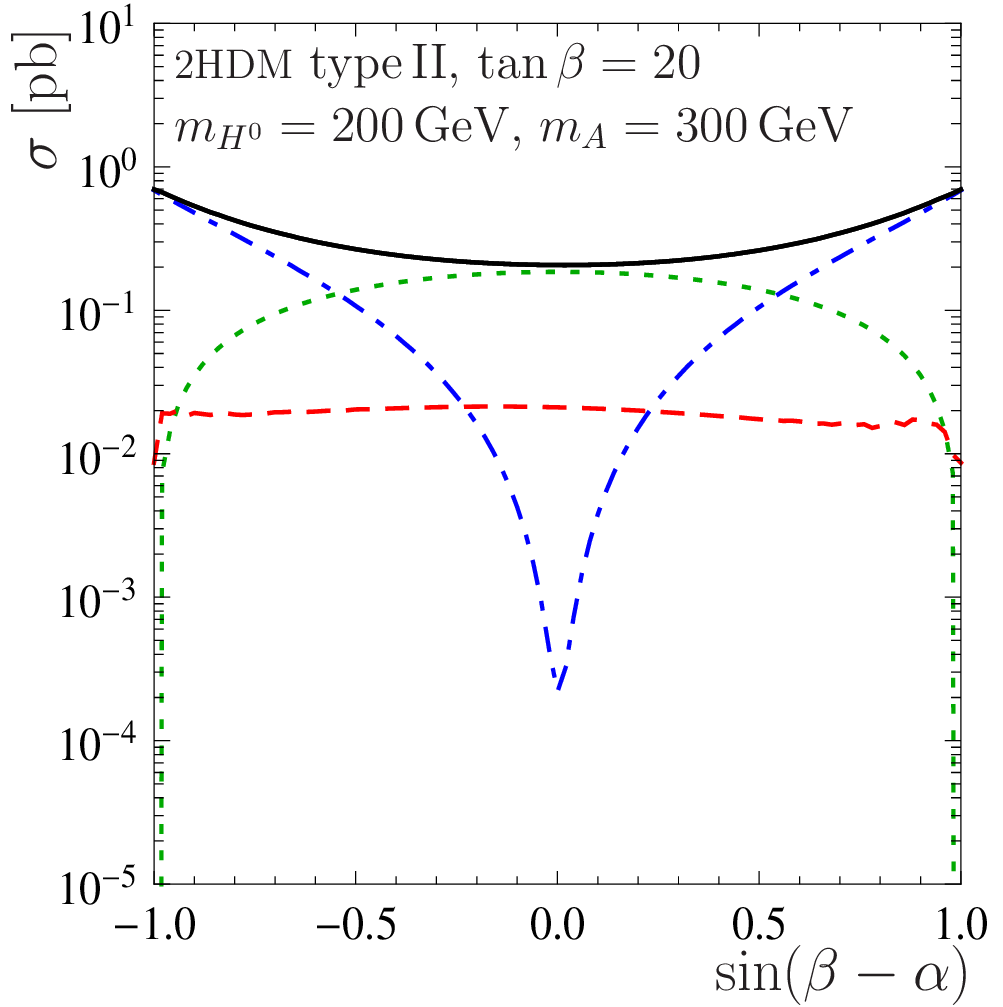} \\[-0.3cm]
(a) & (b) \\
\includegraphics[width=0.3\textwidth]{%
  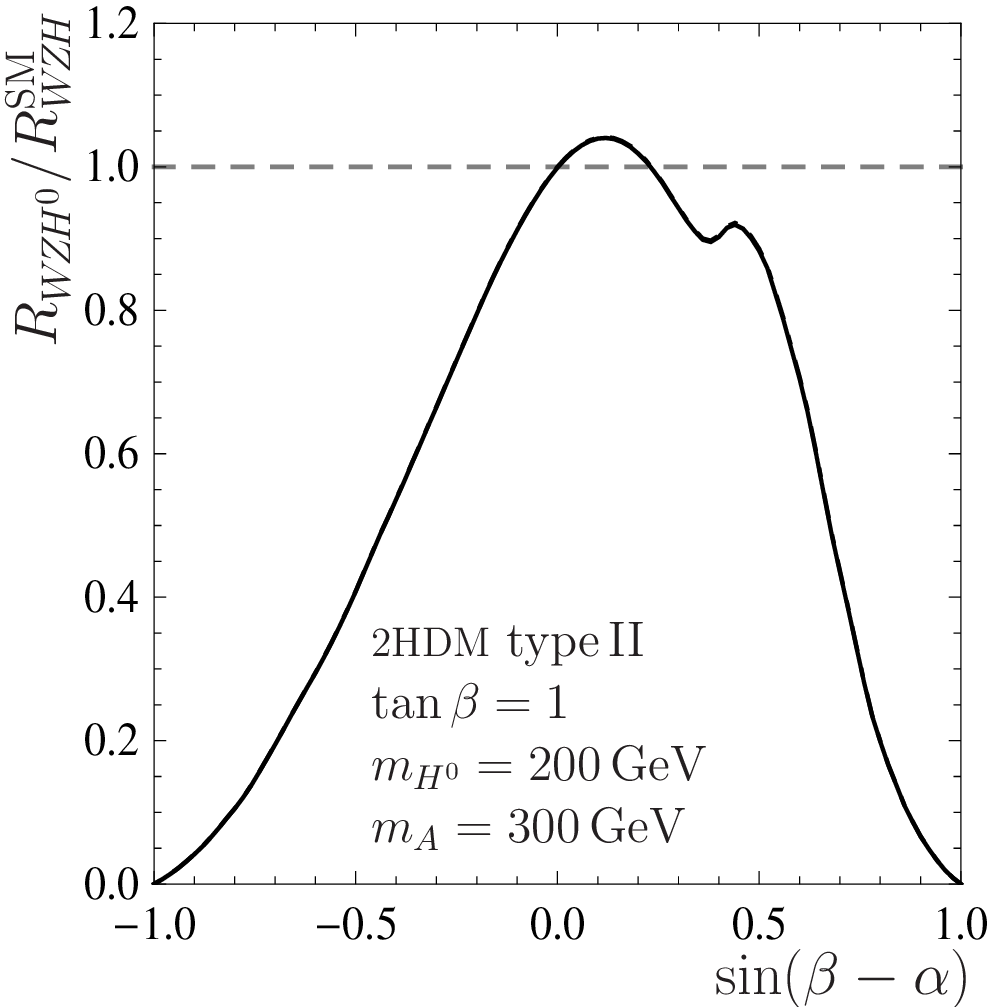} &
\includegraphics[width=0.3\textwidth]{%
  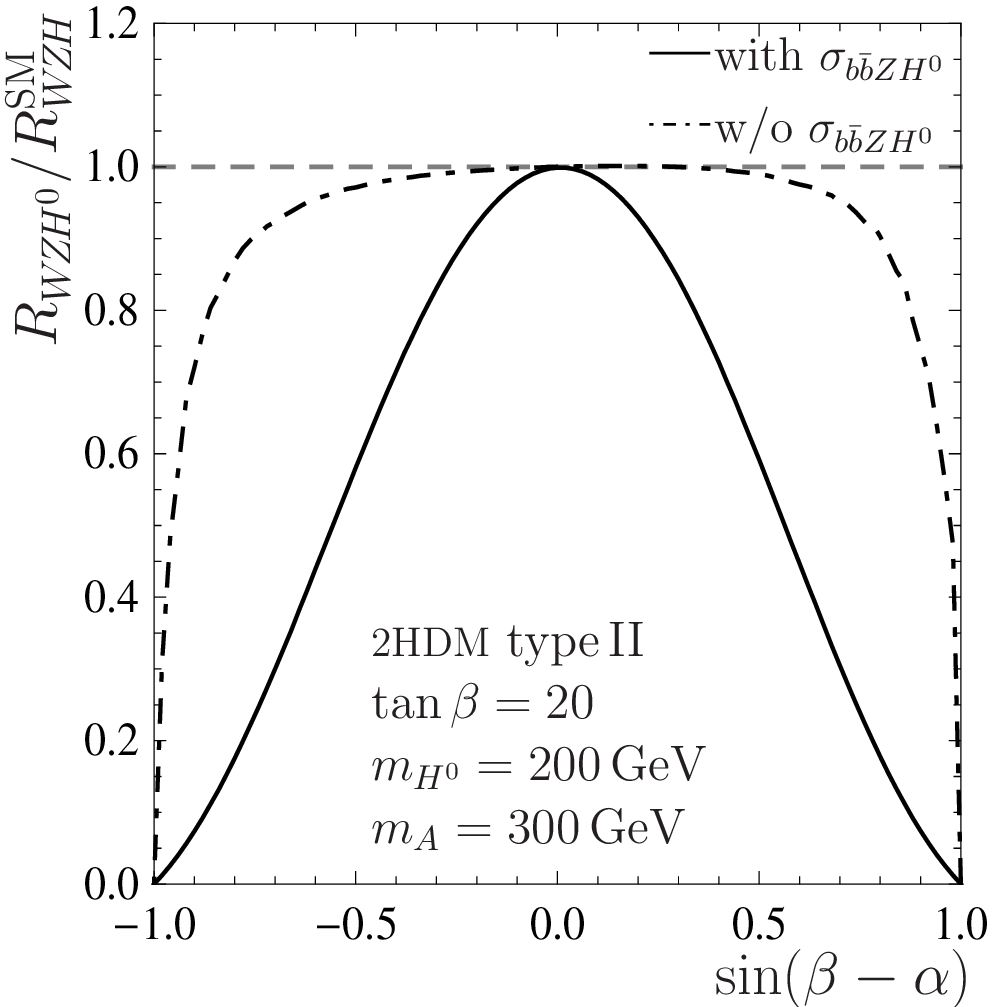} \\[-0.3cm]
(c) & (d) 
\end{tabular}
\end{center}
\vspace{-0.6cm}
\caption{(a-b) $\sigma(pp\to \zHo)$ (black/solid), $\sigma_{\dy}^{\zHo}$
  (green/dotted), $\sigma_{{\ggzHo}}$ (red/dashed) and $\sigma_{\bbzHo}$
  (blue/dash-dotted) in pb for $\sqrt{s}=14$\,TeV and $m_{H^0}=200$\,GeV
  as a function of $\sin(\beta-\alpha)$ for type\,II \thdm{} with
  $m_h=125$\,GeV, $m_A=300$\,GeV, $m_{H^\pm}=200$\,GeV using (a)
  $\tan\beta=1$ and (b) $\tan\beta=20$; (c-d) the ratio
  $\rwzh{H^0}/\rwzh{H}^\sm$ for the cases (a-b), respectively with
  $\sigma_{\bbzHo}$ (solid) and without (dash-dotted).  }
\label{fig:heavyhiggsmA300_2T2}
\end{figure}

\begin{figure}[ht]
\begin{center}
\begin{tabular}{cc}
\includegraphics[width=0.3\textwidth]{%
  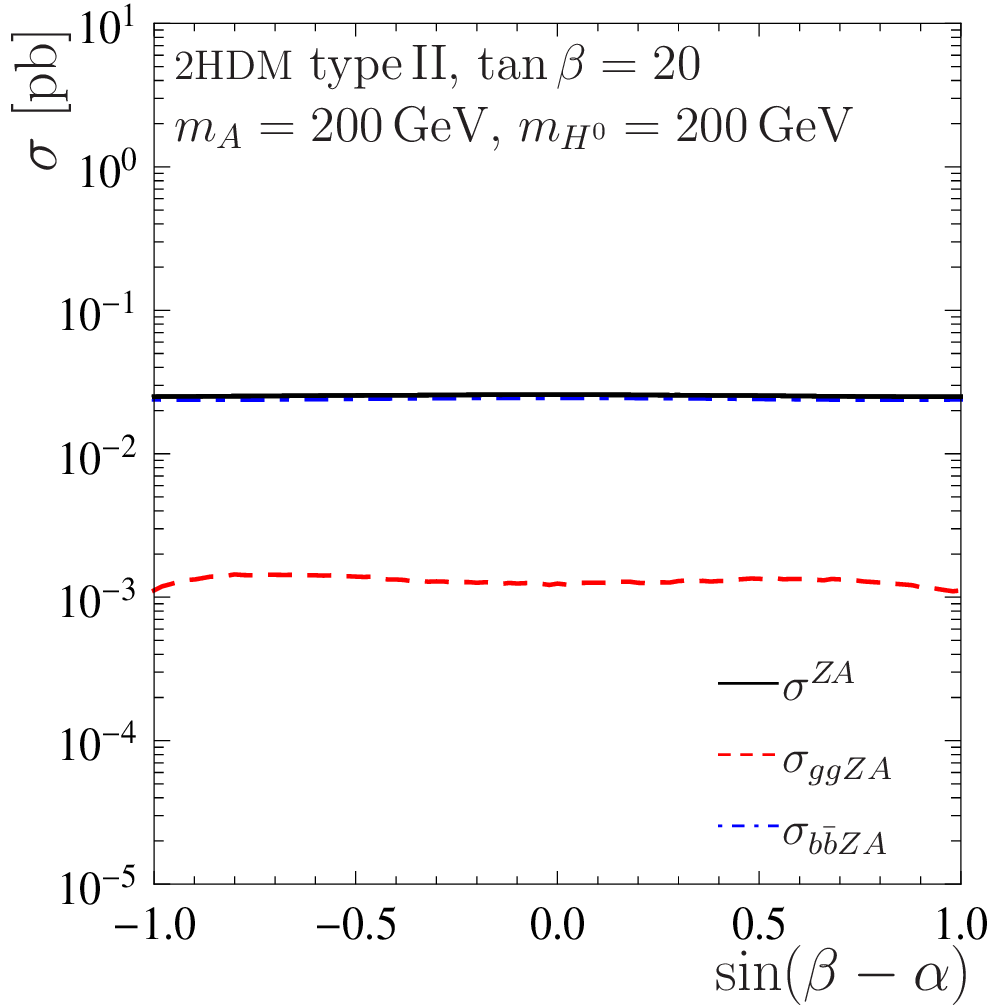} &
\includegraphics[width=0.3\textwidth]{%
  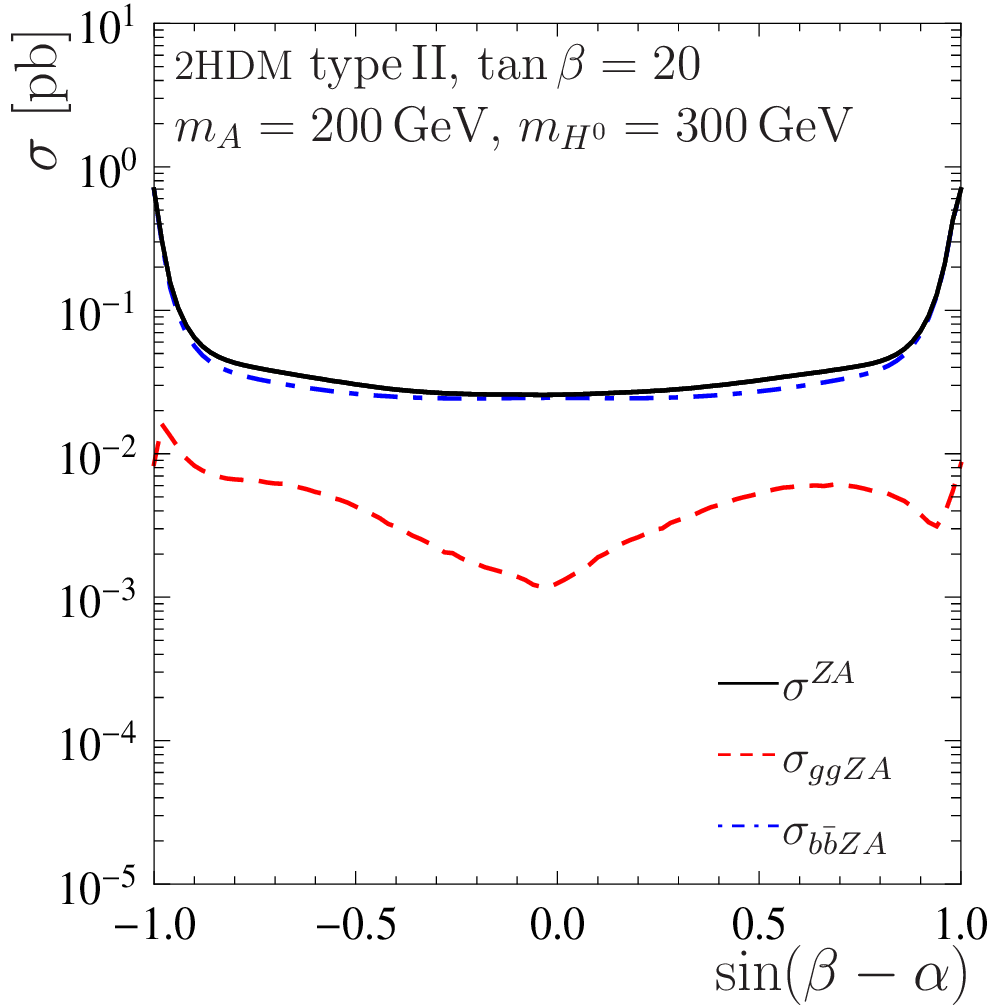} \\[-0.3cm]
(a) & (b)
\end{tabular}
\end{center}
\vspace{-0.6cm}
\caption{(a-b) $\sigma(pp\to \zA)$ (black/solid), $\sigma_{{\ggzA}}$ (red/dashed)
and $\sigma_{\bbzA}$ (blue/dash-dotted) in pb for $\sqrt{s}=14$\,TeV and
$m_{A}=200$\,GeV as a function of
$\sin(\beta-\alpha)$ for type\,II \thdm{}  with $\tan\beta=20$, $m_h=125$\,GeV, $m_{H^\pm}=200$\,GeV
using (a)  $m_{H^0}=200$\,GeV and (b)  $m_{H^0}=300$\,GeV.
}
\label{fig:pseudoscalarmH300_2T2}
\end{figure}


The heavy Higgs couples to the heavy gauge bosons proportional to
$g_{VV}^{H^0}=\cos(\beta-\alpha)$.  Therefore, if the light Higgs
resembles the \sm-like Higgs, the Drell-Yan-like contributions
$\sigma_{\dy}^{\vHo}$ are strongly suppressed. In contrast, the Yukawa
couplings are still sizable and therefore dominate this production
process in the region $|\sin(\beta-\alpha)|\approx 1$.  In
\fig{fig:heavyhiggsmA200_2T2}, we show the different contributions to
the cross section for $pp\to \zHo$ in the type\,II \thdm{} 
when the mass of the pseudoscalar
Higgs is below the $\zHo$ threshold ($m_{H^0}=m_A=200$\,GeV).
\fig{fig:heavyhiggsmA300_2T2}, on the other hand, demonstrates the
presence of a resonance ($m_{H^0} = 200$\,GeV, $m_A=300$\,GeV).  In
contrast to the cases discussed in \sct{sec:light}, the
internal pseudoscalar can then decay via $A\rightarrow Zh$,
$A\rightarrow \zHo$ and $A\rightarrow W^\pm H^\mp$.
Thus, the $\ggzHo$ and $\bbzHo$ contributions are not
constant, but the proportionality to
$(g_{Z}^{AH^0})^2=\sin^2(\beta-\alpha)$ dominates.

\fig{fig:heavyhiggsmA200_2T2} (c-d) and \fig{fig:heavyhiggsmA300_2T2} (c-d)
present the ratios $\rwzh{H^0}$ with respect to the \sm{} ratio.
For $|\sin(\beta-\alpha)|\approx 1$ the cross section $\sigma^{\wHo}$
vanishes, whereas $\sigma^{\zHo}$ is dominated by Yukawa couplings,
such that $\rwzh{H^0}$ tends towards zero.
The detailed measurement of $\sigma^{\zHo}$ therefore provides
important information about the Yukawa couplings of the heavy
Higgs in case the light Higgs resembles the \sm{}-like Higgs.
The latter effect is less important in case of type\,I \thdm{} 
with large values of $\tan\beta$, since all Yukawa couplings
are suppressed.

The pseudoscalar Higgs of the \thdm{} does not couple to the heavy gauge
bosons, since $g_{VV}^A=0$. However, the presence of Yukawa couplings $g_t^A$
and $g_b^A$ allows for large $\ggzA{}$ and $\bbzA{}$ cross section
contributions. \fig{fig:pseudoscalarmH300_2T2}\,(a) demonstrates the
size of both $\ggzA$ and $\bbzA$ cross section contributions to $\zA$
production in the type~II \thdm{}  with $\tan\beta=20$.  As expected, the
cross sections are hardly dependent on the mixing angle $\alpha$.
However, \fig{fig:pseudoscalarmH300_2T2}\,(b) shows the presence of a
heavy scalar $H^0$, which can decay to $\zA$. In this case, the cross
sections are slightly increased and a dependence on the mixing angle
$\alpha$ is induced.
Since we omit $\sigma_{\text{I}}^{\wA}$, the production cross section
$\sigma(pp\rightarrow WA)$ vanishes and a definition of $\rwzh{A}$
is not reasonable for the pseudoscalar.



\section{Effects in the boosted regime}
\label{sec:boost}

\subsection{The gluon-induced component in the \sm{}}

It is well known that the signal-to-background ratio for $\vH{}$
production is significantly enhanced in the so-called boosted
regime~\cite{Butterworth:2008iy}, where the transverse momentum $\ptH{}$
of the Higgs boson is large. In this section, we therefore briefly study
the influence of a lower cut on $\ptH$ on the effects observed in the
previous sections. The discussion in this section will be on a rather
qualitative level; more quantitative studies are beyond the scope of our
paper and will be deferred to a future publication.

First we note that in the \sm{}, the Higgs transverse momentum due to
the $gg\to \zH$ sub-process is peaked at larger values than when the
Higgs is produced through the \dy{}-like process (see, e.g.,
Ref.\,\cite{Englert:2013vua}).  The effect of this is that the relative
$gg\to \zH$ contribution is about twice as large as for the total cross
section if one restricts the Higgs transverse momentum to $\ptH\gtrsim
150$\,GeV\,\cite{Ferrera} (assuming $\mhiggs=125$\,GeV). Increasing the
lower cut on $\ptH$ beyond that value leads again to a decrease of the
relative $gg\to \zH$ portion since the spectrum drops rather sharply
towards large $\ptH$.

The second step is to study the effect of New Physics on the $gg\to
\zphi$ and the $b\bar b\to \zphi$ sub-processes in the boosted
regime. The shape of the $\ptphi$ spectrum is clearly unaffected for the
\dy{}-like terms in the scenarios considered in
Sections~\ref{sec:SMYukawas} and \ref{sec:2HDMdiscussion} of this paper.

\subsection{Modified \sm{} Yukawa couplings}\label{sec:modyuk}

\begin{figure}[ht]
\begin{center}
\begin{tabular}{ccc}
\includegraphics[width=0.3\textwidth]{%
  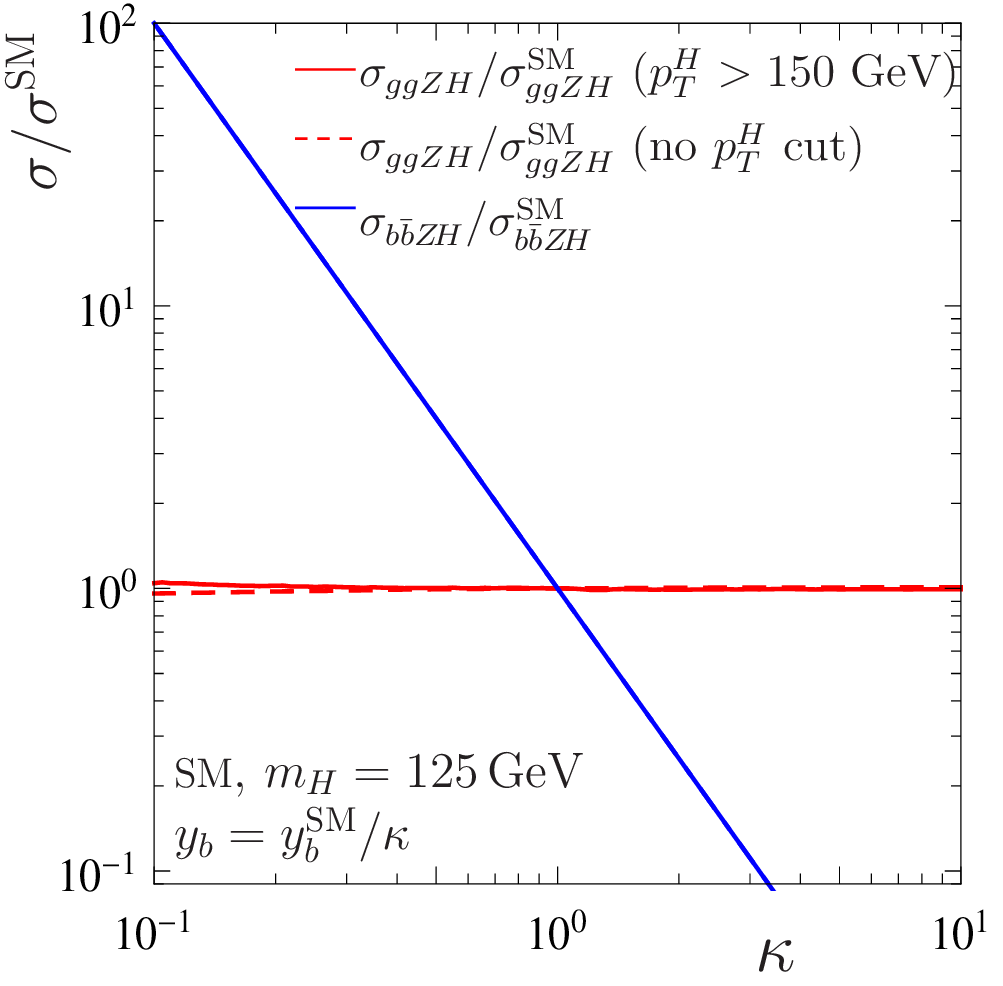} &
\includegraphics[width=0.3\textwidth]{%
  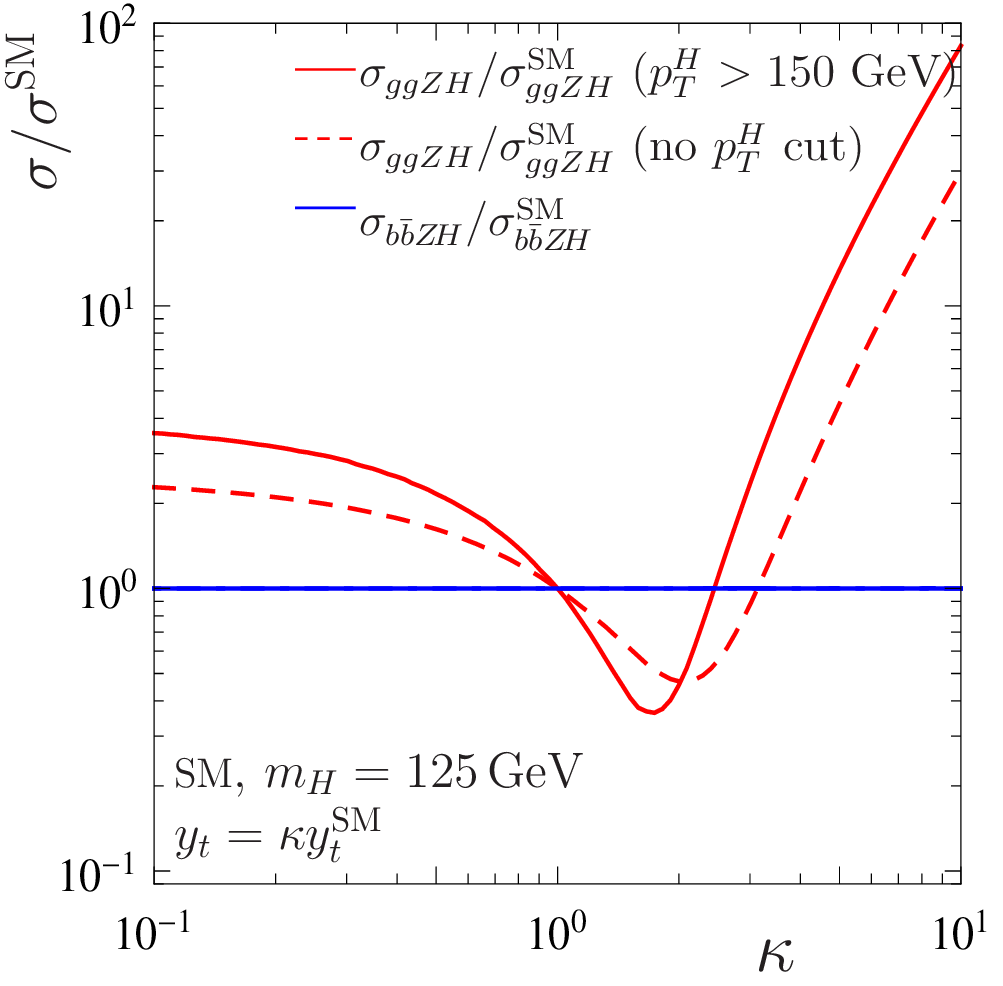} &
\includegraphics[width=0.3\textwidth]{%
  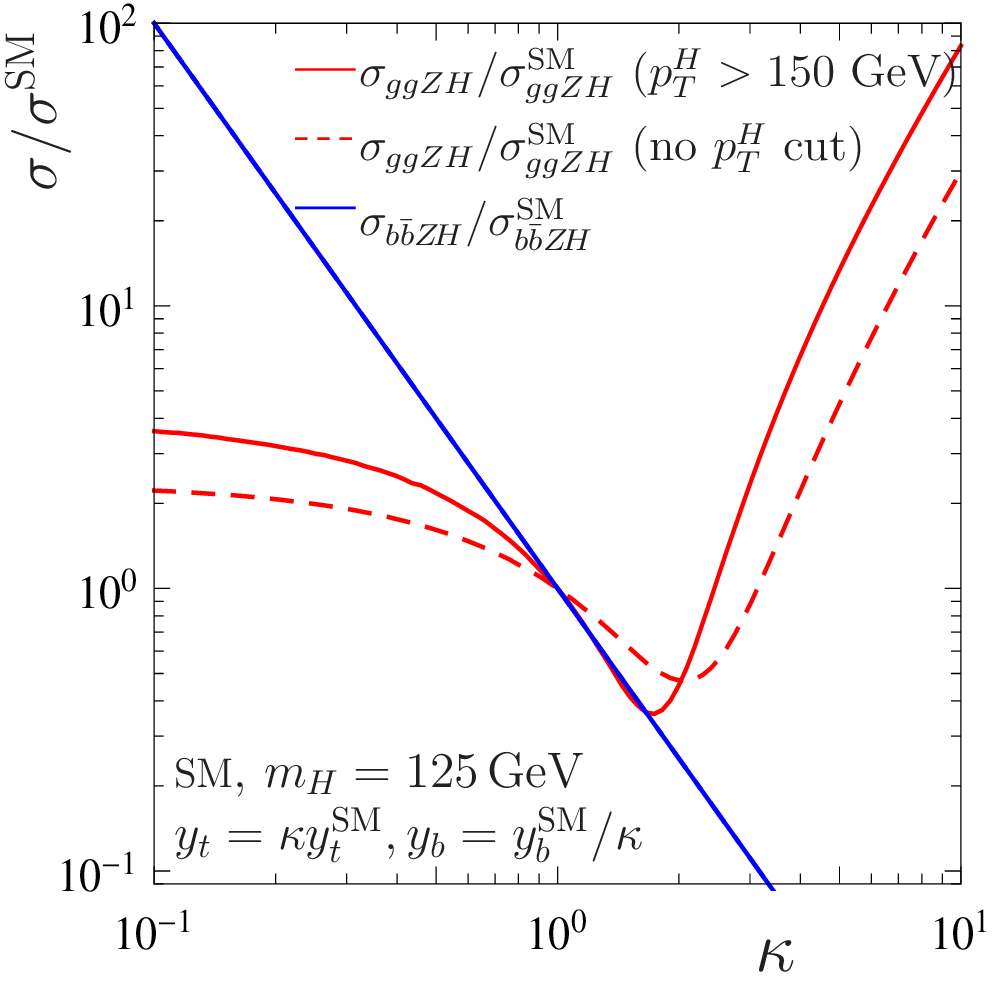} \\[-0.3cm]
(a) & (b) & (c)
\end{tabular}
\end{center}
\vspace{-0.6cm}
\caption[]{\label{fig:SMYukptcut} (a--c) $\sigma_{\ggzH{}}/\sigma_{\ggzH{}}^{\sm}$ (red),
  $\sigma_{\bbzH{}}/\sigma_{\bbzH{}}^{\sm}$ (blue) 
  with $\ptH>150$\,GeV (solid) and without $\ptH$ cut (dashed)
  for $\sqrt{s}=14$\,TeV and $m_H=125$\,GeV as a function of
  $\kappa$, where (a) $y_t= y_t^\sm$ and $y_b = y_b^\sm{}/\kappa$, (b)
  $y_t=\kappa y_t^\sm$ and $y_b = y_b^\sm{}$, (c) $y_t=\kappa y_t^\sm$
  and $y_b = y_b^\sm{}/\kappa$.}
\end{figure}

\fig{fig:SMYukptcut} shows the $gg\to \zH$ (red) and the $b\bar b\to
\zH$ contribution (blue) relative to the \sm{} one for modified top- and
bottom-Yukawa couplings (parameters as in \fig{fig:SMYuk}) without a
$\ptH$ cut (dashed), and when the Higgs transverse momentum is
restricted to $\ptH>150$\,GeV (solid). Since $\sigma_{\bbzH}$ is
proportional to $|y_b|^2$ in the \sm, the corresponding dashed and solid
line are identical.

For $gg\to \zH$, on the other hand, it is remarkable that almost any
modification of the top-Yukawa coupling leads to an increase in the
large-$\ptH$ fraction with respect to the \sm{} one (solid
vs.\ dashed). A decrease is only observed for
$y_t^\sm{}<y_t<2y_t^\sm{}$. Additionally, close to the \sm{} value
$y_t\approx y_t^\sm$, the dependence on $\kappa$ and thus
on the top-Yukawa coupling is increased.  Combined with the observations
of a generally enhanced \sm{} $gg\to\zH$ contribution with respect to
the \dy-terms, we conclude that the $\wH/\zH$ ratio becomes
significantly more sensitive to the top-Yukawa coupling in the boosted
regime.

\subsection{2-Higgs-Doublet Model}

Let us now turn from the simple modification of the Yukawa couplings to
the full \thdm{}. The main new effects here are due to the additional
Higgs bosons $\phi'$ which can occur as virtual particles. Their impact
is particularly large when values $\hat s\approx m_{\phi'}^2$ are
kinematically allowed, see \sct{sec:2HDMtheory}.  Applying a
lower cut on $\ptphi$ restricts the values of $\hat s$ to
\begin{equation}
\begin{split}
\hat s\geq m_{\phi}^2 + m_Z^2 + 2(\ptphi)^2 +
2\sqrt{(m_Z^2+(\ptphi)^2)(m_\phi^2+(\ptphi)^2)}\,.
\end{split}
\end{equation}
For example, while the lower limit on $\sqrt{\hat s}$ for the inclusive
cross section is at $m_Z+m_\phi\approx 216$\,GeV for $m_\phi=125$\,GeV,
it moves up to about $370$\,GeV in the boosted regime with
$\ptphi>150$\,GeV. We therefore expect that the effects on the
$\wphi/\zphi$ ratio observed in \sct{sec:2HDMdiscussion}
decrease in the boosted regime.

\begin{figure}[ht]
\begin{center}
\begin{tabular}{ccc}
\includegraphics[width=0.3\textwidth]{%
  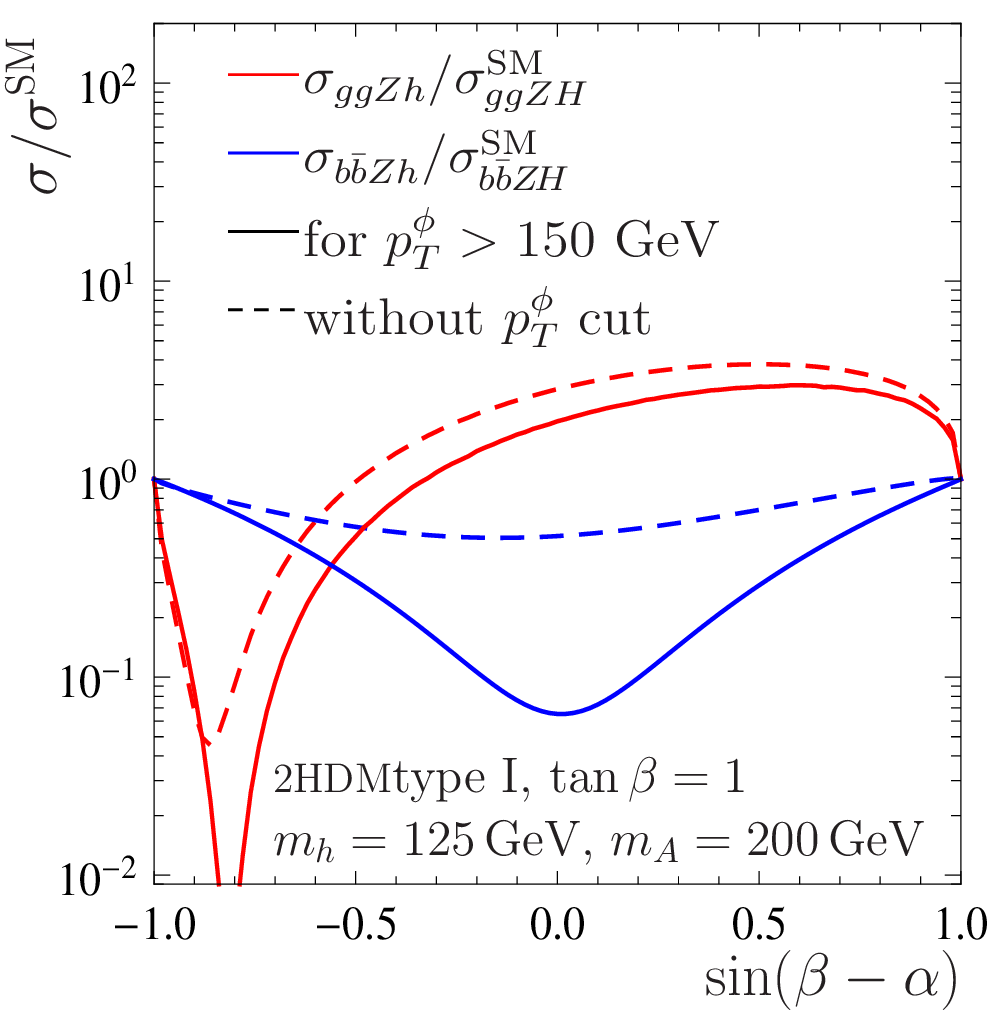} &
\includegraphics[width=0.3\textwidth]{%
  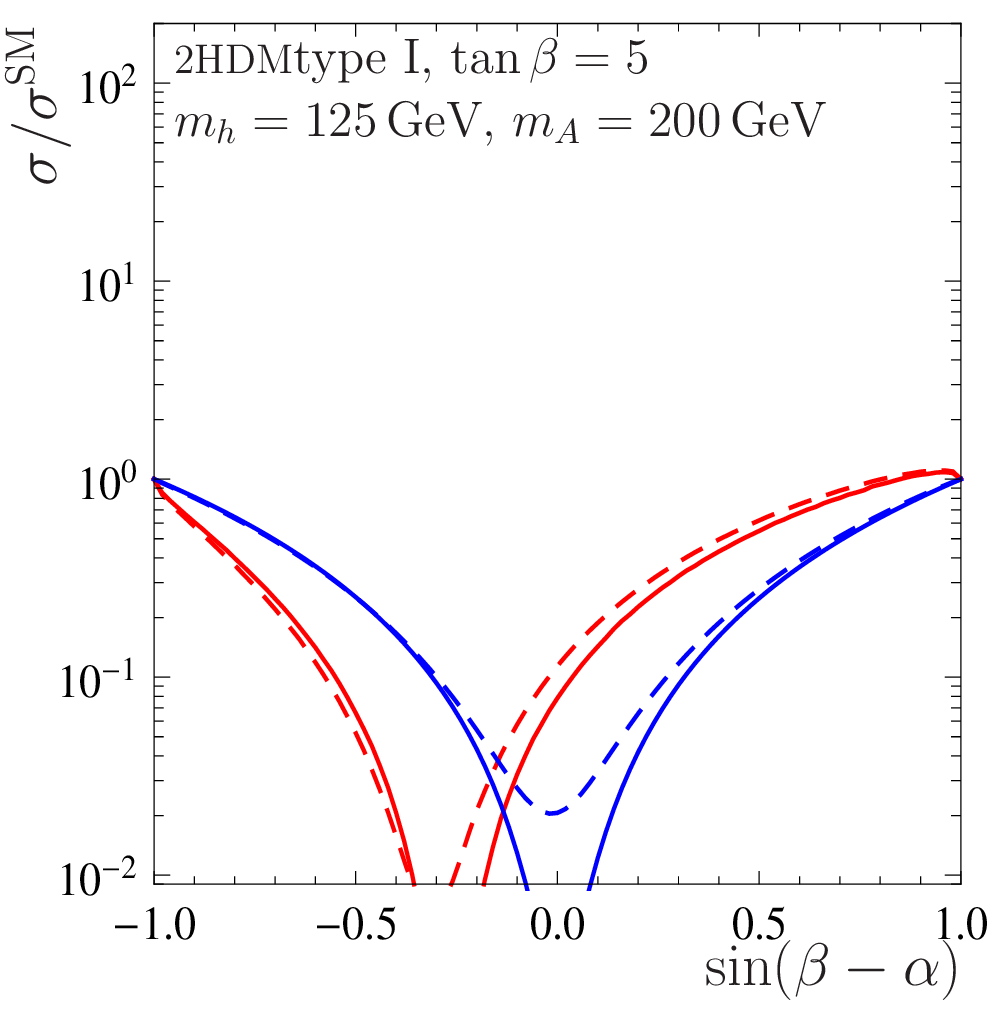} &
\includegraphics[width=0.3\textwidth]{%
  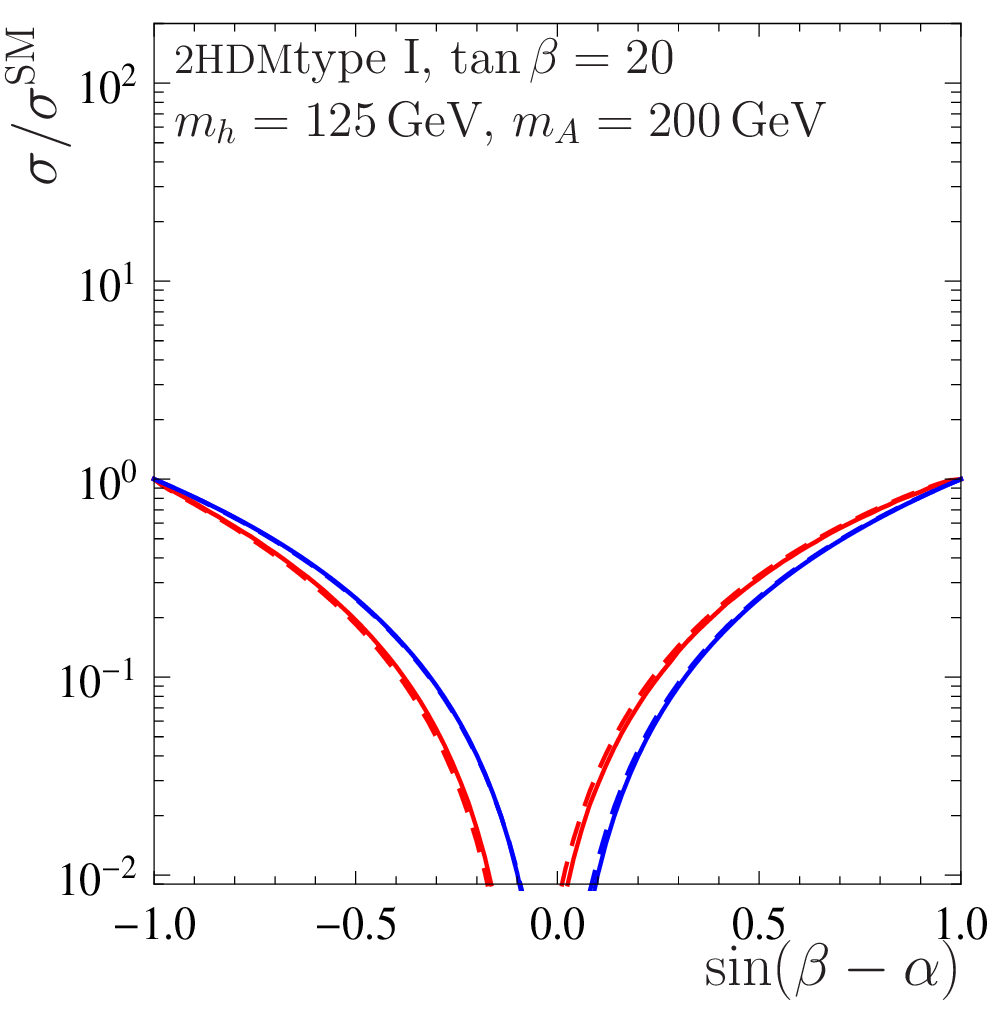} \\[-0.3cm]
(a) & (b) & (c)
\end{tabular}
\end{center}
\vspace{-0.6cm}
\caption[]{\label{fig:2HDMtype1ma200ptcut} (a--c)
  $\sigma_{\ggzh{}}/\sigma_{\ggzH{}}^{\sm}$ (red),
  $\sigma_{\bbzh{}}/\sigma_{\bbzH{}}^{\sm}$ (blue) with
  $\ptphi>150$\,GeV (solid) and without $\ptphi$ cut (dashed) for
  $\sqrt{s}=14$\,TeV in the \thdm{} type\,I with $m_A=200$\,GeV as a
  function of $\sin(\beta-\alpha)$ using (a) $\tan\beta=1$, (b)
  $\tan\beta=5$ and (c) $\tan\beta=20$.}
\end{figure}

First we consider the case corresponding to
\fig{fig:lighthiggsmA200_2T1}, i.e.\ \thdm{} type\,I with
$m_A=200$\,GeV. Note that for the total inclusive cross section, the
pseudoscalar is already slightly below resonant; a lower cut on $\ptphi$
moves the allowed interval for $\hat s$ further away from the
resonance. \fig{fig:2HDMtype1ma200ptcut} shows the contributions of the
subprocesses $gg\to \zh$ (red) and $b\bar b\to \zh$ (blue) in the
\thdm{} relative to the \sm{} quantities in this case. For the solid
lines, a lower cut on $\ptphi$ ($\phi=h$ for the \thdm{}, $\phi=H$ for
the \sm{}) of $150$\,GeV was applied, while this cut is absent for the
dashed lines.  The effect of the cut is non-negligible only for the case
$\tan\beta=1$, where it leads to a moderate decrease of the $gg\to \zh$
fraction relative to the \sm{}.  The $b\bar b\to \zh$ is completely
negligible in this scenario anyway (see \fig{fig:lighthiggsmA200_2T1}),
so the reduction by the $\ptphi$ cut observed in
\fig{fig:2HDMtype1ma200ptcut} is irrelevant.

\begin{figure}[ht]
\begin{center}
\begin{tabular}{ccc}
\includegraphics[width=0.3\textwidth]{%
  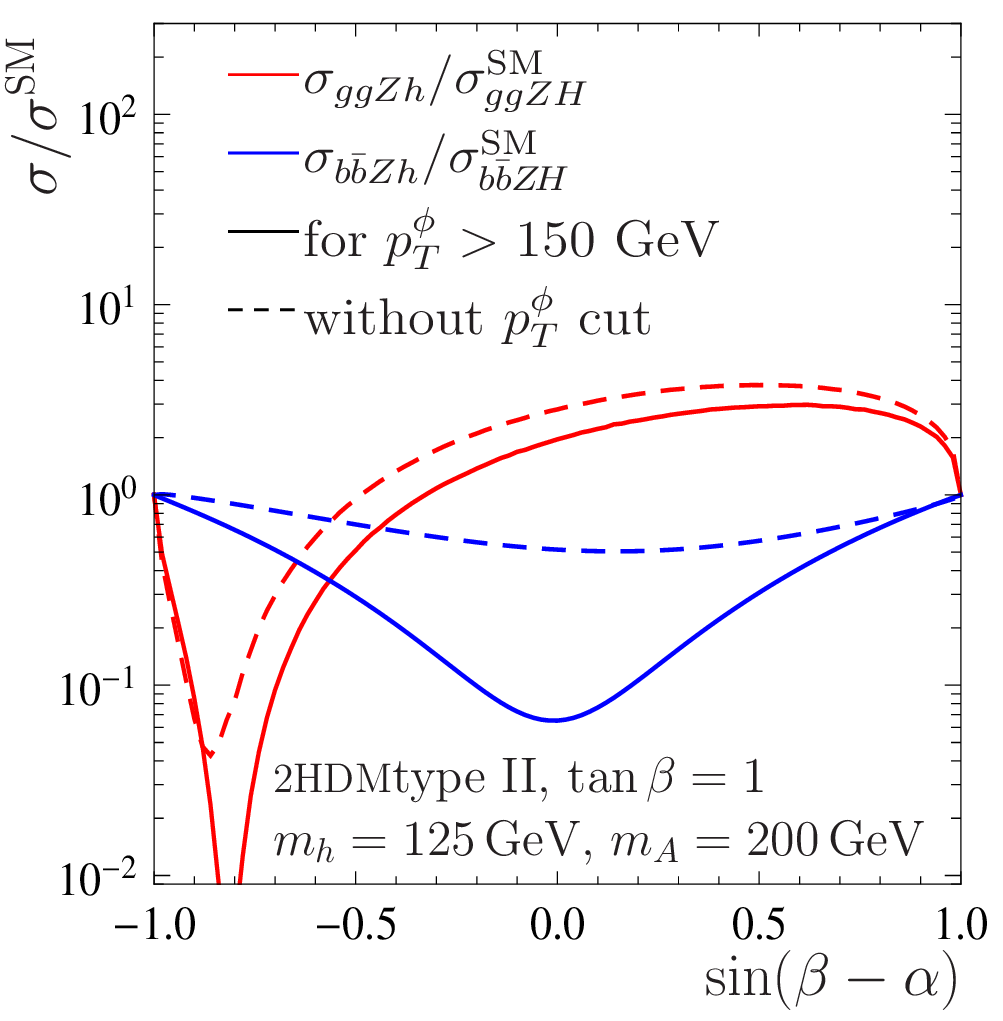} &
\includegraphics[width=0.3\textwidth]{%
  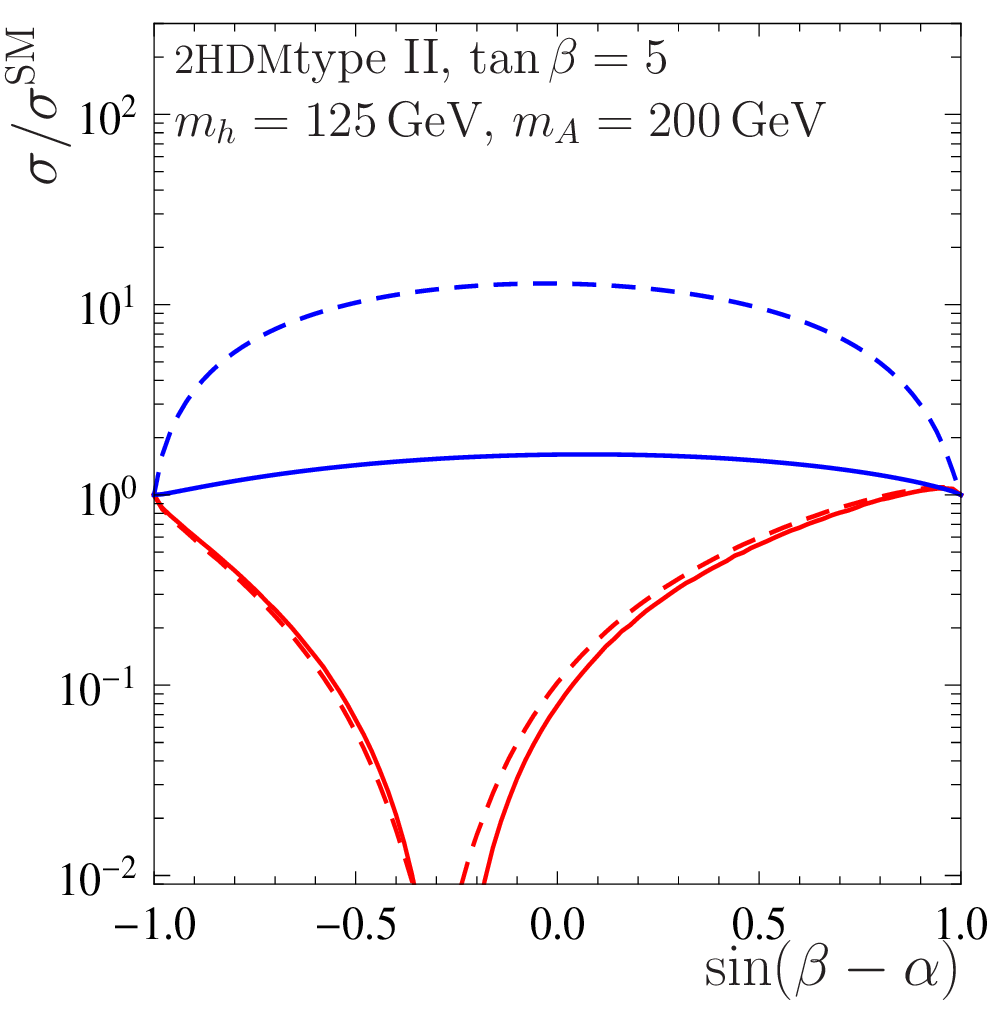} &
\includegraphics[width=0.3\textwidth]{%
  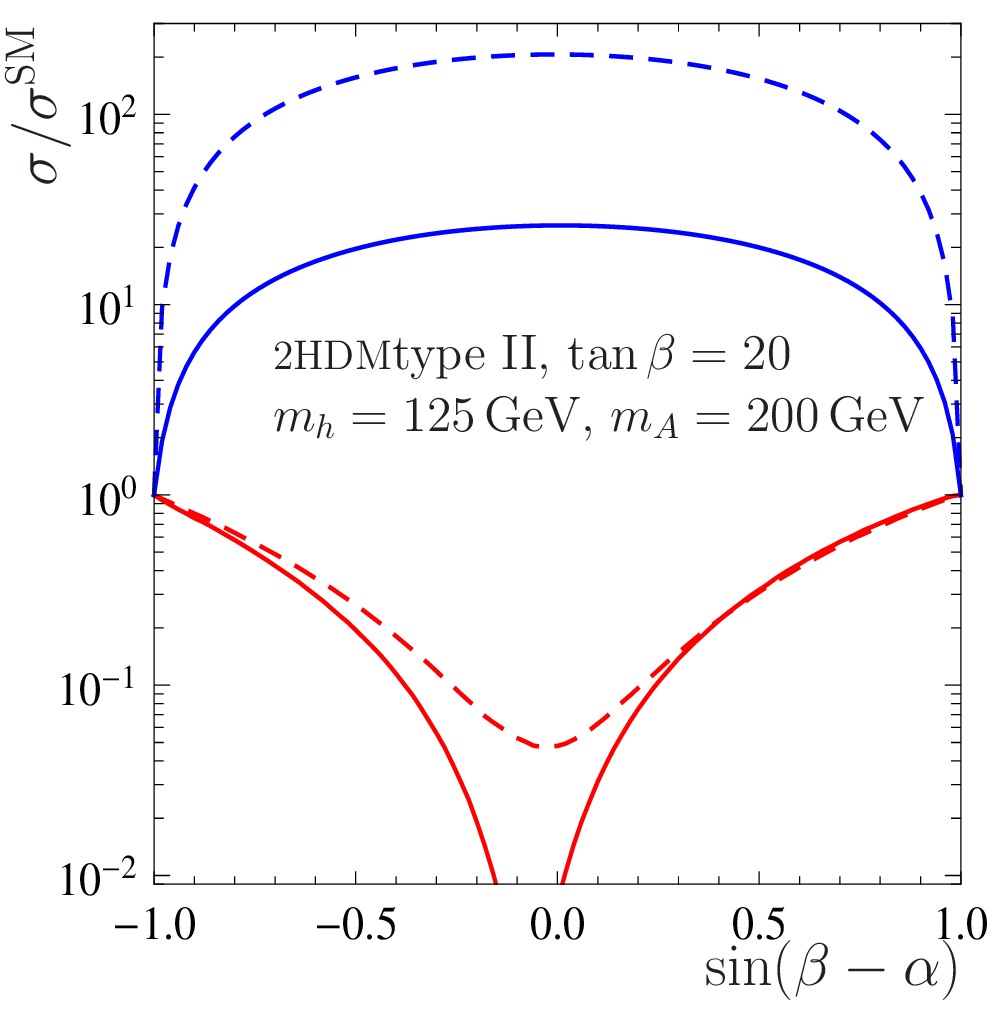} \\[-0.3cm]
(a) & (b) & (c)
\end{tabular}
\end{center}
\vspace{-0.6cm}
\caption[]{\label{fig:2HDMtype2ma200ptcut} (a--c) $\sigma_{\ggzh{}}/\sigma_{\ggzH{}}^{\sm}$ (red),
  $\sigma_{\bbzh{}}/\sigma_{\bbzH{}}^{\sm}$ (blue) 
  with $\ptphi>150$\,GeV (solid) and without $\ptphi$ cut (dashed)
  for $\sqrt{s}=14$\,TeV in the \thdm{} type\,II with $m_A=200$\,GeV as a function of
  $\sin(\beta-\alpha)$ using (a) $\tan\beta=1$, (b) $\tan\beta=5$ and (c) $\tan\beta=20$.}
\end{figure}

For the $gg\to \zh$ component, the situation looks similar also in the
\thdm{} type\,II with non-resonant pseudoscalar (see
\fig{fig:2HDMtype2ma200ptcut}; parameters as in
\fig{fig:lighthiggsmA200_2T2}).  For moderate to large $\tan\beta$, the
effect of the $\ptphi$-cut on the $b\bar b\to \zh$ component, on the
other hand, is quite drastic, leading to a reduction of about an order
of magnitude.

\begin{figure}[ht]
\begin{center}
\begin{tabular}{ccc}
\includegraphics[width=0.3\textwidth]{%
  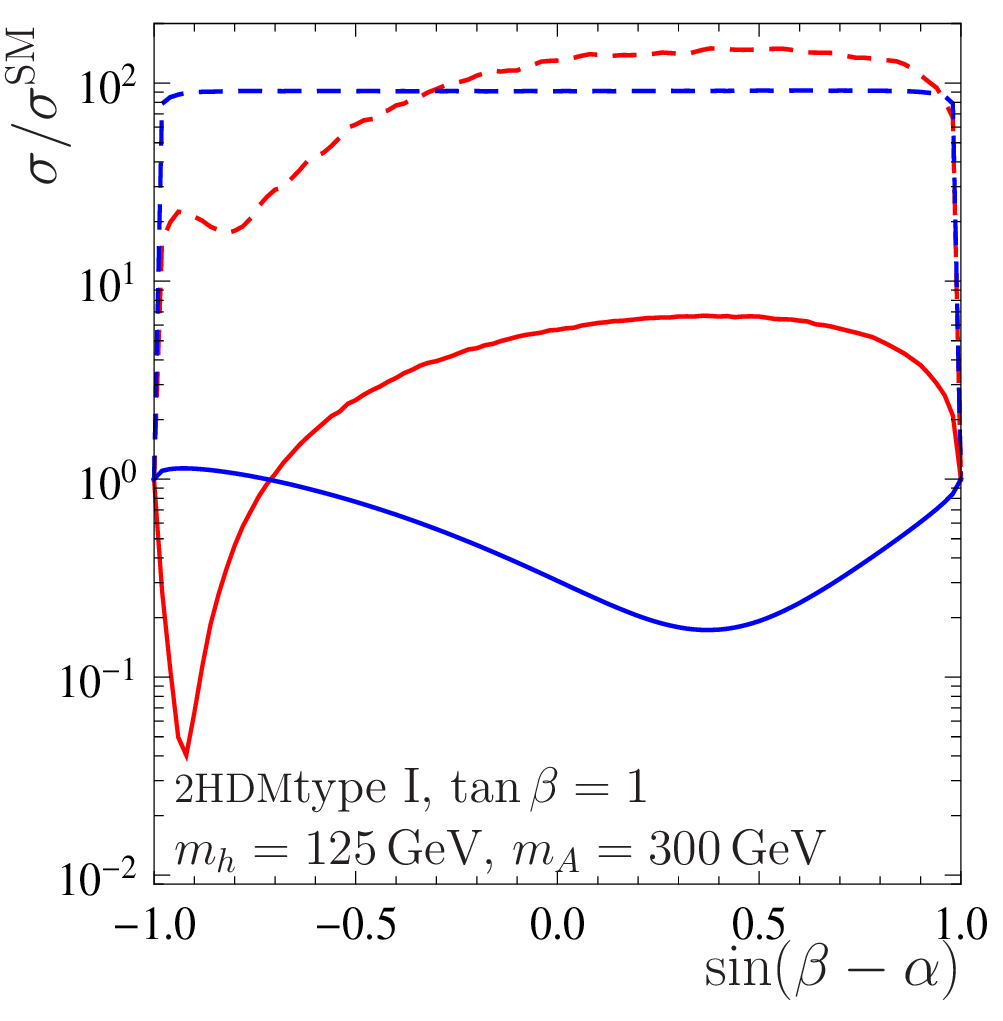} &
\includegraphics[width=0.3\textwidth]{%
  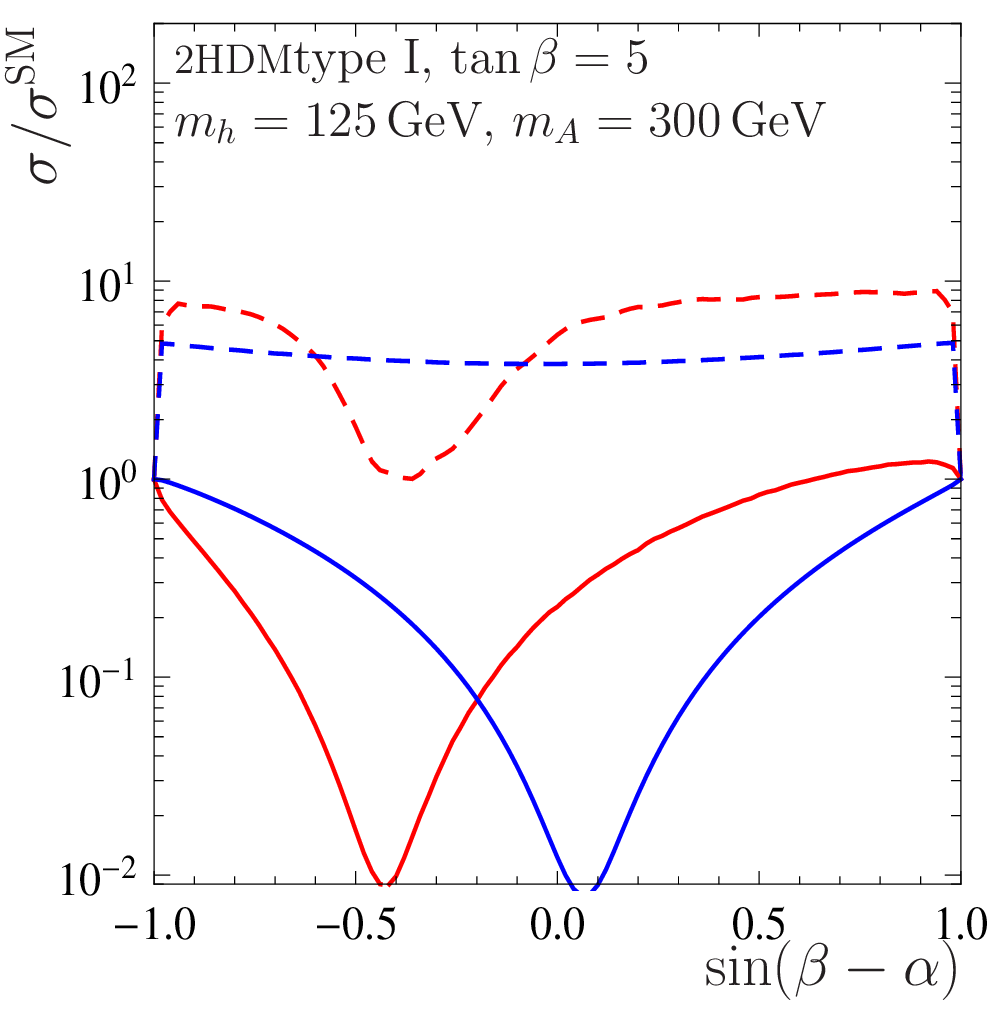} &
\includegraphics[width=0.3\textwidth]{%
  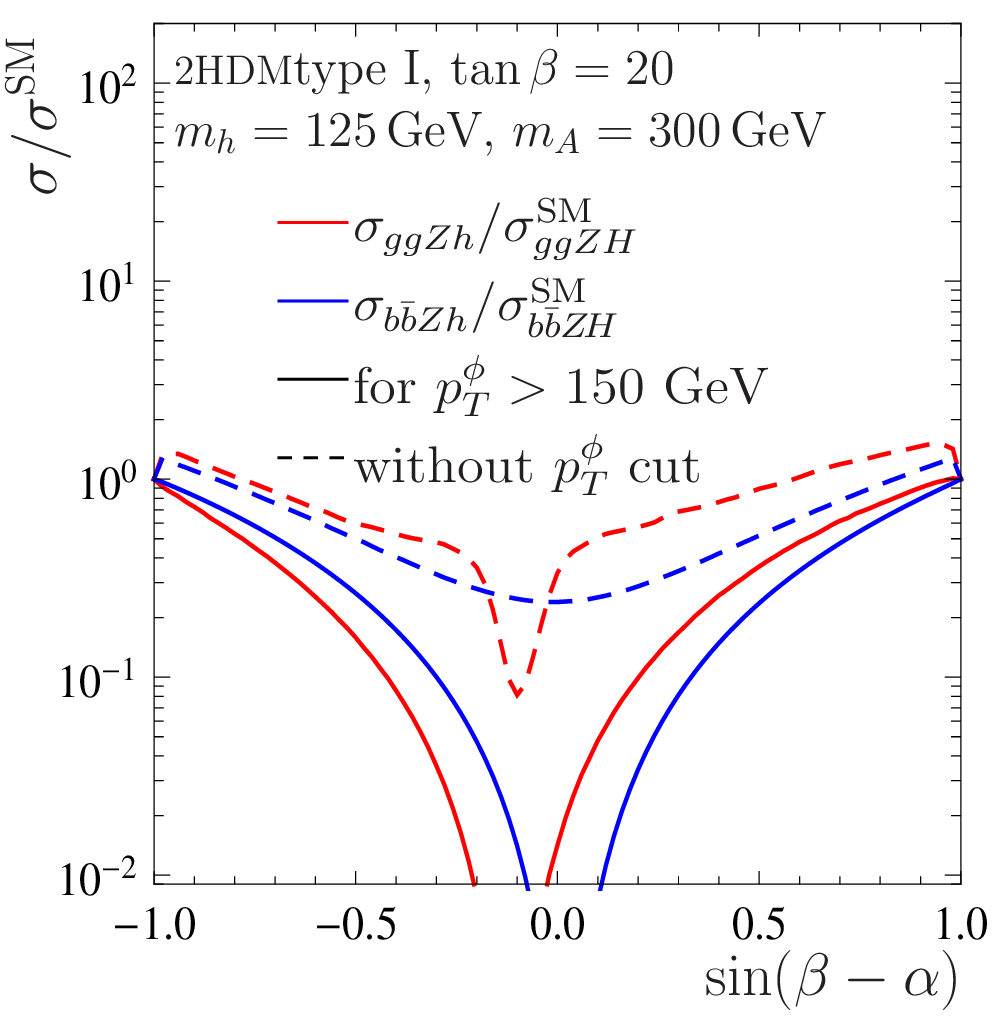} \\[-0.3cm]
(a) & (b) & (c)
\end{tabular}
\end{center}
\vspace{-0.6cm}
\caption[]{\label{fig:2HDMtype1ma300ptcut} (a--c) $\sigma_{\ggzh{}}/\sigma_{\ggzH{}}^{\sm}$ (red),
  $\sigma_{\bbzh{}}/\sigma_{\bbzH{}}^{\sm}$ (blue) 
  with $\ptphi>150$\,GeV (solid) and without $\ptphi$ cut (dashed)
  for $\sqrt{s}=14$\,TeV in the \thdm{} type\,I with $m_A=300$\,GeV as a function of
  $\sin(\beta-\alpha)$ using (a) $\tan\beta=1$, (b) $\tan\beta=5$ and (c) $\tan\beta=20$.}
\end{figure}

\begin{figure}[ht]
\begin{center}
\begin{tabular}{ccc}
\includegraphics[width=0.3\textwidth]{%
  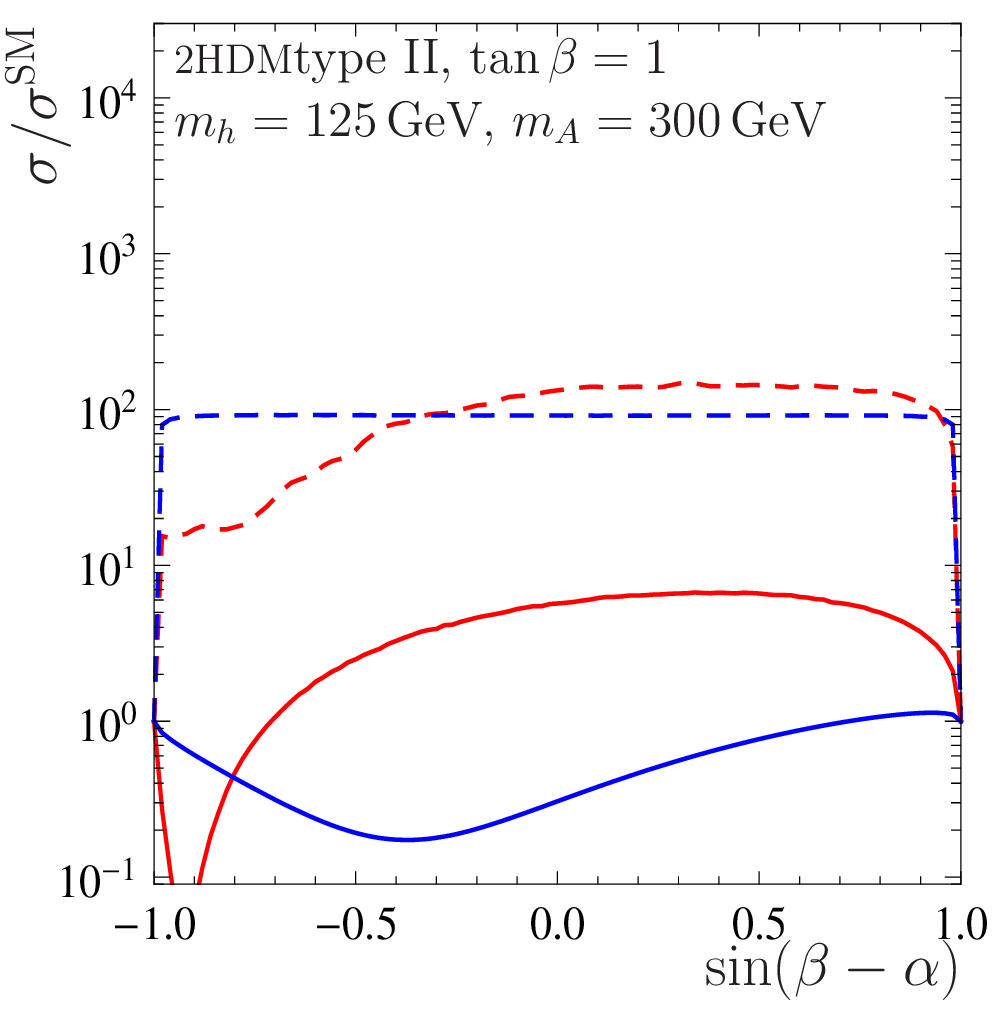} &
\includegraphics[width=0.3\textwidth]{%
  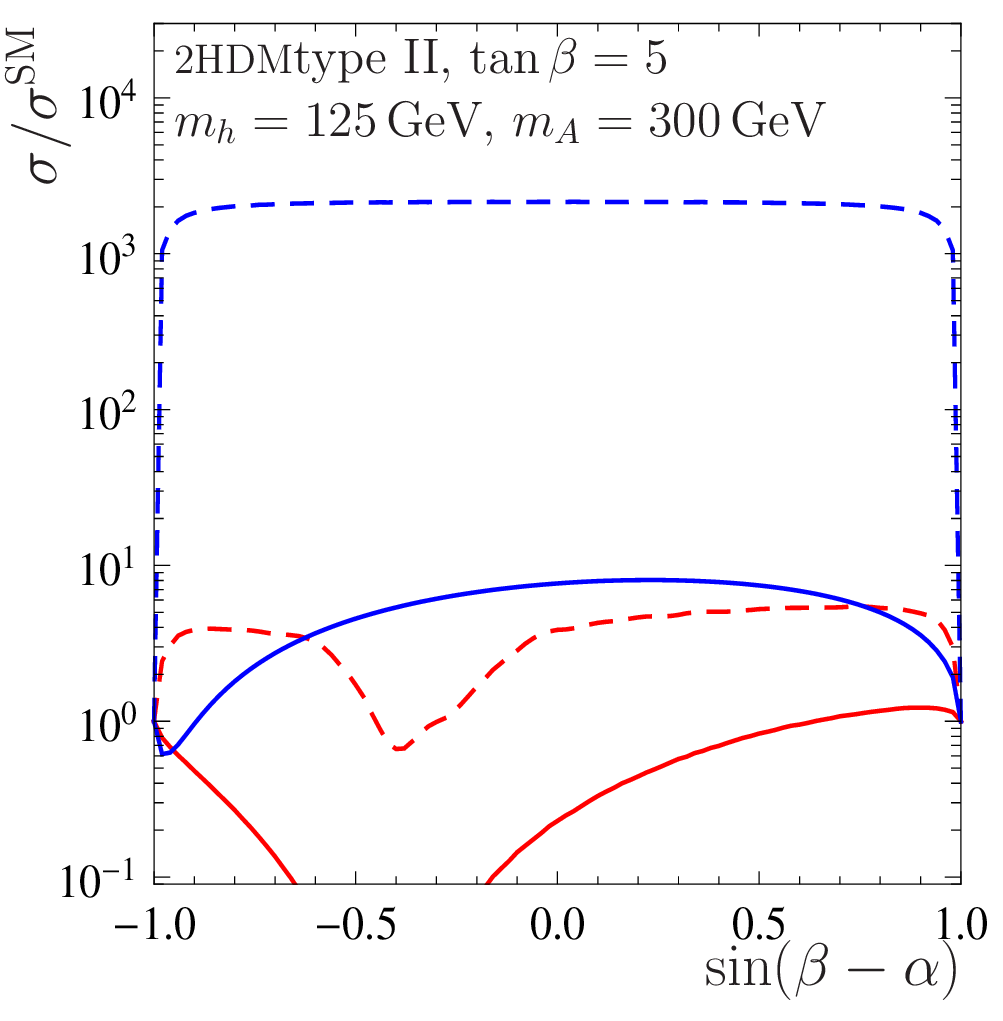} &
\includegraphics[width=0.3\textwidth]{%
  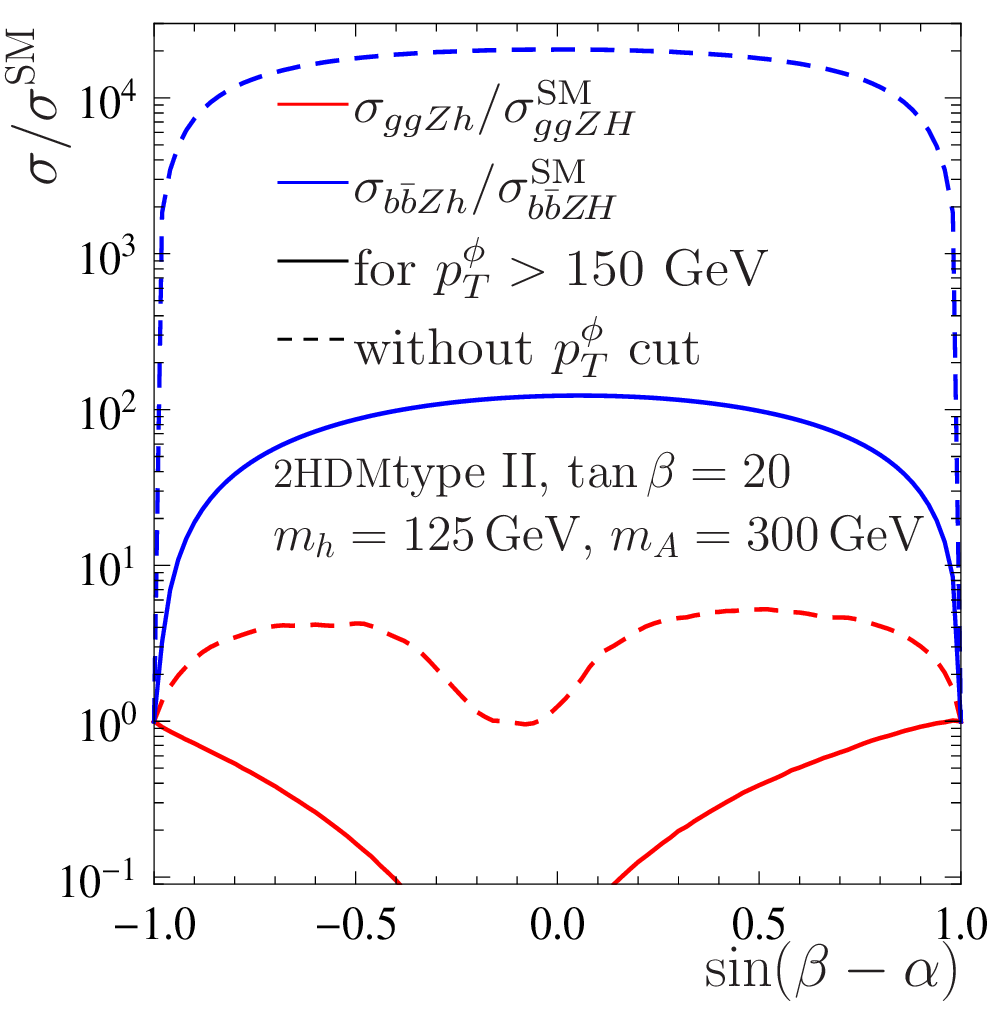} \\[-0.3cm]
(a) & (b) & (c)
\end{tabular}
\end{center}
\vspace{-0.6cm}
\caption[]{\label{fig:2HDMtype2ma300ptcut} (a--c) $\sigma_{\ggzh{}}/\sigma_{\ggzH{}}^{\sm}$ (red),
  $\sigma_{\bbzh{}}/\sigma_{\bbzH{}}^{\sm}$ (blue) 
  with $\ptphi>150$\,GeV (solid) and without $\ptphi$ cut (dashed)
  for $\sqrt{s}=14$\,TeV in the \thdm{} type\,II with $m_A=300$\,GeV as a function of
  $\sin(\beta-\alpha)$ using (a) $\tan\beta=1$, (b) $\tan\beta=5$ and (c) $\tan\beta=20$.}
\end{figure}

The largest impact of a restriction to large $\ptphi$, however, is
expected in scenarios where this cut removes effects from resonant
intermediate particles. In fact, this is what can be observed in
\figs{fig:2HDMtype1ma300ptcut} and \ref{fig:2HDMtype2ma300ptcut}, which
correspond to the parameter sets of \figs{fig:lighthiggsmA300_2T1} and
\ref{fig:lighthiggsmA300_2T2}, respectively, i.e.\ \thdm{} type\,I
and\,II with pseudoscalar mass $m_A=300$\,GeV. Without $\ptphi$ cut, the
pseudoscalar can become resonant, as discussed in
\sct{sec:2HDMdiscussion}. If the events are restricted to
$\ptphi>150$\,GeV, this is no longer the case.  The $gg\to \zh$ and
$b\bar b\to \zh$ fractions are thus diminished, in some cases by more
than two orders of magnitude. The missing resonant contribution also
explains why the respective solid curves for $m_A=200$\,GeV and
$m_A=300$\,GeV in \figs{fig:2HDMtype1ma200ptcut} and
\ref{fig:2HDMtype1ma300ptcut} as well as in
\figs{fig:2HDMtype2ma200ptcut} and \ref{fig:2HDMtype2ma300ptcut}
resemble each other.


Quite generally, we conclude from this discussion that a restriction to
the boosted regime may not be advantageous when searching for effects of
New Physics. A deterioration of the signal-to-background ratio in the
inclusive cross section may well be compensated by effects that are
otherwise cut away.


\newpage
\section{Conclusions}
\label{sec:conclusions}

We provided the \sm{} prediction for the ratio of the cross sections for
Higgs production in association with a $W$ and with a $Z$ boson,
including all available and numerically relevant theoretical
contributions: The Drell-Yan-like terms including \nnlo{} \qcd{} and
\nlo{} electro-weak corrections, the so-called top-loop induced terms,
as well as gluon-initiated effects through \nlo{}. The residual
theoretical uncertainty on this ratio is about 3\%, induced by missing
higher orders, variations in the \pdf{}s, and the experimental error on
$\alpha_s(m_Z)$.

The gluon-induced terms only affect $\zH$ production, and since they are
rather sensitive to New Physics effects, the $\wH/\zH$ ratio provides a
useful test of the Standard Model, once statistics of the collected data
is sufficiently large. As an example, we have considered a general
\thdm{} where it turns out that intermediate Higgs bosons can give a
particularly large contribution to $\zphi$ production. In addition,
bottom quark annihilation becomes numerically relevant and in some cases
even dominant. The consequence is a significant drop of the
$\wphi/\zphi$ ratio $\rwzh{\phi}$ as compared to its \sm{} value, even
close to the value $\sin(\beta-\alpha)=1$ which seems to be preferred by
the recent \lhc{} results.  This is even more so when the intermediate
Higgs can become resonant. Restriction to the ``boosted Higgs''-regime
carries the potential danger of cutting away such resonant
contributions; we therefore suggest to apply dedicated analyses also on
the inclusive cross sections.

There is a number of ways to extend our study, which we leave for future
investigations. Let us name just a few:
\begin{itemize}
\item In this paper, we focussed on total inclusive $\wphi$ and $\zphi$
  production, with only a rather qualitative consideration of the
  boosted-Higgs regime. Quite generally, restriction to particular
  kinematical regions or distributions may further improve the
  experimental and possibly also the theoretical significance of the
  effects observed here.
\item In a more detailed phenomenological analysis, the decay of the
  final state Higgs boson $\phi$ has to be folded in. This does not
  affect the quantity $\rwzh{\phi}$ itself, of course, but its
  experimental sensitivity. Note, however, that for $m_h=125$\,GeV, the
  branching ratio to $b\bar{b}$ is significant unless the bottom Yukawa
  coupling becomes exceptionally small. Therefore, this main decay mode
  should be accessible in most of the \thdm{} parameter space.
\item Clearly, the $\wphi/\zphi$ ratio should be studied also in other
  extensions of the \sm{}, for example in supersymmetric models. Due to
  the tight restrictions on the parameters of the Higgs sector, we
  expect the effects in the \mssm{} to be substantially reduced compared
  to our findings for the \thdm{} though.
\item It was found that the $\order{\alpha_s^3}$ effects to the gluon
  fusion contribution $\sigma_{\ggzH}$ in the \sm{} are quite
  substantial~\cite{Altenkamp:2012sx}. A precise prediction of the
  $\wphi/\zphi$ ratio in other models therefore requires the analogous
  corrections in these models. Similarly, the terms
  $\sigma_\text{I}^{\vphi}$ and $\sigma_\text{II}^{\zphi}$ could receive
  non-negligible contributions from bottom-quark loops if the bottom
  Yukawa coupling is large. The relevant calculations are
  non-trivial, however.
\end{itemize}

Finally, let us point out that a new release of the program \vhnnlo{}
will include all of the \thdm{} effects discussed in this paper and will
allow for a flexible calculation of the Higgs Strahlung cross section in
the \thdm{}. Similarly, for the gluon fusion process $gg\to \phi$ and
bottom quark annihilation $b\bar{b}\to \phi$, a link of
{\tt SusHi}~\cite{Harlander:2012pb} to {\tt 2HDMC} will soon be
available.  Since $t\bar t\phi$ production is typically suppressed in
\thdm{} models, and a rather good approximation for weak boson fusion is
obtained by a simple rescaling of the \sm{} cross section by
$g_{VV}^\phi$, all relevant cross section for Higgs production at the
\lhc{} within a \thdm{} will then be available.


\paragraph{Note added.}

While this paper was in the reviewing process which led us to include
\sct{sec:boost}, Ref.\,\cite{Englert:2013vua} appeared which
touches upon similar issues.


\paragraph{Acknowledgments}

We would like to thank the organizers and the participants of the
informal meetings during the last few months on \thdm{} Higgs production
which has inspired this work. In particular, this includes
S.\,Bolognesi, H.\,Haber, H.\,Logan, M.\,M\"uhlleitner, K.\,Peters,
N.\,Rompotis, O.\,St\aa{}l, and R.\,Tanaka.  In addition, we would like to
thank O.\,St\aa{}l for providing us with a version of {\tt 2HDMC} which
can be linked to {\tt SusHi} and \vhnnlo{}.\\
This work was supported by DFG, contract HA\,2990/5-1.


\begin{appendix}
 
\section{2-Higgs-Doublet Model}
\label{app:2HDM}
 
The 2-Higgs-Doublet Model (\thdm{}), which contains two Higgs doublets
named $H_1$ and $H_2$, can be cast in four types, if \cp{} conservation
and no tree-level flavor-changing neutral currents are demanded. They
differ by the Higgs-fermion Yukawa couplings. By convention, $H_2$
couples to the up-type quarks, whereas the couplings to the down-type
quarks as well as to the leptons depend on the type of \thdm{} and can
be taken from \tab{tab:2hdm}.  Our notation follows
\citeres{Gunion:1989we,Akeroyd:1996he,Akeroyd:1998ui,Aoki:2009ha,Branco:2011iw,Craig:2012vn}.

\begin{table}[htb]
\begin{center}
\begin{tabular}{| c || c | c | c | c |}
\hline
Model & Type I & Type II & Lepton-specific & Flipped\\\hline\hline
up-type quarks     & $H_2$ & $H_2$ & $H_2$ & $H_2$\\
down-type quarks   & $H_2$ & $H_1$ & $H_2$ & $H_1$\\
leptons	  	    & $H_2$ & $H_1$ & $H_1$ & $H_2$\\\hline
\end{tabular}
\end{center}
\vspace{-0.6cm}
\caption{\thdm{} types. In the main text of this paper, we only show
  results for ``type\,I'' and ``type\,II''; the results for
  ``Lepton-specific'' and ``Flipped'' are identical to these,
  respectively.}
\label{tab:2hdm}
\end{table}

Note that type\,I and Lepton-specific as well as type\,II and Flipped
\thdm{} only differ in their lepton couplings. The calculation of our
paper does not involve these couplings, though, which is why it is
sufficient to restrict ourselves to type\,I and type\,II.

Due to \cp{} conservation, the two Higgs doublets form two \cp{}-even
Higgs fields $h$ and $H^0$ and one \cp{}-odd field $A$. In the
``physical basis'', their masses, the ratio of the vacuum expectation
values $\tan\beta=v_2/v_1$, and the \cp{}-even Higgs mixing angle
$\alpha$ are considered as independent input parameters.  In addition,
two charged Higgs fields $H^\pm$ are formed.  The two angles $\alpha$
and $\beta$ determine the relative strength of all couplings with
respect to the \sm{} couplings.  We follow \citere{Gunion:1989we} and
present the Feynman rules required in our calculation:
{\allowdisplaybreaks
\begin{align}
&\parbox{25mm}{\includegraphics[height=0.15\textwidth]{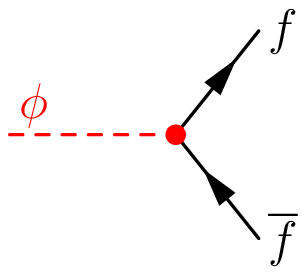}}=-\frac{m_f}{v}g_f^A\gamma_5,
\quad -i\frac{m_f}{v}g_f^\phi \quad \text{for}\quad \phi\in\lbrace h,H\rbrace\\
&\parbox{25mm}{\includegraphics[height=0.15\textwidth]{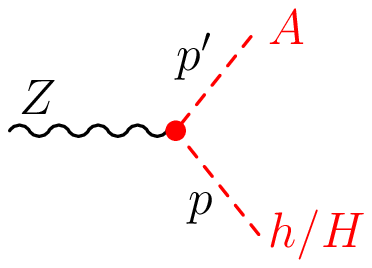}}=\frac{e}{2c_Ws_W}g_Z^{A\phi}(p-p')^\mu\\
&\parbox{25mm}{\includegraphics[height=0.15\textwidth]{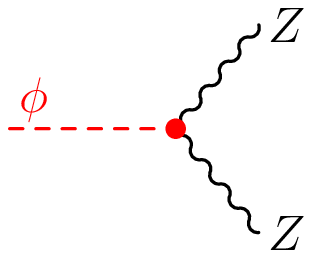}}=i\frac{em_Z}{c_Ws_W}g_{VV}^\phi g^{\mu\nu}\qquad
\parbox{25mm}{\includegraphics[height=0.15\textwidth]{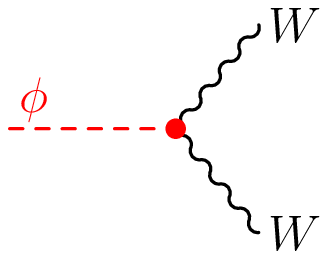}}=i\frac{em_W}{s_W}g_{VV}^\phi g^{\mu\nu}
\end{align}}

All momenta are considered as incoming, $s_W$ and $c_W$ denote the sine
and cosine of the weak mixing angle, $e$ the electromagnetic charge, and
$v=\sqrt{v_1^2+v_2^2}=1/\sqrt{\sqrt{2}G_F}$ the \sm{} vacuum expectation
value.  The relative couplings $g_f^\phi$ which enter the Yukawa
couplings to the quarks, differ in the four \thdm{} types and can be
taken from \tab{tab:2hdmcouplings}.
\begin{table}[htb]
\begin{center}
\begin{tabular}{| c || c | c | c | c |}
\hline
Model          & Type I & Type II & Lepton-specific & Flipped\\\hline\hline
$g_u^h$        & $\cos\alpha/\sin\beta$ & $\cos\alpha/\sin\beta$ & $\cos\alpha/\sin\beta$ & $\cos\alpha/\sin\beta$\\
$g_d^h$        & $\cos\alpha/\sin\beta$ & $-\sin\alpha/\cos\beta$ & $\cos\alpha/\sin\beta$ & $-\sin\alpha/\cos\beta$\\
$g_u^H$        & $\sin\alpha/\sin\beta$ & $\sin\alpha/\sin\beta$ & $\sin\alpha/\sin\beta$ & $\sin\alpha/\sin\beta$\\
$g_d^H$        & $\sin\alpha/\sin\beta$ & $\cos\alpha/\cos\beta$ & $\sin\alpha/\sin\beta$ & $\cos\alpha/\cos\beta$\\
$g_u^A$        & $\cot\beta$ & $\cot\beta$ & $\cot\beta$ & $\cot\beta$\\
$g_d^A$        & $-\cot\beta$ & $\tan\beta$ & $-\cot\beta$ & $\tan\beta$\\\hline
\end{tabular}
\end{center}
\vspace{-0.6cm}
\caption{Relative couplings $g_f^\phi$ with respect to the \sm{} Yukawa coupling for the four \thdm{} types.}
\label{tab:2hdmcouplings}
\end{table}

In contrast, the relative strength of the couplings to the heavy gauge
bosons are independent of the \thdm{} type. They are given by
\begin{align}
 g_{VV}^h = \sin(\beta-\alpha),\qquad g_{VV}^{H^0}=\cos(\beta-\alpha),\qquad g_{VV}^A = 0,
\end{align}
where $V$ represents one of the heavy gauge bosons $V\in\lbrace
W,Z\rbrace$, and by
\begin{align}
 g_{Z}^{Ah} = \cos(\beta-\alpha),\quad g_{Z}^{AH^0} = -\sin(\beta-\alpha)\,.
\end{align}
Obviously, the coupling of the light Higgs $h$ to the gauge bosons
equals the \sm{} coupling for $\sin(\beta-\alpha)=1$, where also the
coupling to the pseudoscalar Higgs vanishes.  For convenience of the
reader, we add the relative Yukawa couplings in terms of
$\sin(\beta-\alpha)$:
\begin{align}
 \frac{\cos\alpha}{\sin\beta}&=\sin(\beta-\alpha)+\cos(\beta-\alpha)\frac{1}{\tan\beta},\quad
 \frac{\sin\alpha}{\sin\beta}=-\sin(\beta-\alpha)\frac{1}{\tan\beta}+\cos(\beta-\alpha)\\
 \frac{\cos\alpha}{\cos\beta}&=\sin(\beta-\alpha)\tan\beta+\cos(\beta-\alpha),\quad
 \frac{\sin\alpha}{\cos\beta}=-\sin(\beta-\alpha)+\cos(\beta-\alpha)\tan\beta.
\end{align}


\end{appendix}


\def\app#1#2#3{{\it Act.~Phys.~Pol.~}\jref{\bf B #1}{#2}{#3}}
\def\apa#1#2#3{{\it Act.~Phys.~Austr.~}\jref{\bf#1}{#2}{#3}}
\def\annphys#1#2#3{{\it Ann.~Phys.~}\jref{\bf #1}{#2}{#3}}
\def\cmp#1#2#3{{\it Comm.~Math.~Phys.~}\jref{\bf #1}{#2}{#3}}
\def\cpc#1#2#3{{\it Comp.~Phys.~Commun.~}\jref{\bf #1}{#2}{#3}}
\def\epjc#1#2#3{{\it Eur.\ Phys.\ J.\ }\jref{\bf C #1}{#2}{#3}}
\def\fortp#1#2#3{{\it Fortschr.~Phys.~}\jref{\bf#1}{#2}{#3}}
\def\ijmpc#1#2#3{{\it Int.~J.~Mod.~Phys.~}\jref{\bf C #1}{#2}{#3}}
\def\ijmpa#1#2#3{{\it Int.~J.~Mod.~Phys.~}\jref{\bf A #1}{#2}{#3}}
\def\jcp#1#2#3{{\it J.~Comp.~Phys.~}\jref{\bf #1}{#2}{#3}}
\def\jetp#1#2#3{{\it JETP~Lett.~}\jref{\bf #1}{#2}{#3}}
\def\jphysg#1#2#3{{\small\it J.~Phys.~G~}\jref{\bf #1}{#2}{#3}}
\def\jhep#1#2#3{{\small\it JHEP~}\jref{\bf #1}{#2}{#3}}
\def\mpl#1#2#3{{\it Mod.~Phys.~Lett.~}\jref{\bf A #1}{#2}{#3}}
\def\nima#1#2#3{{\it Nucl.~Inst.~Meth.~}\jref{\bf A #1}{#2}{#3}}
\def\npb#1#2#3{{\it Nucl.~Phys.~}\jref{\bf B #1}{#2}{#3}}
\def\nca#1#2#3{{\it Nuovo~Cim.~}\jref{\bf #1A}{#2}{#3}}
\def\plb#1#2#3{{\it Phys.~Lett.~}\jref{\bf B #1}{#2}{#3}}
\def\prc#1#2#3{{\it Phys.~Reports }\jref{\bf #1}{#2}{#3}}
\def\prd#1#2#3{{\it Phys.~Rev.~}\jref{\bf D #1}{#2}{#3}}
\def\pR#1#2#3{{\it Phys.~Rev.~}\jref{\bf #1}{#2}{#3}}
\def\prl#1#2#3{{\it Phys.~Rev.~Lett.~}\jref{\bf #1}{#2}{#3}}
\def\pr#1#2#3{{\it Phys.~Reports }\jref{\bf #1}{#2}{#3}}
\def\ptp#1#2#3{{\it Prog.~Theor.~Phys.~}\jref{\bf #1}{#2}{#3}}
\def\ppnp#1#2#3{{\it Prog.~Part.~Nucl.~Phys.~}\jref{\bf #1}{#2}{#3}}
\def\rmp#1#2#3{{\it Rev.~Mod.~Phys.~}\jref{\bf #1}{#2}{#3}}
\def\sovnp#1#2#3{{\it Sov.~J.~Nucl.~Phys.~}\jref{\bf #1}{#2}{#3}}
\def\sovus#1#2#3{{\it Sov.~Phys.~Usp.~}\jref{\bf #1}{#2}{#3}}
\def\tmf#1#2#3{{\it Teor.~Mat.~Fiz.~}\jref{\bf #1}{#2}{#3}}
\def\tmp#1#2#3{{\it Theor.~Math.~Phys.~}\jref{\bf #1}{#2}{#3}}
\def\yadfiz#1#2#3{{\it Yad.~Fiz.~}\jref{\bf #1}{#2}{#3}}
\def\zpc#1#2#3{{\it Z.~Phys.~}\jref{\bf C #1}{#2}{#3}}
\def\ibid#1#2#3{{ibid.~}\jref{\bf #1}{#2}{#3}}
\def\otherjournal#1#2#3#4{{\it #1}\jref{\bf #2}{#3}{#4}}
\newcommand{\jref}[3]{{\bf #1}, #3 (#2)}
\newcommand{\hepph}[1]{[hep-ph/#1]}
\newcommand{\mathph}[1]{[math-ph/#1]}
\newcommand{\arxiv}[2]{[arXiv:#1]}
\newcommand{\bibentry}[4]{#1, ``#2'', #3\ifthenelse{\equal{#4}{}}{}{ }#4.}


\end{document}